\documentclass[
superscriptaddress,
reprint,
nofootinbib,
 amsmath,amssymb,
aps,
prx,
]{revtex4-1}

\usepackage{natbib}
\usepackage{graphicx}
\usepackage{dcolumn}
\usepackage{bm}

\usepackage{enumerate}
\usepackage{graphicx}
\usepackage{dcolumn}
\usepackage{bm}
\usepackage{microtype}
\usepackage{xcolor}
\definecolor{navyblue}{rgb}{0.0, 0.0, 0.5}
\usepackage{hyperref}
\hypersetup{
    colorlinks=true,
    linktoc=all,
    allcolors=navyblue
}

\usepackage[capitalize]{cleveref}
\crefname{section}{Sec.}{Secs.}
\Crefname{section}{Sec.}{Secs.}
\crefname{appendix}{App.}{Apps.}
\Crefname{appendix}{App.}{Apps.}
\Crefname{figure}{Fig.}{Figs.}
\crefname{figure}{Fig.}{Figs.}
\creflabelformat{equation}{(#2#1#3)}

\usepackage{color,environ}
\usepackage{fontawesome5}
\definecolor{orcidlogocol}{rgb}{0.65, 0.807, 0.223}
\newcommand{\orcid}[1]{$\,$\href{https://orcid.org/#1}{\textcolor{orcidlogocol}{\footnotesize\faOrcid}}}
\usepackage{etaremune}
\usepackage{comment}

\usepackage{xcolor}

\usepackage{cleveref} 

\usepackage{cleveref}

\newcommand{\GeV}{\mathrm{GeV}}
\newcommand{\MeV}{\mathrm{MeV}}

\newcommand{\eV}{\mathrm{eV}}

\newcommand{\mpl}{m_\mathrm{pl}}

\newcommand{\kpc}{\mathrm{kpc}}
\newcommand{\Mpc}{\mathrm{Mpc}}
\newcommand{\Gyr}{ \mathrm{Gyr}}
\newcommand{\Myr}{ \mathrm{Myr}}
\newcommand{\Hz}{ \mathrm{Hz}}
\newcommand{\cm}{\mathrm{cm}}
\newcommand{\km}{\mathrm{km}}
\newcommand{\s}{\mathrm{s}}
\newcommand{\G}{\mathrm{G}}
\newcommand{\kg}{\mathrm{kg}}
\newcommand{\pc}{\mathrm{pc}}

\newcommand\blfootnote[1]{%
  \begingroup
  \renewcommand\thefootnote{}\footnote{\hspace{-1.5em}#1}%
  \addtocounter{footnote}{-1}%
  \endgroup
}
\begin{document}

\title{Black hole scalar sirens in the Milky Way}
\author{Daniel Gavilan-Martin}
\email{gaviland@uni-mainz.de}
\affiliation{Johannes Gutenberg-Universit\"at Mainz, 55128 Mainz, Germany}
\affiliation{Helmholtz Institute Mainz, 55099 Mainz, Germany} 
\affiliation{GSI Helmholtzzentrum für Schwerionenforschung GmbH, 64291 Darmstadt, Germany}
\author{Olivier Simon}
\email{osimon@princeton.edu}
\affiliation{Princeton Center for Theoretical Science, Princeton University, Princeton, NJ 08544, USA}
\affiliation{Department of Physics, Princeton University, Princeton, NJ 08544, USA}

\author{Dhashin Krishna}
\affiliation{Johannes Gutenberg-Universit\"at Mainz, 55128 Mainz, Germany}
\affiliation{Helmholtz Institute Mainz, 55099 Mainz, Germany}

\author{Derek F. Jackson Kimball}
\affiliation{Department of Physics, California State University – East Bay, Hayward, CA 94542, USA}

\author{Dmitry Budker}
\affiliation{Johannes Gutenberg-Universit\"at Mainz, 55128 Mainz, Germany}
\affiliation{Helmholtz Institute Mainz, 55099 Mainz, Germany}
\affiliation{GSI Helmholtzzentrum für Schwerionenforschung GmbH, 64291 Darmstadt, Germany}
\affiliation{Department of Physics, University of California, Berkeley, CA 94720, USA}
 
\author{Arne Wickenbrock}
\affiliation{Johannes Gutenberg-Universit\"at Mainz, 55128 Mainz, Germany}
\affiliation{Helmholtz Institute Mainz, 55099 Mainz, Germany}
\affiliation{GSI Helmholtzzentrum für Schwerionenforschung GmbH, 64291 Darmstadt, Germany}
\blfootnote{$^{*,\dagger}$\,These authors contributed equally to this work.}
\begin{abstract}
Hypothetical light scalar particles trigger the superradiant instability around spinning black holes (BHs), causing clouds of scalars to grow around the BH. In the presence of sufficiently strong particle self-interactions (characterized by the decay constant $f$), scalars are ejected from BH orbits, resulting in coherent, non-relativistic emissions that continuously carry away the BH's angular momentum. Parameters exist for which cloud growth is much faster, and scalar depletion is much slower, than the age of the Galaxy. This defines a distinct class of astrophysical sources of scalars, which we call \emph{BH scalar sirens} -- BHs that persistently emit scalars effectively forever. We compute the scalar background from  the expected population of $N_\text{BH}\sim 10^{8}$  isolated stellar-mass BHs in the Milky Way, which are sirens for scalars in the mass range $10^{-13}$--$10^{-11}\,\eV$ and $f\lesssim 10^{14}$--$10^{9}\,\GeV$. This provides a detection target independent of early-universe scalar production or cosmological initial conditions. The generated observable signals are up to two orders-of-magnitude larger than those expected from a misaligned cosmic scalar in this mass range. The energy spectrum of emitted scalars is distinctly broader and at higher velocities (up to $\sim 10^{-1}c$) than that of virialized dark matter, and encodes the mass and spin distributions of the BH population. While stellar-mass Milky Way BHs are our primary target, our framework extends to supermassive, intermediate-mass and light BHs. Given the difficulty of directly observing populations of isolated BHs, scalar emissions offer a novel probe of these otherwise invisible objects, highlighting the potential for joint discovery between scalars and BHs, and broadly motivating searches for scalars over many orders-of-magnitude in mass.
\end{abstract}

\maketitle

\tableofcontents

\section{Introduction}
Over the last decade, the observable implications of black hole superradiance (BHSR) \cite{Zeldovich1971,Zeldovich1972,Starobinskii:1973vzb,Detweiler:1980uk,Zouros:1979iw,Press:1972zz,Teukolsky:1973,Misner:1972kx,Bekenstein:1998nt} have emerged as compelling avenues for probing new bosonic particles beyond the Standard Model (BSM) \cite{Arvanitaki:2010sy,Arvanitaki:2014wva,Dolan:2012yt,Rosa:2011my,Pani:2012bp,Pani:2012vp,East:2017mrj,Baryakhtar:2017ngi,Baryakhtar:2020gao,Siemonsen:2022ivj,Siemonsen:2019ebd,Hoof:2022xbe,Hoof:2024quk,Mehta:2021pwf}. In particular, a wide range of new scalar and pseudoscalar\footnote{Unless explicitly stated, scalar refers to both pseudoscalar and scalar particles.} particles are well motivated on theoretical grounds \cite{Peccei:1977hh,Weinberg:1977ma,Wilczek:1977pj,Dimopoulos:1996kp,Svrcek:2006yi,Arvanitaki:2009fg,
Witten:1984dg,Ringwald:2014vqa,marsh_axion_2016,Hui:2016ltb, Hui:2021tkt,preskill_cosmology_1983,dine_not-so-harmless_1983,abbott_cosmological_1983,Linde:1987bx}, and if their masses fall within one of two ultralight windows, $10^{-13}-10^{-11}\,\eV$ or $10^{-19}-10^{-16}\,\eV$, they are expected to trigger the BHSR process for stellar-mass and supermassive black holes (BHs), respectively \cite{Arvanitaki:2010sy,Arvanitaki:2014wva}. During the BHSR process, the angular momentum of the BH is converted into the angular momentum of new excitations of the scalar field. In this way, BHs can act simultaneously as factories and detectors for such new scalar particles.

Since superradiance relies solely on the gravitational coupling of the new scalar field to the BH, the earliest studies of BHSR for the particle phenomenology of BSM scalars focused on the signatures expected in this minimal, purely gravitational framework. 
Extensive efforts have since developed primarily along three directions: 1) constraining the parameter space of scalar particles from the measured spin of astrophysical BHs \cite{Arvanitaki:2010sy,Arvanitaki:2014wva, Baryakhtar:2020gao,Stott:2018opm,Fernandez:2019qbj,Ng:2019jsx,Mehta:2021pwf,Hoof:2024quk, Witte:2024drg, Caputo:2025oap}, 2) evaluating prospects for directly detecting the gravitational waves (GWs) emitted by the cloud of scalar particles which builds up around the BH during the BHSR process \cite{Arvanitaki:2014wva,Zhu:2020tht,Collaviti:2024mvh,Mirasola:2025car,Brito:2017zvb,Brito:2017wnc,Yuan:2021ebu,Omiya:2024xlz}, and 3) characterizing and evaluating prospects for detecting the imprint of the cloud on the GW emitted during binary inspirals \cite{Baumann:2018vus,Baumann:2019ztm,Baumann:2022pkl,Tomaselli:2024bdd,Tomaselli:2024dbw,Boskovic:2024fga,Boskovic:2025ixx,Tomaselli:2024faa,Zhang:2019eid,Takahashi:2024fyq}. However, these phenomenological signatures are suppressed when the new scalar has significant particle interactions beyond gravity.

The well-motivated, minimal form of interaction for new BSM scalars is a quartic self-interaction \cite{GrillidiCortona:2015jxo,Reece:2023czb}.  Building on the toy-model exploration of \cite{Gruzinov:2016hcq}, the authors of \cite{Baryakhtar:2020gao} performed the first systematic analysis of how quartic self-interactions modify the dynamics of BHSR. In the presence of sufficiently large self-couplings, a new phenomenon emerges: the black hole emits nearly monochromatic, non-relativistic scalar waves that play a dominant role in the evolution of the system. These emissions substantially slow the rate of angular-momentum extraction from the BH and suppress the amplitude of gravitational-wave signals from the superradiant cloud. Thus, while strong self-interactions tend to quench the emission of gravitational waves, they open up a new observational channel through the direct detection of the emitted scalars, thereby allowing strongly self-interacting BSM scalars to evade existing GW-based detection.

While \cite{Baryakhtar:2020gao} quantitatively situates the afforementioned monochromatic scalar emission regime within the broader landscape of BHSR phenomenology by identifying the self-interaction strength (equivalently, the decay constant $f$ of the scalar) at which quartic effects begin to modify the standard picture, most subsequent studies have restricted themselves to model parameters \emph{below} that threshold, precisely because this is where the familiar observational effects, e.g., constraints from BH spin measurements, are obtained.

In contrast, in this paper, we focus on the regime well \emph{above} this threshold, that of \emph{large} self-interactions (i.e.\,$f$ \emph{smaller} than some threshold value $f_\text{siren}$ defined later in the text). Building on the analysis in \cite{Baryakhtar:2020gao}, we identify, define and characterize a new class of long-lived astrophysical emitters of scalar radiation, which we call BH scalar sirens. From the characterization of isolated individual sirens, we show how to compute ensemble signals from a population of BH scalar sirens. We ground our discussion by computing the ensemble signal for the large ($N_\text{BH}\sim 10^{8}-10^9$) population of ancient ($\mathcal O(1-10\,\Gyr)$) stellar-mass BHs expected towards the Galactic Center (GC) of the Milky Way 
\cite{Olejak:2019pln,Timmes:1995kp,Shapiro:1983du,brown12003scenario,Wiktorowicz:2019dil,Bambi:2025rod}.

\begin{figure}[htb]
    \centering
        \includegraphics[width=0.5\textwidth]{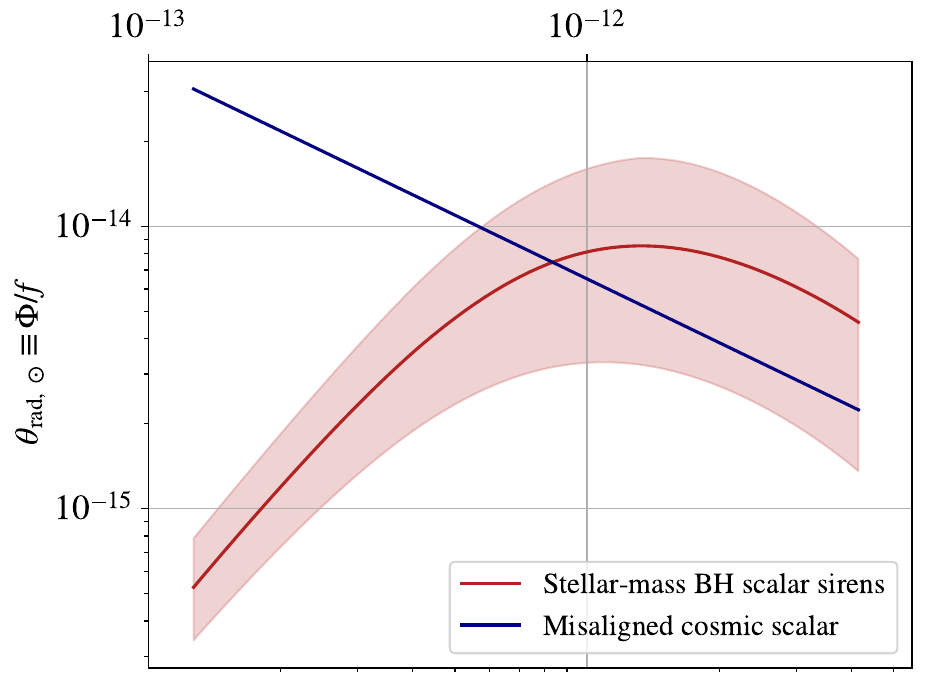}
        \includegraphics[width=0.5\textwidth]{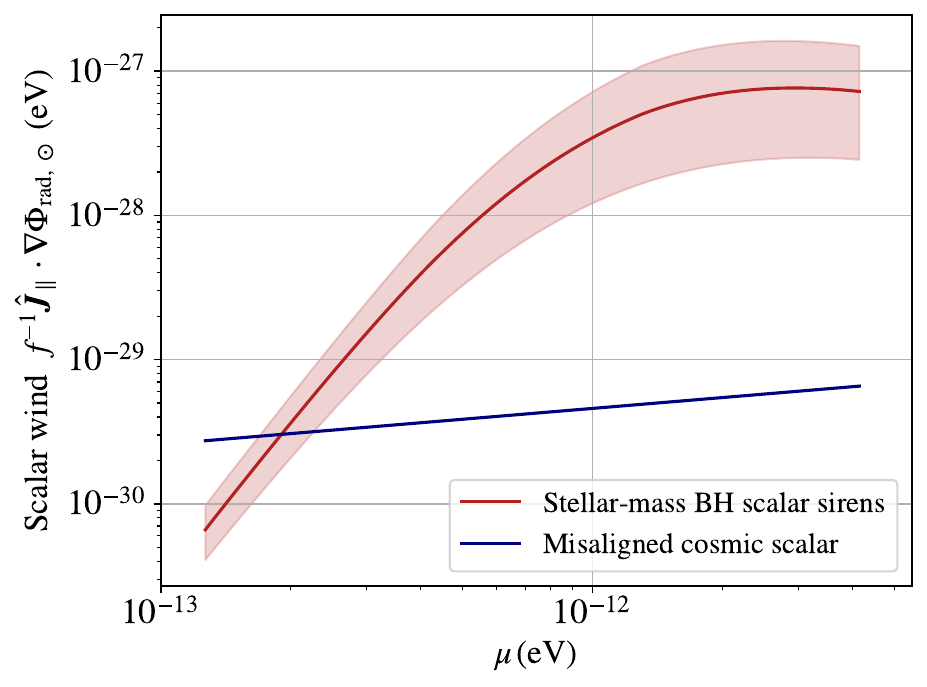}

        \caption{Amplitude at the Sun-Earth location of the two local scalar observables most relevant to direct experimental detection: the dimensionless angle $\theta \equiv \Phi/f$ (top), and the scalar wind $f^{-1}\hat{\bm{J}}_\parallel\cdot \bm \nabla \Phi$ (anti)parallel to the line-of-sight to the Galactic Center (bottom), for a population ($N_\text{BH} = 10^8$) of stellar-mass BH scalar sirens in the Milky Way, assumed to be exponentially distributed with characteristic spread $R_s=3\, \kpc$ from the Galactic Center.
        Due to the increased velocity of the scalars, the wind from the MW stellar-mass BH siren population is up to two orders of magnitude larger than that of a scalar relic produced through a pre-inflationary misalignment mechanism, \cref{eq:misalignment}, with initial misalignment angle $\theta_i \approx 1$.
        We assume a characteristic BH mass scale $M_s=9.5\,\mathrm{M}_\odot$ and a unitless BH spin parameters $a_\star$ distributed according to a power law with ``pitch'' $\beta$ (Eq.\,\ref{eq:spindistro}). The solid red line assumes $\beta=0.65$, the best fit to existing spin measurements of stellar-mass BHs \cite{Draghis:2023vzj}; the translucent red region spans from  $\beta=0$ to $\beta=0.9$.
        }
    \label{fig:sirensvsrelic}
\end{figure}

Scalars emitted from BHs provide an alternative target for the vast experimental program underway to look for a local abundance of non-relativistic ultralight scalars with primordial origin \cite{JacksonKimball:2017elr,Garcon:2019inh,Kimball:2023vxk,Garcon:2017ixh,Jiang:2021dby,Walter:2025ldb,Gavilan-Martin:2024nlo,GNOME:2023rpz,Safronova:2017xyt,Tretiak:2022ndx,Bloch:2019lcy,Bloch:2023wfz,Sikivie:1983ip,ADMX:2019uok,ADMX:2018gho,Sushkov:2023fjw,DMRadio:2022pkf,Caldwell:2016dcw, Millar:2016cjp,Baryakhtar:2018doz,Arvanitaki:2024taq}. Importantly, this is an experimental target even if the scalar is only a trace component of the cosmic\footnote{In this work, we separate primordial (or ``cosmic'') dark matter originating from \emph{before} the time of galaxy formation and still present today as the primary constituent of galactic halos, from dark matter that is merely \emph{astrophysical in origin}, present today in galaxies, but produced inside galaxies, \emph{after} their formation.} mass-energy budget (\cref{fig:sirensvsrelic}). Additionally, we compare the expected size of fluctuations in local scalar observables due to BH scalar sirens to the conservative expectation of a cosmically misaligned scalar relic. Because the local energy density from both sources scale $\propto f^2$, we estimate that the scalar background expected from stellar-mass BH sirens in the MW can be significantly more observable than that associated with a cosmic relic in the corresponding mass range for equal values of $f<f_\text{siren}$. 

Haloscope experiments typically target the new scalar as the totality or a subcomponent of primordial dark matter (DM), and therefore search for signals whose lineshape correspond to a virialized (locally Maxwell-Boltzmann) population of scalar particles moving at the virial velocity $\sim 300\,\km/\s.$  In contrast, scalars from BH sirens are also non-relativistic, but travel at speeds up to two orders-of-magnitude greater. Because the velocity of the scalars emitted by BH sirens is set by the BH mass, the velocity fluctuations in the observed flux of scalars closely mirror the spread of BH masses in the population (Sec.\,\ref{sec:self-interactions}). Therefore, while virialization under gravity has long since scrambled the phase space of scalars with primordial, early-universe origins, the lineshape of scalars from BH sirens observed today is directly tied to the mass spectrum of the progenitor BH population.
 Beyond predicting the spectral lineshape, the same ensemble formalism determines the spatial distribution of the scalar energy density in the Milky Way: the signal traces the underlying BH population and is therefore strongly concentrated toward the Galactic Center while remaining measurable at the Sun's location. 
Summing over the $\sim 10^8 - 10^9$ ancient stellar-mass BHs then yields a collective, narrow-band ``scalar background'' whose magnitude can be quantified both in the field amplitude and in its spatial gradient. 

Especially interesting for experimental searches is the corresponding scalar-wind observable: for derivative couplings it is related to the momentum flux and is effectively velocity-enhanced (as compared to virialized scalar dark matter), while its directionality (predominantly pointing back to the Galactic Center with a calculable angular spread) provides a powerful discriminator and predicts a characteristic daily modulation as the Earth rotates. 
To connect these signatures to realistic Galactic substructure, we complement analytic population models with results from a state-of-the-art synthetic BH catalog \cite{Olejak:2019pln}. A summary recasting of existing limits on scalar DM from existing haloscope experiments suggests sensitivity to scalar-to-nucleon-spin couplings as small as $\sim 3 \times 10^{-7}\,\GeV^{-1}$ for $f \sim 10^{14} \,\GeV$. While this corresponds to scalars with exceptionally large couplings to the Standard Model (SM), near-future experiments should reach below astrophysically excluded parameter space, and within only a few orders-of-magnitude of conservative model-building parameter space (SM couplings $\sim f^{-1}$).

A detection of the scalar background from stellar-mass MW BHs would not only constitute the discovery of a new scalar, but also inform our knowledge of the underlying BH population. While expected stellar-mass sirens are the most conservative targets, our formalism extends, with caveats, to supermassive (SMBHs) and intermediate-mass (IMBHs), and light BHs in the MW, which would source lighter and heavier scalars, respectively. We demonstrate that the way local observables from BH sirens scale with the characteristic BH mass of the target population naturally motivates a mass-dependent detection strategy. Within the context of models predicting a dense spectrum of fundamental scalar masses, such as string axiverse constructions \cite{Arvanitaki:2009fg,Demirtas:2021gsq,Gendler:2023kjt}, BH scalar siren phenomenology represents exciting prospects for joint BH-scalar phenomenology.

This work is structured as follows. First, in \cref{sec:sirens}, we review the BHSR of self-interacting scalar particles, particularly in the pertinent regime of large self-interactions. We define and characterize the BH scalar siren in its rest frame. Then, in \cref{sec:MW sirens}, we explain how to compute the ensemble signal for a population of such sirens. While much of the formalism for computing ensemble signals remains general, we ground the analysis by explicitly evaluating the (power) spectral density of scalars from the population of ancient stellar-mass BHs expected at the GC, considering both the magnitude of the field and its spatial gradient. In \cref{sec:observational}, we discuss the observational implications of populations of BH sirens, in terms of Galactic energy density profile and potential signatures at haloscope experiments in the presence of couplings to Standard-Model (SM) particles at low energies. We conclude in \cref{sec:conclusion} with a summary of our findings and a commentary on their implication for the joint phenomenology of scalars and BHs.

In this work, $G$ denotes the gravitational constant and $\mpl = \sqrt{1/G} \approx 1.22\times 10^{19}\,\GeV$ is the Planck mass.  The solar mass unit is $M_\odot \approx 2 \times 10^{30}\,\kg$. Unless otherwise specified, we use natural units $\hbar = c = 1.$ 

Because our analysis naturally spans both astrophysical and microphysical regimes, we report dimensionful quantities in the time or energy units most appropriate to the context: the cosmological/Galactic scale ($\Gyr$), laboratory frequencies ($\Hz$), particle-physics energies ($\eV$), or the astronomical black-hole scale  ($GM_\odot \approx 5\times 10^{-6}\,\s \approx 1.5\,\km$), in the way that most facilitates relevant comparisons. 

At some points of the analysis, it is important to distinguish between three kinds of reference frames: an individual BH's rest frame, an observer's rest frame, and the Galactocentric rest frame (hereafter Galactic frame). When this distinction is relevant, we use subscripts to indicate the object being designated, and superscripts to designate the frame. For example, $\bm v_\text{o}^\text{G}$ is the velocity vector of the observer in Galactic frame coordinates.

\section{Scalar sirens from black hole superradiance}
\label{sec:sirens}

In this section, we review how the BHSR of ultralight scalar fields can generate long-lived, compact clouds of scalar particles around rotating astrophysical black holes (BHs). For scalars endowed with suitably large self-interactions, these superradiant (SR) clouds continuously emit coherent scalar radiation escaping the BH. We refer to such systems as black hole scalar sirens: a distinct class of long-lived astrophysical sources of coherent scalar emission.

We begin by summarizing the essentials of the BHSR process in Sec.\,\ref{sec:BH-superradiance}, with emphasis on scenarios involving light scalar fields with quartic self-interactions (Sec.\,\ref{sec:self-interactions}), and explain how black hole scalar sirens arise. We then characterize an individual siren and its scalar emission in its rest frame in Sec.\,\ref{sec:sirens_signal}.

\subsection{Gravitational black hole superradiance}
\label{sec:BH-superradiance}

BHSR, broadly defined, is a process by which rotating (i.e.,\,Kerr) BHs enhance the amplitude of incident \emph{bosonic} fields \cite{Zeldovich1971,Zeldovich1972,Starobinskii:1973vzb,Bekenstein:1998nt,Brito:2015oca,Press:1972zz,Bekenstein:1998nt}. The Pauli exclusion principle ultimately forbids BHSR of fermionic radiation, at least in minimal settings \cite{Unruh:1973bda,Unruh:1974bw,Iyer:1978du,Martellini:1977qf,Kim:1997hy,Dai:2023zcj}.  As such, the BHSR process applies to the scattering of both massless and massive bosonic fields. However, only massive bosonic fields can be trapped around the BH. When it is possible to confine the field near the BH environment, the BHSR process can be spontaneously triggered, leading to repeated amplifications and the efficient growth of a field profile around the BH, referred to as the \emph{superradiant (SR) cloud} \cite{Press:1972zz,Zouros:1979iw,Detweiler:1980uk,Arvanitaki:2010sy}. 

Confinement of a bosonic field near a BH occurs most minimally through gravitational effects, when the associated field particle (i.e.,\,the quantum of the field) has a small, but non-zero mass\footnote{There is no stable, massive, fundamental boson in the Standard Model. The lightest massive \emph{composite} SM bosons are the pions. The BHSR process applied to pions was considered as early as \cite{Detweiler:1980uk}. While pions are unstable, the SR rate may still be fast relative to the pion lifetime, leading to the phenomenological possibility of observing pion clouds around putative light (primordial) BHs with sizes comparable to $m_\pi^{-1}\sim (100 \,\MeV)^{-1}$ (see e.g.\,\cite{Ferraz:2020zgi}). Crucially, this is only possible because the QCD confinement scale $\Lambda_\text{QCD}\sim 300\,\MeV$ is slightly \emph{larger} than $m_\pi$, such that pions can be considered composite on the scale of the corresponding BH. This is \emph{not true} of other composite bosons in the SM, namely some bosonic nuclei and atomic states, whose scale of compositeness (the nuclear Fermi or atomic Bohr scales) is \emph{smaller} than their masses.} $\mu$. In this case, the spacetime far enough outside the BH is well approximated by a Newtonian gravitational potential
\begin{equation}
\label{eq:newton_potential}
U_\text{G}(r) \approx - \frac{\alpha}{r},
\end{equation}
where 
\begin{align}
\begin{split}
\label{eq:alpha}
\alpha \equiv{} G \mu M_\text{BH} ={} \frac{\mu}{\mpl}\frac{M_\text{BH}}{\mpl}
\end{split}
\end{align}
is akin to a dimensionless \emph{gravitational fine-structure constant} characterizing the system. The potential \cref{eq:newton_potential} confines the particles to orbital motion around the BH, thereby enabling the BHSR process.

More quantitively, when non-gravitational dynamics are negligible, a massive scalar field $\phi$ obeys the covariant Klein-Gordon equation
\begin{align}
\begin{split}
0={}& \left(g^{\sigma\nu}\nabla_\sigma \nabla_\nu + \mu^2\right)\phi \\
\approx{}&\left(\frac{\partial}{\partial t^2} - \bm\nabla^2 +2\mu U_\text{G}(r) +\mu^2\right)\phi,
\end{split}
\end{align}
where $\nabla_\nu$, $\nabla_\sigma$ designates the covariant derivatives and $g^{\sigma\nu}$ designates the metric tensor, and $\bm\nabla^2$ is the Laplace operator. For $g^{\sigma\nu}$ corresponding to the spacetime of a rotating BH, this equation admits solutions dominated by non-relativistic motion in the part of the BH's gravitational potential well approximated by the Newtonian potential of \cref{eq:newton_potential}. 
Solving for the scalar field modes in this regime
approximates closely the problem of finding the non-relativistic quantum mechanical wavefunctions of a particle in a $1/r$ potential. For this reason, the term ``gravitational atom'' is sometimes used to refer to the BH-SR-cloud system \cite{Arvanitaki:2010sy,Arvanitaki:2014wva}. One finds that the spectrum of bound waves approximately matches the spectral pattern of hydrogenic wavefunctions and is therefore labeled by the usual hydrogenic spectral quantum numbers $n$, $\ell$, and $m$, namely the principal, azimuthal, and ``magnetic'' quantum numbers, respectively. Similarly, the eigenfrequencies of the bound states are approximately \cite{Detweiler:1980uk,Arvanitaki:2010sy,Dolan:2012yt,Arvanitaki:2014wva,Baumann:2019eav} $\omega_{n\ell m} \approx E_n + i\Gamma^\text{SR}_{n\ell m}/2$, 
\begin{equation}
\label{eq:spectrum}
    E_n \approx \mu \left(1-\frac{\alpha^2}{2n^2}\right).
\end{equation}
 Crucially, $\Gamma^\text{SR}_{n\ell m}$ designates the small imaginary part of the frequency which arises because of \emph{near} horizon effects, namely the imposition of strictly ingoing boundary conditions \cite{Detweiler:1980uk,Bekenstein:1998nt,Arvanitaki:2010sy,Dolan:2012yt,Brito:2015oca,Teukolsky:1973}. Importantly,  
\begin{equation}
\label{eq:SR_condition}
    \text{sign}\left(\Gamma^\text{SR}_{n\ell m}\right) = \text{sign}\left(m \Omega_\text{BH}-E_n\right),
\end{equation}
where $\Omega_\text{BH}$ is a parameter of the Kerr BH metric which can be roughly thought of as the ``angular velocity'' of the BH. The limit $\Omega_\text{BH} \rightarrow 0$ of a non-rotating BH (i.e.\,Schwarzschild) geometry is intuitive: the imaginary part is always negative because a non-rotating BH strictly \emph{damps} any oscillation with spatial support at the horizon, and can only \emph{absorb} incident energy. The defining, remarkable feature of BHSR is that, when $\Omega_\text{BH}$ is non-zero, the condition $E_n < m\Omega_\text{BH}$ may be realized for some modes, which then experience spontaneous, exponential \emph{growth}, rather than damping. Equation \eqref{eq:SR_condition} thus defines the so-called \emph{superradiance condition} \cite{Zeldovich1971,Zeldovich1972,Detweiler:1980uk,Arvanitaki:2010sy,Arvanitaki:2014wva,Brito:2015oca,Press:1972zz,Teukolsky:1973}.

While the sign condition of \cref{eq:SR_condition} dictates whether SR growth is possible for a bound state or not, the magnitude of the instability rate $|\Gamma^\text{SR}_{n\ell m}|$ grows with the spatial overlap between the bound state wavefunction and the BH. The parameter $\alpha$ dictates the degree of gravitational binding, as is evident either from the expression for the gravitational binding energy $\sim \mu\alpha^2/2n^2$ or the spatial extent of the ``Bohr'' radius of the gravitational atom $r_\text{Bohr}^{-1} \propto \alpha \mu$.  A larger $\alpha$ leads to a more spatially compact cloud, closer to the BH, and therefore results in larger SR instability rates. 

From Eq.\,\eqref{eq:SR_condition}, the \textit{superradiant condition} can be written as
\begin{equation}
\label{eq:SR_condition_2}
m a_\star - 2\alpha \left(1+\sqrt{1-a_\star^2}\right) > 0\,,
\end{equation}
where 
\begin{equation}
    a_\star = J/(GM_\text{BH}^2) \in [0,1]
    \label{eq:astar definition}
\end{equation}
is the dimensionless Kerr parameter characterizing the BH's angular momentum $J$. Thus, achieving the SR conditions limits the size of $\alpha$ for which a BH of fixed spin $a_\star$ can undergo SR. Equivalently, only BHs with spins larger than 
\begin{equation}
\label{eq:a crit}
a_\text{crit}(\alpha) = \frac{4(\alpha/m)}{1+4(\alpha/m)^2}\,.
\end{equation}
have superradiant $m$ levels.

Thinking of the Compton wavelength $\mu^{-1}$ as the impact parameter of a classical particle provides a useful heuristic for the parametrics of SR. Particles whose impact parameter $\mu^{-1}$ is smaller relative to the size of the BH $\sim GM_\text{BH}$ (larger $\alpha$) are more likely to interact with the BH's near-horizon region, where amplification takes place in proportion to the spin of the BH. However, if the impact parameter is \emph{too small} the particle will  fall altogether into the BH.

Per the condition \eqref{eq:SR_condition_2}, levels with $m=0$ are not superradiant. The rewritten condition also shows that, for fixed $m$, the SR condition is not fulfilled for $\alpha$ large enough. In particular, for $m=1$, we have at most $\alpha \leq a_\star/2(1+\sqrt{1-a_\star})\leq 0.5$. Of course, one might consider $m>1$; however, the overlap between hydrogenic waveforms and the origin (the approximate location of the BH horizon) decreases exponentially with increasing quantum number $\ell \leq m$, due to the increased angular momentum barrier. Consequently, the fastest SR growth is achieved for the hydrogenic 211 level, when the SR condition is near saturation. Thus, the reach of SR as a probe for new scalars peaks for parameters for which $\alpha \simeq 0.5$ or
\begin{equation}
\mu\simeq \frac{10\,M_\odot}{M_\text{BH}}\times (7\times 10^{-12}\,\eV)\,,
\end{equation}
namely for new scalars whose Compton wavelength is comparable to the size of the BH event horizon.

An attractive feature of the SR process is that it arises purely through gravitational effects, without necessitating any extension of the underlying physical framework other than the inclusion of a minimally coupled scalar field. Because the SR process obeys conservation laws, the BH's angular momentum and a comparably minute amount of its mass are transferred from the BH to the SR cloud. The maximum mass/energy that can be extracted by the SR process can be assessed from \cref{eq:a crit}. For example, for $m=1$, each particle added to the cloud carries $\hbar$ of angular momentum. The number of particles $\Delta N$ created through SR to $m=1$ levels is therefore
\begin{equation}
    \nonumber
    \Delta N = \Delta J \approx \Delta a_\star GM_\text{BH}^2 + 2a_\star GM_\text{BH}\Delta M_\text{BH}\,,
\end{equation}
 using \cref{eq:astar definition}. Since each particle carries energy $\sim\mu$, the relative change in mass of the BH obeys 
 \begin{equation}
     \nonumber
     \Delta M_\text{BH}/M_\text{BH} \approx \alpha (\Delta a_\star + 2a_*\Delta M_\text{BH}/M_\text{BH})\,,
 \end{equation}
  or 
  \begin{equation}
      \nonumber
      \Delta M_\text{BH}/M_\text{BH}\lesssim \alpha[1-4\alpha/(1+4\alpha^2)][1-2\alpha]^{-1}\,.
  \end{equation}
  This is $\approx \alpha$ for $\alpha \lesssim 0.1$ and peaks at $\sim 10\%$ for $\alpha\approx 0.2$.

\subsection{Effects of  self-interactions}
\label{sec:self-interactions}

In the absence of other significant non-gravitational processes, the SR cloud reaches its maximum potential occupation number $N_\text{cloud}^{\text{max}}= \mathcal O\left(G M_\text{BH}^2\right)$. At fixed $\alpha$,  the timescale of the SR instability scales as the BH light-crossing time $\sim GM_\text{BH}$:
\begin{equation}
\label{eq:SR timescale}
\tau_\text{SR} \approx 0.04\,\,\Gyr \times\left(\frac{10^{-2}}{\alpha}\right)^{9}\left(\frac{M_\text{BH}}{10\,M_\odot}\right)\left(\frac{1}{a_\star}\right)
\end{equation}
for superradiance to the 211 level \cite{Detweiler:1980uk,Arvanitaki:2010sy}. The unimpeded growth of the cloud is an exponential process, which is therefore completed in $T_\text{SR} \approx \ln\left(N_\text{cloud}^\text{max}\right)\,\tau_\text{SR}$.

The fast scale, compared to the age of the galaxy, and the potential observability of such effects has led in the last decade to the birth of a rich experimental and theoretical search program in which BHs are used as detectors of new light bosons, most notably the so-called axion-like particles (ALPs) \cite{Svrcek:2006yi,Arvanitaki:2009fg}. Because BHSR is a purely gravitational process, existence of a new scalar minimally coupled to gravity can be excluded by observations of a sufficiently old, fast rotating BH \cite{Arvanitaki:2009fg,Baryakhtar:2020gao,Hoof:2024quk,Stott:2018opm,Witte:2024drg,Fernandez:2019qbj}. Gravitational wave emissions from a SR cloud, as well as the imprint of the cloud on gravitational waves emitted during binary inspirals, can be searched for using current and near-future gravitational wave detectors 
\cite{Baumann:2018vus,Baumann:2019ztm,Baumann:2022pkl,Tomaselli:2024bdd,Tomaselli:2024dbw,Boskovic:2024fga,Boskovic:2025ixx,Tomaselli:2024faa,Takahashi:2024fyq,Arvanitaki:2014wva,Zhu:2020tht,Collaviti:2024mvh,Mirasola:2025car,Brito:2017zvb,Brito:2017wnc,Yuan:2021ebu,Omiya:2024xlz,Kim:2025izt}.

Many BSM scalar candidates, however, are generically expected to have some level of particle self interactions. For example, the simple paradigmatic cosine Lagrangian for an ALP 
\begin{align}
\begin{split}
\mathcal L_\text{ALP} \approx{}&-\frac{1}{2}(\nabla_\nu \phi)(\nabla^\nu \phi) - 
\mu^2f^2\left[1-\cos\left(\frac{\phi}{f}\right)\right]\\
={}& -\frac{1}{2}(\nabla_\nu \phi)(\nabla^\nu \phi) - \frac{1}{2}\mu^2 \phi^2 + \frac{\lambda}{4!}\phi^4 + \dots\,,
\end{split}
\end{align}
where $f$ is the so-called ``decay constant'', has quartic interactions. The coefficient of the quartic self-interaction is then identified to be $\lambda =\mu^2/f^2$, while the  coefficients of higher self-interactions are similarly fully fixed by $f$ and the structure of the full potential (the cosine in the above). In a bottom-up effective field theory perspective, one remains agnostic about the structure of the full potential and \emph{defines} the auxiliary scale 
\begin{equation}
\label{eq:definition of f}
f^2 \equiv \frac{\mu^2}{|\lambda|}
\end{equation}
outright, without regard for a particular form of the full Lagrangian. We allow $\lambda$ to be a signed quantity. Physically, quartic terms $\sim \phi^4$ represent a repulsive ($\lambda < 0$) or attractive $(\lambda > 0)$ contact (``billiard-ball'') interactions between the quanta of the field $\phi$. Within this bottom-up context, which we adopt for this work, the dimensionless $\lambda$ and the dimensionful $f$ should be viewed as two equivalent ways of quantifying the strength of self-interactions, related through \cref{eq:definition of f}. On the other hand, the dimensionless combination 
\begin{equation}
\label{eq:theta definition}
    \theta \equiv \frac{\phi}{f}
\end{equation}
can be defined as the ``theta angle'' of the general scalar, in \emph{analogy} to $\theta_\text{QCD}$ when $\phi$ is a QCD axion. Note that the defining properties of the QCD axion \emph{specifically} \cite{ParticleDataGroup:2024cfk} place it outside the siren regime in the mass range corresponding to the SR instability of stellar-mass BHs \cite{Baryakhtar:2020gao}.

The presence of contact self-interactions allows for the scalars in the cloud to scatter off one another. In the wave mechanical picture, mixing is introduced between the hydrogenic states the SR cloud, such that an abundance of particles in one hydrogenic level can now drive up the amplitude of the field in some other level.

\begin{figure}[h]
    \centering
        \includegraphics[width=0.5\textwidth]{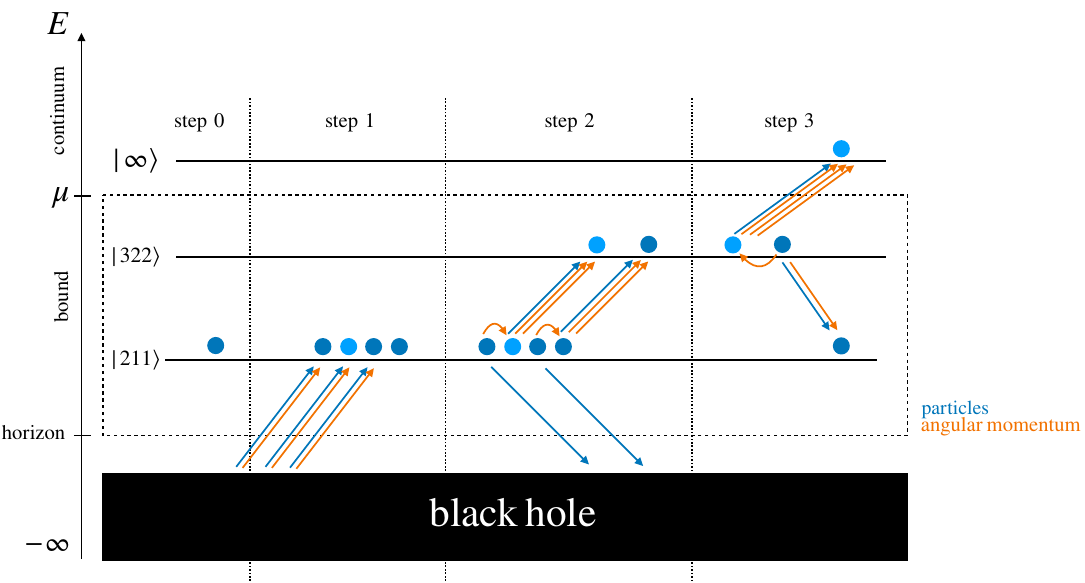}
        \caption{Schematic depiction of the ``autoionization'' of the gravitational atom and the flow of particles (blue arrows) and of the component of angular momentum parallel to the BH spin axis (orange arrows) in a steady-state BH scalar siren. The dashed box corresponds to the SR cloud at quasi-equilibrium. One particle (light blue) can be thought of as ``climbing out'' of the BH-cloud system in a three-step process assisted by ``auxiliary'' cloud particles (dark blue), carrying $3\hbar$ units of angular momentum with them. Once the process reaches equilibrium, the cloud is (very nearly) unchanged at the end and the process repeats. While we heuristically represent the emission of a ``classical'' point particle here, the emission is really a continuous wave process. The ``phase information'' (which the heuristic picture above fails to capture) carried by every particle is such that the emission takes place coherently over the volume of the cloud.
        }
    \label{fig:self ionization}
\end{figure}

The authors of Ref.\,\cite{Baryakhtar:2020gao} performed the first systematic study of self-interactions in the scalar SR system (see also \cite{Omiya:2022gwu}). It was found that, in addition to the SR process $\text{BH}\rightarrow 211$, key roles are played by two-particle scatterings: the process $211\times 211 \rightarrow 322 \times \text{BH}$, denoting  the upward scattering of particles from the SR state 211 to state 322; and $322 \times 322\rightarrow 211\times \infty$, denoting the down-scattering of a particle from 322 to 211, accompanied by the ejection of a particle partner to infinity as free, outgoing radiation. Together, these processes set up an almost-closed, quasi-equilibrium configuration between the BH and hydrogenic levels $211$ and $322$. This ``autoionization'' of the gravitational atom is schematically depicted in \cref{fig:self ionization}.

From energy conservation, the radiation released by the $322\times 322\rightarrow 211\times \infty$ process has a frequency (one-particle energy)
\begin{equation}
\label{eq:BH_frame_frequency}
\omega_\phi^\text{BH} =2E_3-E_2 \approx \mu\left[1+\frac{1}{2}\left(\frac{\alpha}{6}\right)^2\right]
\end{equation}
in the BH frame. By conservation of angular momentum, the radiation from the $322\times 322\rightarrow 211\times \infty$ process carries angular momentum in the $m=3$ and $\ell = 3,5$ spherical harmonics.

In the particle picture, writing $\omega_\phi^\text{BH} \approx \mu+ \mu(v_\phi^\text{BH})^2/2$, where $v_\phi^\text{BH}$ is the velocity of the emitted scalars at infinity, reveals that the radiation can be viewed semi-classically as a stream of scalars with non-relativistic velocities
\begin{equation}
\label{eq:velocity}
v^\text{BH}_\phi \approx \frac{\alpha}{6}\, .
\end{equation}
Note that the velocity of the emitted scalars is of the same order as the characteristic \emph{orbital} velocity scale $\alpha$ of the bosons around the BH.

Thus, once the 211 level, growing first according to the short timescale \cref{eq:SR timescale}, reaches sufficient occupation, the quasi-equilibrium regime is established and the growth is cut off. The overall number of particles in the SR cloud is therefore \emph{reduced} greatly compared to the case with negligible self-interactions. Parametrically, 
\begin{equation}
\label{eq:occupation}
N_\text{cloud}^\text{eq} = \left(\frac{f}{f_\text{eq}}\right)^2N_\text{cloud}^\text{max}\,,
\end{equation}
where $f_\text{eq}$ is a critical value of $f$ defined in \cite{Baryakhtar:2020gao}. 

The subsequent extraction of angular momentum from the BH is then slowed down dramatically. Rather than storing the angular momentum, the small equilibrium cloud serves as an auxiliary that slowly and continuously circulates angular momentum directly from the BH to spatial infinity in the form of scalar radiation. Demanding that less than $\mathcal O(1)$ of the spin be extracted in time $T$ yields $f < f_\text{siren}$ where
\begin{align}
\begin{split}
\label{eq:threshold f}
f_\text{siren}\approx{}& 2\times 10^{11}\,\GeV\\\times{}&\left(\frac{M_\text{BH}}{10\,M_\odot}\frac{10
\,\Gyr}{T}\right)^{1/2}\left(\frac{0.1}{\alpha}\right)^3\left(\frac{0.9}{a_\star}\right)^{1/4}.
\end{split}
\end{align}
On the other hand, perturbative unitarity (i.e. the requirement that the cross section for the particle physics process $\phi\phi\rightarrow \phi\phi$ not be so large as to break conservation of probability) sets a limit on the size of self-interactions $\lambda$, namely $ f\gtrsim \mu$. For $T \sim 10\,\Gyr$ and $\alpha \sim 0.1$, this implies the siren regime can be reached for $\mu \lesssim \mathcal O(10)\,\GeV$.

Because the presence of self-interaction suppresses the number of particles in the cloud, \cref{eq:occupation}, GW emission represents a small correction to the quasi-equilibrium dynamics of the cloud, and the power of this emission is greatly suppressed relative to the scenario of pure gravitational dynamics with negligible self-interactions \cite{Baryakhtar:2020gao}.

Altogether, when a spinning BH is born, it may reach quasi-equilibrium effectively instantaneously relative to astrophysical and/or Galactic timescales, \cref{eq:SR timescale,eq:occupation}. Moreover, when $f<f_\text{siren}$, the BH has its spin virtually unchanged from birth and is still emitting scalar radiation today. This picture evokes the definition of a new effective, composite, stable (i.e.\,effectively infinitely-long lived) and steadily radiating object which we call a \emph{black hole scalar siren}.

\subsection{Scalar sirens}
\label{sec:sirens_signal}
In this section, we characterize the radiation field of an individual black hole scalar siren. For the moment, we neglect any external gravitational field, such as that of the Galaxy and defer our discussion of its effects to the next section.

\begin{figure}[h!]
    \centering
     \includegraphics[width=0.7\linewidth]{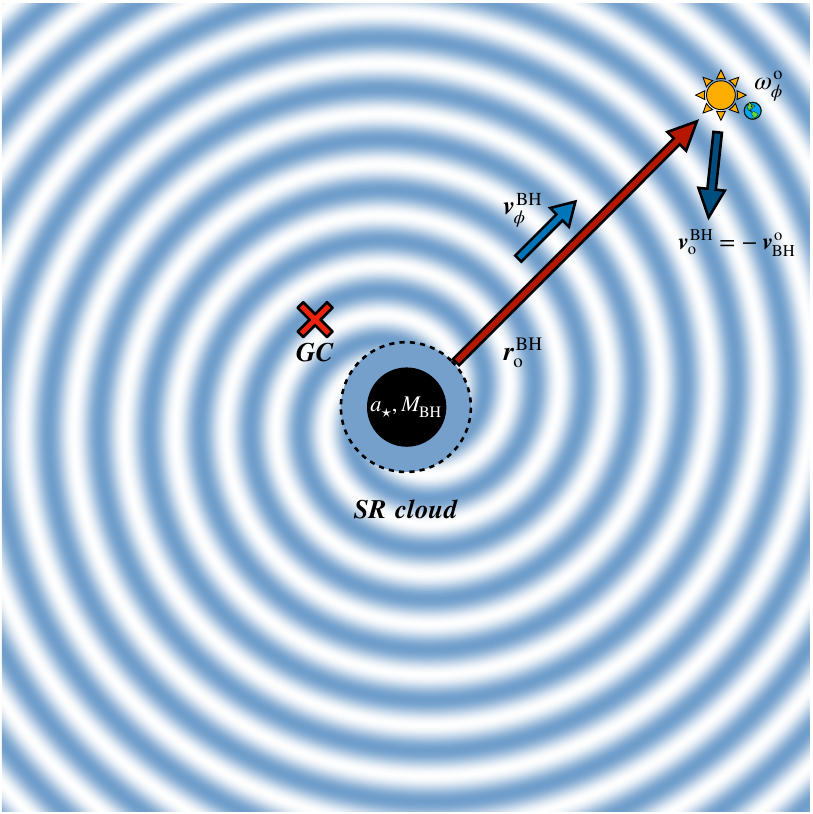}
    \caption{Schematic depiction of the radiated scalar field (light blue) in the \emph{non-rotating rest frame} of a spinning BH scalar siren (center), with some generic offset from the Galactic Center (GC; red cross). The angular momentum vector of the BH points out of the page. Effects of external gravitational potentials (such as that of the galaxy) on the propagating wave are neglected. As per \cref{eq:radiation}, when viewed from along the spin axis, points of equal phase are outwardly traveling spirals: $\text{Re}[e^{-i(\omega_\phi^\text{BH} t-\mu v_\phi^\text{BH} r-3\varphi)}] = \cos(\omega_\phi^\text{BH} t-v_\phi^\text{BH} r-3\varphi)$. In the distant radiation zone, the wavefronts appear as outgoing shells, \cref{eq:radiation}. An observer traveling in this field measures the scalar values, \cref{eq:observed_field}, and the local gradient, \cref{eq:gradient}. 
    }
    \label{fig:BH rest frame}
\end{figure}

We work first in a non-rotating coordinate system centered on the rotating BH siren, translationally at rest, and whose $z$-axis points along the angular momentum vector of the BH (\cref{fig:BH rest frame}). In the radiation zone at distances far greater than the cloud radius, the radiated scalar field is then
\begin{align}
\begin{split}
\label{eq:radiation}
{}&\phi_\text{rad}(\bm r,t) \\={}& \frac{A(\mu,M_\text{BH},f;r)}{2} \mathcal A(\hat r)S(\alpha,a_\star) e^{-i\left(\omega_\phi^\text{BH}t-\mu \bm v_\phi^\text{BH}\cdot \bm r - 3\varphi+\delta'_\text{BH}\right)}\\+{}& \text{ c.c.}\,,
\end{split}
\end{align}
where $\{r,\vartheta,\varphi\}$ denote polar coordinates in the frame of the siren, $\hat r = \bm r/r$, $\bm v_\phi^\text{BH} = v_\phi^\text{BH}\hat r$ and c.c.\,denotes the complex conjugate of the first term. The field of \cref{eq:radiation} accounts for the gravitational effects of the BH siren itself on the outgoing waveform, but assumes that a flat geometry is recovered sufficiently far away from the BH at the observation point, $r/GM_\text{BH}\rightarrow\infty.$

We now define and discuss the factors in \cref{eq:radiation}.
\begin{enumerate}
    \item The amplitude is (\cref{app:power waveform})
\begin{align}
\begin{split}
\label{eq:amplitude}
A(\mu,M_\text{BH},f;r) \approx 2.1 \,\,\frac{\alpha^2GM_\text{BH}}{r} f\,.
\end{split}
\end{align}
At fixed $\alpha$, it scales with the Newtonian dimensionless gravitational potential $GM_\text{BH}/r$ at a distance $r$ away from the siren. That radiated \emph{field} (as opposed to the radiated \emph{power}) is proportional to the total siren mass $M_\text{BH}$.
The emission takes place coherently over the entire volume of the cloud \cite{Arvanitaki:2010sy,Arvanitaki:2014wva,Brito:2015oca} so that, at fixed $\alpha$, the siren acts as a single source with total ``gravitational charge'' $\propto GM_\text{BH}$. This, in addition to the fact that the emitted scalars are non-relativistic, is a unique feature of black hole scalar sirens relative to most astrophysical sources of ultralight scalar particles.
\item
The angular dependence is
\begin{align}
\begin{split}
\label{eq:angular_dependence}
{}&e^{i3\varphi}\mathcal A(\vartheta,\varphi) 
\\={}& e^{-i\sigma_3}\sqrt{\frac{44}{45}}Y_3^3(\vartheta,\varphi) + e^{-i\sigma_5}\sqrt{\frac{1}{45}}Y^3_5(\vartheta,\varphi)\,,
\end{split}
\end{align}
where $Y^m_\ell(\vartheta,\varphi)$ are the spherical harmonics. The terms in \cref{eq:angular_dependence} are normalized such that $\int \mathcal |\mathcal A(\vartheta,\varphi)|^2 \mathrm d^2\Omega  =1$. Note that because $e^{-i3\varphi}Y^3_3 \sim \sin^3\vartheta$, the emissions peak towards the equatorial plane. The variable $\sigma_\ell \sim \alpha v_\phi^\text{BH}$ designates the ``Coulomb phase shift.'' It corresponds to the phase accumulated
by the emitted radiation as it travels out of the $1/r$ gravitational potential of the BH, while carrying total angular momentum $\ell$. 
\item 
The ``spin'' factor
\begin{align}
\begin{split}
\label{eq:spin}
S(\alpha,a_\star) = \frac{[a_\star-2\alpha(1+\sqrt{1-a_\star^2})]^{3/4}}{\sqrt{1+\sqrt{1-a_\star^2}}}\,
\end{split}
\end{align}
encapsulates the dependence of the signal on the SR condition. It is
real-valued only for combinations of $(\alpha, a_\star)$ for which
the SR condition ($m = 1$) is satisfied and zero otherwise.
\item The steady-state constant phase factor $\delta'_\text{BH}$ is a real phase factor tied, among other things, to the origin of the coordinate time relative to the start of the steady-state, quasi-equilibrium signal. Additionally, the onset of the SR process itself is inherently quantum mechanical, as the growth of the SR cloud proceeds from zero-point fluctuations \cite{Arvanitaki:2010sy,Arvanitaki:2014wva,Press:1972zz,Fu:2025ztk}. This can roughly be understood as the $211$ superradiant level being ``born'' with an unknown arbitrary phase factor. This, in turn, introduces an uncertainty regarding the phase of the bound wave components of the SR cloud themselves, which ultimately propagates to the radiated emission. There, $\delta'_\text{BH}$ is a unique draw from a uniform distribution: $\delta'_\text{BH}\sim U(0,2\pi)$.
\end{enumerate}
Next, consider the value of the radiation field evaluated along a short segment of an observer's worldline $x_\text{o}^\mu=\left(t,\bm r^\text{BH}_\text{o}(t)\right)$ whose 4-velocity is 
\begin{equation*}
    u_\text{o}^\mu = (1-\dot{\bm r}_\text{o}^\text{BH}\cdot\dot{\bm r}_\text{o}^\text{BH} )^{-1/2}\left(1,\dot{\bm r}_\text{o}^\text{BH}\right)\,.
\end{equation*}
 
Assuming the wordline segment is sufficiently short so that the observer's coordinate velocity $ \bm{v}_\text{o}^\text{BH}=\dot{\bm r}_\text{o}(t)$ does not change appreciably during the observation:
\begin{align}
\begin{split}
\label{eq:observed_field}
\Phi_\text{rad,o}(t) {}&\approx \frac{A(\mu, M_\text{BH},f;r^\text{BH}_\text{o}(t))}{2}\mathcal A\left(\hat r_\text{o}^\text{BH}(t)\right) S(\alpha,a_\star)\\{}&\times e^{-i\left(\omega_\phi^\text{o} (\tau-\tau_i) - \delta_\text{BH}\right)},
\end{split}
\end{align}
where $\delta_\text{BH} = \delta'_\text{BH}+\omega_\phi^\text{BH}t_i-\mu \bm v_\phi^\text{BH}\cdot\bm r_\text{o}(t_i)- 3\varphi_\text{o}(t_i)$, $t_i$ is the initial time of observation, $\tau - \tau_i = (1-\bm v^\text{BH}_\text{o}\cdot \bm v^\text{BH}_\text{o})^{1/2}(t-t_i)$ is the proper time interval elapsed for the observer and
\begin{align}
\begin{split}
\label{eq:Doppler_formula}
\omega^\text{o}_\phi =  \frac{\omega_\phi^\text{BH}-\mu  \bm v_\phi^\text{BH}  \cdot \bm v_\text{o}^\text{BH}}{\sqrt{1-\bm v^\text{BH}_\text{o}\cdot \bm v^\text{BH}_\text{o}}}
\end{split}
\end{align}
is the Doppler-shifted frequency.
As long as $v_\text{o}^\text{BH}, v_\phi^\text{BH} \ll 1$, the kinetic energy of the scalar quanta measured by the observer is 
\begin{align}
\begin{split}
\tilde\omega^\text{o}_\phi\equiv {}& \omega_\phi^\text{o}-\mu \\\approx{}& \mu\left[\frac{\bm v_\phi^\text{BH}\cdot \bm v_\phi^\text{BH}}{2} - \bm v_\phi^\text{BH}  \cdot \bm v_\text{o}^\text{BH}+\frac{\bm v_\text{o}^\text{BH}\cdot \bm v_\text{o}^\text{BH}}{2}\right] \\\approx{}& \frac{1}{2}\mu \bm v_\phi^\text{o}\cdot \bm v_\phi^\text{o}\,,
\end{split}
\end{align}
corresponding to the Galilean addition of velocities
\begin{align}
\begin{split}
\bm v_\phi^\text{o}  = \bm v_\phi^\text{BH} - \bm v_\mathrm{o}^\text{BH}.
\end{split}
\end{align}
Another quantity of interest is the spatial gradient of the field along the observer's worldine. It is related to the derivative in the BH's frame through a projector:
\begin{align}
\begin{split}
    \vec\nabla_\text{obs} \Phi_\text{rad,o}\approx \left(\vec\nabla_\text{BH}+\bm v^\text{BH}_\text{o}\partial_t\right)\phi_\text{rad}\,. 
\end{split}
\end{align}
Note that, in the frame of the BH, $\vec \nabla_\text{BH}\phi_\text{rad}$ is nearly but not exactly parallel to $\bm{v}_\phi^\text{BH}$. This is because the radiation carries angular momentum. In the radiation zone $\mu r \gg 1$, however, the lever arm $r$ is large and, to a good approximation,
\begin{align}
\begin{split}
\label{eq:gradient}
{}&\vec \nabla_\text{obs} \Phi_\text{rad,o} \approx \mu \bm{v}^\text{o}_\phi \\{}&\Big[\frac{iA(\mu,M_\text{BH},f;r^\text{BH}_\text{o}(t))}{2} \mathcal A(\hat r_\text{o}^\text{BH})S(\alpha,a_\star) e^{-i\left(\omega_\phi^\text{o}(\tau-\tau_i)-\delta_\text{BH}\right)}\\+{}&\text{ c.c.}\Big]\,.
\end{split}
\end{align}
As can most easily be seen in the frame of the BH, the \emph{lookback time} (i.e.\, the time between emission and detection) for the observation of a non-relativistic particle is 
\begin{equation}
\label{eq:lookback time}
t_\text{lb}(\alpha) \approx 13\,\Myr \left(\frac{r_\text{o}^\text{BH}}{8\,\kpc} \right)\left(\frac{600\,\km/\s}{v^\text{BH}_\phi}\right),
\end{equation}
which is parametrically short relative to the age of the Galaxy.

Finally, note that the rate of angular momentum extraction itself sets the rate of change for all other quantities once the quasi-equilibrium regime is reached. Drifts in the frequency of emissions caused by the slow change in BH parameters 
over any experimentally relevant periods of observation can be neglected in the siren regime. We explain this in more detail in \cref{sec:drift}. 

\subsubsection{Sirens at larger $\alpha$}
\label{sec:larger alpha sirens}

Having established the properties of the two-level siren, we briefly comment on how this picture is modified at larger values of $\alpha$, before turning to the ensemble signal.

As discussed, the parameter $\alpha$ controls the gravitational binding of the scalar cloud to the BH geometry and thus sets the characteristic velocities and energy scales of the (non-relativistic) siren dynamics. Higher-order corrections to siren processes therefore correspond to relativistic effects associated with near-horizon physics, including horizon absorption and corrections to the real gravitational potential. The analytic computation of superradiant (SR) growth rates by Detweiler \cite{Detweiler:1980uk}, summarized in \cref{eq:SR timescale} for $n\ell m =211$, remains accurate up to moderately large values of $\alpha/\ell$. More precise determinations at larger $\alpha$ have since been obtained in \cite{Dolan:2012yt,Siemonsen:2022yyf}.

The two-level siren is not merely an approximation. As shown in \cite{Baryakhtar:2020gao}, for sufficiently small $\alpha$ only the $211$ and $322$ levels acquire significant occupation. This is a consequence of the $322$ level being fratricidal once equilibrium is reached: attempts to populate an additional level $n'\ell'm'$, whether it be through SR or self-interactions, are suppressed by the process $n'\ell'm'\times 322 \rightarrow 211\times \infty$.

At larger $\alpha$, modifications to BH interaction rates affect both SR growth and self-interaction-mediated upward scattering processes involving re-absorption of an intermediate $s$-state by the BH. For $\alpha$ large enough, the effective growth rate of some higher $n'\ell'm'$ levels can exceed the fratricidal depletion rate. The corresponding threshold is estimated as $\alpha \sim 0.2$ in \cite{Baryakhtar:2020gao}, with later works reporting $\alpha \sim 0.15$ \cite{Takahashi:2024fyq,Witte:2024drg}.

Above this threshold, multiple hydrogenic levels may grow simultaneously. The existence of a closed quasi-equilibrium configuration involving a finite (possibly large) number of levels is plausible. Although progress have been made in this direction \cite{Omiya:2022gwu,Takahashi:2024fyq,Witte:2024drg}, a converging procedure for the study of high-$\alpha$ sirens remains to be developed. In such systems, emission may proceed through processes of the form $n_1\ell_1 m_1\times n_2\ell_2m_2\rightarrow n_3 \ell_3 m_3 \times \infty$, with $E_{n_1}+E_{n_2}-E_{n_3} >\mu$, resulting in a polychromatic siren spectrum with potentially many \emph{sharp} emission lines.

The effects of this added complexity on the ensemble-level analysis remain nonetheless tractable, and we can proceed in a way indicative of how future improvements in understanding the spectrum of individual sirens can straightforwardly be factored into our analysis. We return in \cref{sec:larger alpha ensemble} to the impact of the polychromatic spectra of single sirens with $0.15 \lesssim \alpha < 0.5$ on the observed ensemble spectrum.

For $\alpha > 0.5$, the hydrogenic $211$ level is no longer superradiantly unstable, and equilibria must instead originate from states with $m>1$. In this regime, the SR timescale, \cref{eq:SR timescale}, should be replaced by that of the lowest-$m$ level satisfying the SR condition, \cref{eq:SR_condition}. However, due to the increasingly large kinetic barrier, the maximum SR rates achievable for given $m$ quickly become exponentially suppressed. For $\alpha \gg 1$, a WKB approximation yields $\tau_\text{SR} \approx 10^7\,
GM_\text{BH}e^{1.8\alpha}$ \cite{Arvanitaki:2010sy,Zouros:1979iw}. Because the SR rates set the overall rate of both growth and emission processes in a siren, sirens with larger $\alpha$ are effectively stretched in time, making them slower to grow, longer lived, and therefore dimmer.

Moreover, such sirens lie at the boundary between non-relativistic and relativistic regimes: while the SR levels remain quasi-non-relativistic, the intermediate $s$-states relevant for upward scattering are significantly deformed, requiring accurate BH interaction rates as computed in, e.g., \cite{Dolan:2012yt,Siemonsen:2022yyf}. 

For all these reasons, we restrict our analysis to $\alpha < 0.5$.

\section{Black-hole scalar sirens in the Milky Way}
\label{sec:MW sirens}
After characterizing the emission of an isolated BH scalar siren in section \cref{sec:sirens}, we now endeavor to study phenomenologically interesting astrophysical signals.

It is evident, both from intuition and from \cref{eq:amplitude}, that the amplitude of the signal from an individual BH scalar siren decreases rapidly with distance. In fact, from \cref{eq:amplitude}, one sees that the signal scales as the gravitational influence of the BH at the observer's location. 

Thus, only nearby, fast-spinning black holes would produce strong signals. Additionally, the angular dependence factor described by \cref{eq:angular_dependence} shows that emission is peaked in the equatorial plane, suppressing the signal from an individual BH siren whose spin axis faces the observer ``head on.'' The detectability of a signal from some known fast-spinning astrophysical BHs was summarily explored in \cite{Baryakhtar:2020gao}, ignoring the dependence on the angle of emission.

As explained in \cref{sec:sirens}, for scalars with suitably small decay scales, BH scalar sirens  are effectively infinitely long lived. In other words, if a scalar particle exists such that a single BH in a BH population with representative mass scale $M_s$ is a siren, then \emph{all} BHs in that same population are likely to be sirens, provided that they have sufficient spin at birth. This makes it interesting to leverage \emph{large-number populations} of potential BH scalar sirens, and look at aggregated, \emph{ensemble} signals. By leveraging a large population, this method increases the likelihood that high-spin, favorably oriented systems contribute appreciably to the measured signal.

We therefore turn our attention to the expected large population of isolated stellar-mass BHs in the Milky Way. 

\begin{figure}
    \centering
    \includegraphics[width=0.49\linewidth]{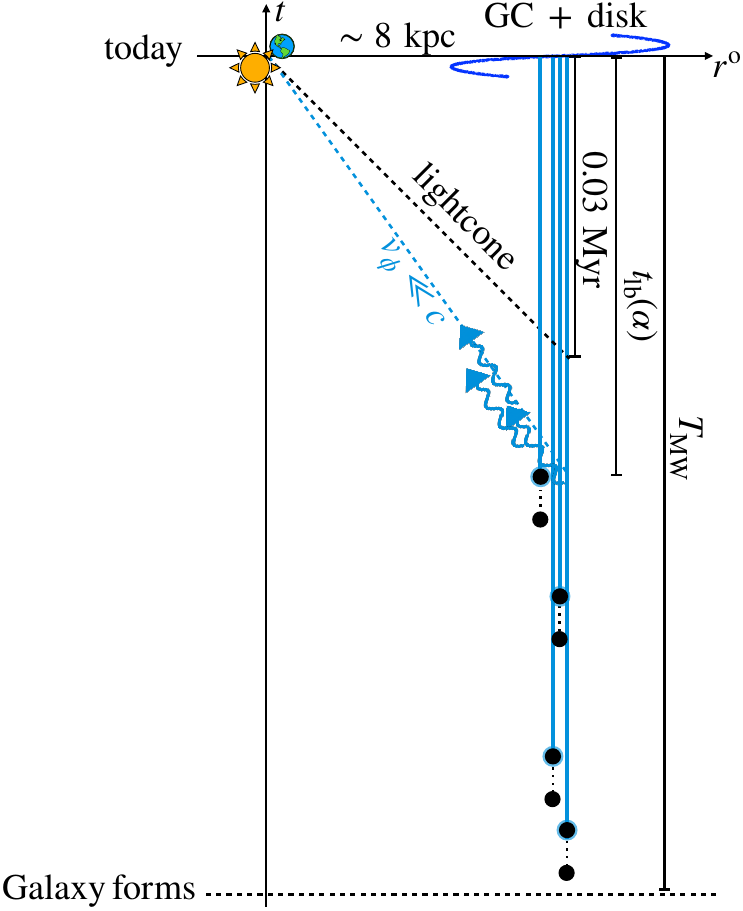}
    \includegraphics[width=0.49\linewidth]{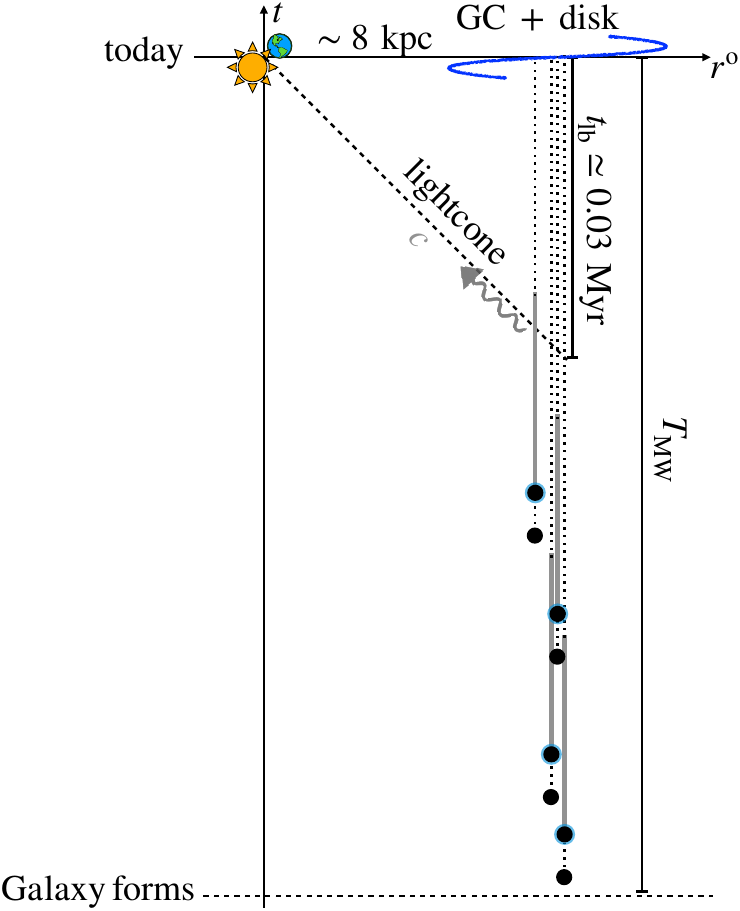}
    \caption{\emph{Left}: Spacetime diagram for the observation of non-relativistic scalar radiation from the worldlines of BH scalar sirens (strong self-interactions, small $f$) in the Galactic Center and disk.
    \emph{Right}: Corresponding diagram for massless gravitational waves emitted by superradiant clouds in the regime of negligible self-interactions (large $f$). The distance to the Galactic Center (GC) is approximately $8\,\kpc$ or $0.03\,\Myr$.
    Dotted worldline segments denote inert periods; emission begins after a delay set by the superradiance timescale. In the weak-interaction regime, individual BHs may emit GWs only for a finite time that can be short compared to the age of the MW, so only sources emitting at the retarded time $t=t_\text{today}-r_\text{BH}^\text{o}/c$ contribute to the signal. In contrast, in the scalar-siren regime the emission of massive scalars is effectively eternal after cloud growth. Because scalar propagation is timelike, the relevant lookback time \cref{eq:lookback time} satisfies $t_\text{lb}(\alpha) = r_\text{o}^\text{G}/v_\phi> r_\text{o}^\text{G}/c$ is parametrically controlled by $\alpha$, yet remains short compared to the Galactic age. Consequently, the scalar signal can be estimated directly from the total number of BH sirens today, with greatly reduced sensitivity to the detailed formation history of Galactic black holes. The two regimes are mutually exclusive in the sense that angular-momentum extraction is dominated either by GW emission (negligible self-interactions) or by scalar radiation (strong self-interactions). Finally, unlike massless gravitational waves, massive scalar radiation may remain gravitationally bound to the MW and accumulate over time, although this cannot be achieved for stellar-mass sirens. (\cref{sec:low_alpha_bound}).}
    \label{fig:sirensvsGW}
\end{figure}

The ensemble GW signal from scalar clouds around MW BHs has been considered in \cite{Zhu:2020tht}. In that case, the finite emission times complicate the analysis. In contrast, for the scalar-siren signal, the possibility of leveraging small $f$ and the quasi-equilibrium regime of the self-interacting SR cloud to define a BH scalar siren as an effectively stable object greatly simplifies the analysis. These two scenarios are displayed in \cref{fig:sirensvsGW}.

Just like the SR phenomenon itself applies to BHs of arbitrary mass, a BH scalar siren need not, in principle, be an astrophysical BH (i.e.\,stellar-mass or supermassive). 
The study of signals from astrophysical BH populations is, of course, best motivated, since such BHs are known to exist. Although many of our considerations apply to arbitrary BH populations, we ground our analysis in astrophysical BH sirens; we comment on sirens from exotic BH populations in \cref{sec:exotic}.

\subsection{Milky Way black holes}
\label{sec:galacitc_center_BH}
From both standard stellar evolution theory and numerical simulations, it is believed that the Milky Way is home to between $10^7$ and $ 10^9$ isolated stellar-mass black holes \cite{Olejak:2019pln,Timmes:1995kp,Shapiro:1983du,brown12003scenario,Wiktorowicz:2019dil,Bambi:2025rod}. Because they constitute the stellar graveyard of heavy, short-lived ($\sim \text{Myr}$) stars formed early in the history of the Galaxy \cite{matteucci2019origin,bensby2013chemical,Clarkson:2008ha,Clarkson:2011aw}, the oldest of the BHs are expected to be comparable in age to the Milky Way itself,  $\mathcal O\left(1-10\,\Gyr\right)$. Since BH sirens are effectively infinitely long-lived, the exact age of the BH is mostly irrelevant for the study of the ensemble scalar emission, as long as it is longer than the (parametrically short) SR timescale and the lookback time ($\ll \Gyr$). 

Direct observation of this large BH population is difficult and has yet to be realized. These are largely undetectable through conventional electromagnetic observations because they emit no light unless they are accreting. Astrometric microlensing observables represent a promising avenue \cite{OGLE:2022gdj,Sweeney:2022fxx}. Given the challenge of direct observation, artificial population synthesis and dynamical studies, such as the state-of-the-art synthetic catalog \cite{Olejak:2019pln}, provide precious insights.

\subsubsection{Mass distribution}
The mass of a BH remnant differs from that of the progenitor star(s) due to significant mass ejection (wind mass loss) during the supernova process.
Theoretical modeling of stellar core collapse, along with priors from the distribution of stellar masses in the MW, which can be studied both theoretically \emph{and} observationally, point to an exponential distribution of stellar-mass BH remnants \cite{Fryer:1999ht}: 
\begin{equation}
\label{eq:mass_distribution}
p_M(M_n)= \begin{cases}
M_s^{-1} \exp\left[-\frac{(M-M_\text{min})}{M_s}\right],& M>M_\text{min}\,,\\
0, & \text{otherwise}\,,
\end{cases}
\end{equation}
where $M_\text{min}$ is the minimum mass leading to total gravitational stellar collapse, and $M_s$ is a mass scale for the distribution. 

The exponential form is also found to be the best fit out of five distribution functional forms for the mass of 15 observed X-ray binary systems \cite{farr2011mass}, yielding $M_{\text {min }}=5.3_{-1.2}^{+0.9}\,M_{\odot}$  and $M_s=4.7_{-1.9}^{+3.2}\,M_{\odot}$. Fitting the synthetic catalog of \cite{Olejak:2019pln} one obtains $M_s=\,9.5 M_{\odot}$. We consider both scenarios as benchmarks for our analysis.

\subsubsection{Spatial distribution}
Because they are compact stellar remnants, stellar-mass MW BHs are expected to follow the spatial distribution of stars and star-formation profile in the MW, itself estimated to drop off exponentially from the Galactic Center, with more complicated tri-axial modeling in the bulge  \cite{sale2010structure, wegg2013mapping}. For concreteness, we follow \cite{Arvanitaki:2014wva} and use a planar distribution in the Galactic disk,
\begin{equation}
\label{eq:spatial_distribution}
p_{r^\text{G}_\text{BH}}\left(r^\text{G}_n\right) = r_s^{-1}\exp\left[-\frac{r^\text{G}_n}{r_s}\right],
\end{equation}
with a scale length of $r_s \approx 3.2\,\kpc$ \cite{sale2010structure}. This distribution is approximately matched by the BHs in the synthetic catalog \cite{Olejak:2019pln}. 

\subsubsection{Spin distribution}
Little is known about the distribution of BH spins at birth, either from first principles, or observational information, the latter being ultimately limited by the overall relatively small number of MW BHs that have been directly observed. 

Continuing observations of gravitational waves from BH-BH and BH-neutron-star binary mergers by the LIGO-Virgo-KAGRA collaboration, as well as the upcoming LISA observatory, promise to eventually provide powerful probes into the spin and mass distribution of the population of MW BHs. However, the inherent degeneracies within GW observables between BH spin magnitude and direction relative to the orbital angular momentum and angle of observation render an unambiguous determination of BH spin magnitudes from individual events difficult at present.

Spins of BHs in X-ray binaries can also be obtained via both continuum fitting \cite{McClintock:2013vwa} and ``relativistic reflection'' \cite{Reynolds:2013qqa,Reynolds:2020jwt} methods. These techniques have so far generated spin measurements for 36 Galactic BHs \cite{Draghis:2023vzj}, yielding a distribution that is strongly peaked at very high spins, with $89\%$ of spins being above $a_\star = 0.8$. It is believed, however, that X-ray methods suffer from a selection bias towards high-spin systems for supermassive BHs \cite{Vasudevan:2015qfa}. This bias should not be so prominent for stellar-mass BHs, as the sample is volume-limited, not luminosity-limited \cite{Reynolds:2020jwt}.

In light of this, we consider a normalizable one-parameter family of BH spin distributions:
\begin{equation}
p_{a_\star}(a_n) = \frac{(1-\beta)}{(1-a_n)^\beta}, \quad 0 \leq \beta < 1,
\label{eq:spindistro}
\end{equation}
where $\beta$ controls the ``pitch'' of the distribution towards high spin. We find that $\beta = 0.65$ gives the best fit to the data of \cite{Draghis:2023vzj}.

\subsubsection{Binary partners and accretion}

Binary companions to superradiant BHs can act as gravitational perturbers. This has been studied both  without
\cite{Baumann:2018vus,Baumann:2019ztm,Baumann:2021fkf,Baumann:2022pkl,Tomaselli:2024bdd,Tomaselli:2024dbw,Boskovic:2024fga,Boskovic:2025ixx,Tomaselli:2024faa,Zhang:2019eid} and with \cite{Takahashi:2024fyq} significant quartic self-interactions. A binary partner can lead to level mixing, imprint rich features on the GWs emitted during binary inspirals, and potentially collapse or ionize the SR cloud. 

The fraction of stars in binary systems in the Milky Way is estimated to be around 50\% or higher, with some predictions as high as 85\% for all systems (binary, triple, etc.), varying significantly by stellar mass \cite{Duchene:2013cba,chini2013stellar}.\footnote{Based on observations and modeling \cite{Duchene:2013cba}, for stars similar to our Sun (G-type), the binary frequency is about 44-50\%. The fraction increases significantly with mass: nearly all O-type and B-type stars (the most massive) are in binary or multiple systems ($\gtrsim$\,70\%).  The fraction decreases for less massive stars, with only about 25\% of red dwarfs (M-type stars, which are the most common star type in the galaxy) having a companion.} However, the fraction of black holes in binary systems is much lower, estimated to be just a few percent of the total black hole population \cite{wiktorowicz2019populations,lamberts2018predicting,Olejak:2019pln}. This is because the violent processes involved in stellar mass black hole formation, specifically core-collapse supernovae, tend to destroy or kick out any binary partner.

Furthermore, one expects that the effects of a companion are parametrized by the distance of the companion $|d_\text{comp}|$ relative to the spatial extent of the cloud $r_c \sim \alpha^{-2}\,GM_\text{BH}$. Typically, in the Milky Way $|d_\text{comp}|\approx 10^5\,R_\odot$ \cite{Olejak:2019pln}, yielding the condition $10^8 \left(\frac{\alpha}{0.1} \right)^2\left( \frac{ M_\odot}{M_\text{BH}}\right)\sim 1$.  While the effects of a binary companion can in fact be appreciable even when $|d_\text{comp}|\gg r_c$ \cite{Baumann:2018vus,Baumann:2019ztm,Baumann:2021fkf,Baumann:2022pkl,Tomaselli:2024bdd,Tomaselli:2024dbw,Boskovic:2024fga,Boskovic:2025ixx,Tomaselli:2024faa,Zhang:2019eid,Takahashi:2024fyq}, this intuition appropriately captures the fact that companion effects are more pronounced for small $\alpha$.

A key difference between GW and scalar emissions from the SR cloud which \cref{fig:sirensvsGW} cannot convey is that, while GW emissions proceed from a fully grown cloud, quasi-equilibrium BH scalar sirens have a cloud filled only to a fraction of its maximum capacity. Crucially, the particle cloud of a siren does not store a significant fraction of the angular momentum of the system at any time. If a gravitational scattering event were to strip the full cloud, the resulting ``bare'' BH would still contain essentially all of the angular momentum it had prior to the event, and the siren would be ``reborn'' once the perturber has left.

The situation can be more complicated in a binary inspiral. The authors of \cite{Takahashi:2024fyq} find that the $322$ level of a siren, having a larger spatial extent, is more susceptible to perturbations from a companion and can be tidally stripped first, thereby disrupting the quasi-equilibrium regime. Depending on the binary mass ratio and the value of $\alpha$, this may allow the $211$ level to resume growth even at values of $f$ that would otherwise lead to quasi-equilibrium in an isolated system. Even in this case, the growth of the cloud, and the corresponding extraction of angular momentum from the BH, is ultimately stalled by the self-collapse of the cloud under attractive self-interactions (a so-called ``bosenova'')\footnote{The analysis of \cite{Baryakhtar:2020gao} and the numerical results of \cite{Omiya:2022gwu} suggest that a bosenova likely does not happen for isolated BHs, being replaced by the self-saturating quasi-equilibrium regime described in \cref{sec:self-interactions} instead. Conversely, the results of \cite{Takahashi:2024fyq} suggest that a transition from a siren regime to a bosenova may be possible in a binary system. In any case however, particle self-interactions impede the extraction of angular momentum and prolong the timescale associated with BHSR. Thus, there is a sense in which self-interactions increases the likelihood of observing a signal, both the lifetime of continuous scalar emissions, and by allowing more time for bosenova events to be triggered in inspirals. Transient burst emission of scalars in bosenovae could be searched for using networks of quantum sensors like atomic clocks or magnetometers \cite{dailey2021quantum,khamis2025multimessenger,arakawa2026multimessenger,sen2026gps}.}. It is therefore possible that an actively disturbed sirens enter a ``repeater'' period; whether repeated cycles of growth and collapse can efficiently extract angular momentum remains unclear, and the continuous emission regime may be restored once the companion moves through some critical region. Finally, a completed merger would produce a new BH which, if highly spinning, woverould itself become a new siren.

Evaluating the long-term stability of BH scalar sirens, even relatively isolated ones, within the Galactic environment remains an important task beyond our present scope. If sirens are gradually degraded in this way, the effect could be incorporated by adopting a spin distribution $p_{a_*}$ tilted toward low spins, which would itself signal the action of the BHSR process. In any case, for the mostly isolated stellar-mass BH population in the Milky Way, the effective lifetime is bounded from below by the characteristic timescale between gravitational encounters.

Similarly, the formation of an accretion disk around a stellar-mass BH almost always requires mass transfer from a sufficiently close binary companion; isolated black holes accreting from the interstellar medium rarely develop disks \cite{beskin2005low,tripathi2025disk,Tsuna:2019kny}\footnote{Isolated stellar-mass BHs traversing the diffuse interstellar medium (ISM) are expected to accrete primarily in the Bondi–Hoyle–Lyttleton (BHL) regime \cite{Bondi:1944rnk}, which typically yields very low accretion rates and does not generically provide enough net angular momentum for a long-lived, rotationally supported thin disk to form; instead, the flow is expected to be largely quasi-spherical and radiatively inefficient \cite{beskin2005low,Yuan:2014gma}. While disk-like structures can form transiently in the presence of ISM inhomogeneities/turbulence, numerical studies of BHL-like flows indicate that such configurations are often unstable or highly time-variable rather than settling into a persistent thin disk \cite{tripathi2025disk}.}.

As we discuss subsequently, the signals from individual BHs sirens add incoherently. One may therefore fold uncertainties surrounding the effects of a binary partner into an $\mathcal O(\text{few}\,\%)$ uncertainty on $N_\text{BH}$, understood to be the number of individual, isolated sirens. Doing this should provide a lower bound on the signal.

\subsection{Effects of the Galactic potential}
As discussed in \cref{sec:sirens}, the scalar radiation field around an individual siren is most easily described in the non-rotating rest frame of the BH, and by considering the value of field observables along an observer's worldline taken to be in motion in the siren's frame. Of course, both the siren and an Earth-bound observer are in motion relative to the Galactic frame. By the equivalence principle, the frame of any individual siren can be considered inertial, even as the siren itself undergoes accelerated motion in the gravitational potential of the Galaxy. The action of gravity on the siren, the observer, and the propagation of the scalar radiation can all, in principle, be systematically accounted for by considering the propagation of the waveform in the gravitational curvature (i.e.\,the gravitational potential) of the Galaxy, expressed in the siren's frame. If these corrections are important, \cref{eq:radiation} becomes invalid.

Physically, this mathematical criterion of small curvature potentials corresponds to the requirement that the emitted scalar particles emerge from the BH siren with velocities well above the local Galactic escape velocity\footnote{This can also be understood from the perspective of the eikonal/WKB approximation \cite{MCP}. Wave points propagate along the rays, or classical trajectories of potential. Well above the Galactic escape velocity, any gravitational deflection is small and the rays are nearly straight and radially outgoing from the source.}: $v_\phi^\text{G} \gg v_\text{esc}(\bm r_\text{BH}^\text{G})$. As long as $v_\text{esc}(\bm r^\text{G}_\text{BH}) \gg v_\text{BH}^\text{G}$, this requirement unambiguously amounts to
\begin{equation}
\label{eq:escape_velocity}
v^\text{BH}_\phi \gg v_\text{esc}(\bm r^\text{G}_\text{BH})\,.
\end{equation}
Scalar-siren emissions for which this requirement is not fulfilled remain trapped and accumulate in the Galaxy (see \cref{sec:low_alpha_bound}). More precisely, the waveform (Eq.\,\ref{eq:radiation}) is significantly altered by the presence of reflecting boundary conditions at the turnaround point of the trapped particle.

Precisely determining the escape velocity profile of the Milky Way is  challenging. Most recent studies that exploit Gaia data \cite{Monari:2018ckf,Roche:2024gcl} to map the escape velocity from $\sim 4$ to $\sim 10\,\kpc$ away from the GC place the escape velocity anywhere between $\sim 600$ and $750$ km/s. In light of \cref{eq:velocity}, this translates to $\alpha \gtrsim 1.5\times 10^{-2}$.

Of course, both the BH population and the Earth are themselves bound to the Galaxy. The requirement, \cref{eq:escape_velocity}, that the velocity scale of the scalars be above the Galactic escape velocity implies that Doppler shifts due to the relative motion of the Earth (Eq.\,\ref{eq:Doppler_formula}) on the emission spectra are negligible. This is further quantified in \cref{sec:population variables}.

At such velocities, the lookback time, \cref{eq:lookback time}, to a source in the Galactic disk is at most $\mathcal O(10\,\Myr)$ and does not significantly impact the Milky Way BH-siren signal.

\subsection{Stochastic signals from scalar sirens}

We now characterize the \emph{collective} signal from the large population of ancient BHs in the Galactic Center discussed in \cref{sec:galacitc_center_BH}. We ground our analysis in the regime of validity of \cref{eq:radiation}, where the radiation velocity is larger than the galactic escape velocity. For a given scalar mass $\mu$, the kinetic energy (and therefore frequency and velocity) of the escaping radiation is set by the mass of each BH siren. We therefore expect the inherent spread in BH masses, \cref{eq:mass_distribution}, to result in a corresponding spread of frequencies around the primary frequency $\mu$. The amplitude of the signal at each experimentally binned frequency should then be a function of the number of BHs expected to produce radiation within that bin, accounting for any Doppler shifts if necessary, as well as the expected distance, orientation, and spin of those BHs. 

We can formalize this discussion through an analysis similar to that in Refs.\,\cite{Foster:2017hbq,Gramolin:2021mqv,Cheong:2024ose,Bao:2025nsd} for the computation of signal from a virialized scalar component of the Galactic dark matter halo. The first step is to index each of the $N_\text{BH}$ BHs in the Galactic Center population and introduce a corresponding set of indexed variables: $\{M_\text{BH},a_\star,\bm r_\text{o}^\text{BH},\alpha,\delta_\text{BH}\} \rightarrow \{M_n,a_n,\bm r_\text{o}^n,\alpha_n,\delta_n\}$. The total collective field signal measured by the observer is then
\begin{align}
\begin{split}
\label{eq:stochastic_series}
\Phi_\text{rad,o}(t) ={}& \sum^{N_\text{BH}}_{n\,\in\,\{\text{galactic BHs}\}}\phi_{\text{o},n}\,,
\end{split}
\end{align}
where $\phi_{\text{o},n}$ from \cref{eq:observed_field} is evaluated with the indexed variables corresponding to the $n^\text{th}$ BH. 

The observer phase $\delta_n$ associated with each BH is effectively random for a number of reasons. As discussed in \cref{sec:sirens}, $\delta_n$ is randomized first by the inherently quantum mechanical nature of the onset of the SR process. Additionally, the relationship between the spacetime onset of the quasi-equilibrium signal of each BH relative to the spacetime onset of the period of observation is effectively random. As a result, we can think of each $\delta_n$ as randomly and independently drawn from a uniform distribution: $\delta_n \sim U(0,2\pi)$. Because $\delta_n$ is a random variable, $\Phi_\text{rad,o}(t)$ is a time-dependent random variable or, in other words, a (stationary) stochastic time series. 

An experiment with a duration $T_\text{exp}$ can only resolve frequency widths $\Delta \omega \simeq 2\pi/T_\text{exp}$. Thus, we partition the total sum of \cref{eq:stochastic_series} into discrete frequency bins around $\mu$, namely $\tilde \omega_j = 2\pi j/ T_\text{exp} = \omega_j-\mu$:
\begin{align}
\begin{split}
\label{eq:spectral_series}
\Phi_\text{rad,o}(t)= \sum_j \left[\sum^{N_\text{BH}(\omega_j)}_n s_n \cos(\omega_j t +\delta_n)\right],
\end{split}
\end{align}
where
\begin{align}
\begin{split}
s_n \equiv A(\mu, M_n,f,r^n_\text{o}) \mathcal A(\hat r_\text{o}^n) S(\alpha_n,a_n)\,.
\end{split}
\end{align}
Here, $N_\text{BH}(\omega_j)$ is the number of BHs whose emission falls within the $j^\text{th}$ frequency bin. Note that $N_\text{BH}(\omega_j)$ is itself a random variable subject only to the normalization constraint that $\sum_j N_\text{BH}(\omega_j)$ be equal to the total number of Galactic BHs. 

Each term in the squared brackets of \cref{eq:spectral_series} is a sum over coherent emitters with a fixed frequency $\omega_j$. For $N_\text{BH}(\omega_j)$ large enough, the amplitudes $s_n$ can be thought of as independently sampled from a restricted distribution of sirens whose parameters are constrained to yield emission in the $j^\text{th}$ bin. The phases $\delta_n$ are also assumed to be independent of $s_n$. The calculation then reduces to a classic problem in the theory of random processes \cite{Rayleigh01081880,Rayleigh01041919,RevModPhys.15.1,PapoulisPillai2002}, the study of which dates back to Rayleigh. The spectral decomposition for 
$\Phi_\text{rad,o}(t)$ itself is that of a Rayleigh-distributed random process\footnote{As explained in \cite{Rayleigh01081880,Rayleigh01041919,RevModPhys.15.1,PapoulisPillai2002,Foster:2017hbq,Gramolin:2021mqv,Cheong:2024ose,Bao:2025nsd}, formally,  $\Phi_\text{rad,o}(t) = \sum_j \sqrt{|\tilde \Phi(\omega_j)|^2} R_j \cos(\omega_j t + \delta_j)$, where $\delta_i$ is a random variable distributed uniformly between $0$ and $2\pi$, while $R_i$ is sampled from a Rayleigh distribution with unit variance.}, while its spectral power density is
\begin{subequations}
\label{eq:power_spectral_density}
\begin{align}
\langle\Phi^2_\text{o}(t)\rangle =\sum_j \Delta \omega_j \frac{|\tilde \Phi_\text{rad,o}(\omega_j) |^2}{\Delta \omega_j}
\end{align} 
and 
\begin{align}
\frac{|\tilde \Phi_\text{rad,o}(\omega_j)|^2}{\Delta \omega_j} = \frac{ N_\text{BH}E_{\omega_j}\left(s_n^2\right)}{2}\,,
\end{align}
where
\begin{align}
\begin{split}
\label{eq:conditional_expectation_value}
{}&E_{\omega_j}\left(s_n^2\right) = \int  \mathrm dM_n  \mathrm da_n \mathrm d r_\text{o}^n \mathrm d^2 \Omega_\text{o}^n \mathrm d^3 \bm v^n_\text{o}\delta\left(\tilde\omega^\text{o}_\phi -\tilde\omega_j\right)\,s_n^2\\{}&\times  p_M(M_n)p_{a_*}(a_n)p_{r_\text{o}^\text{BH}}(r_\text{o}^n)p_{\hat r_\text{o}^\text{BH}}\left(\hat r_\text{o}^n\right) p_{\bm v_\text{o}^\text{BH}}(\bm v^n_\text{o})\,
\end{split}
\end{align}
\end{subequations}
is an expectation value computed over the restricted ensemble with $\tilde\omega^\text{o}_\phi =\tilde\omega_j$. Here, the various $p_x(x^n)$ are the unit-normalized distributions for the BH mass, spin magnitude, observer distance, angular position, and velocity, respectively. We have additionally assumed that each of these parameters is independently distributed. 
Equation \eqref{eq:conditional_expectation_value} effectively counts the number of ways a BH from the ensemble can have emission falling in the frequency bin $\omega_j$, as enforced by the delta function $\delta(\tilde \omega_\phi^\text{o}-\tilde \omega_j)$, weighted by the amplitude squared of the emitted field when those parameters are obtained.

It is sometimes useful to think in terms of velocity bins $\Delta v_j$, rather than frequency bins $\Delta \omega_j$, through the kinetic energy relation
\begin{equation}
\tilde \omega_j \equiv \mu v_j^2/2\,.
\end{equation}
Equivalently, we can define a power spectral density per unit \emph{velocity} 
\begin{equation}
\label{eq:velocity to frequency}
E_{v_j}(s_n^2)\equiv\mu v_jE_{\omega_j} (s_n^2)\,.
\end{equation}
While, as stressed in \cref{sec:sirens_signal}, each siren individually emits coherently over its volume, the emissions of the $N_\text{BH}$ Galactic BHs are uncorrelated among them. Equation \eqref{eq:power_spectral_density} formalizes the intuition that the \emph{power} emitted by $N_\text{BH}$ uncorrelated BH sirens should scale \emph{linearly} with $N_\text{BH}$.

As written, \cref{eq:power_spectral_density} is in terms of variables defined in the rest frame of a single siren and can be thought of as an average over a distribution of \emph{observer}, rather than BH, parameters. It is instructive to rewrite the result from the point of view of the observer. First, because $\bm v^n_\text{o} = -\bm v_n^\text{o}$, one can re-express things in terms of the velocity of the BHs as viewed by the observer, $\int [\dots]\,p_{\bm v^\text{BH}_\text{o}}(\bm v^n_\text{o}) \mathrm d^3 \bm v^n_\text{o}$ = $\int [\dots]\,p_{\bm v^\text{o}_\text{BH}}(\bm v_n^\text{o}) \mathrm d^3 \bm v_n^\text{o}$. 
Second, the integration over the observer's distance and angular position is recognized as a joint integration over both the position $\bm r^\text{o}_n$ of the BHs in the observer's frame and the ``intrinsic'' directed angular momentum vector $\bm a_n$ of the BHs: $\int [\dots]\,p_{a_*}(a_n)p_{\hat r_\text{o}^\text{BH}}\left(\hat r_\text{o}^n\right)p_{r_\text{o}^\text{BH}}(r_\text{o}^n) \mathrm da_n \mathrm d^2 \Omega_\text{o}^n\mathrm d r_\text{o}^n   = \int [\dots]\,p_{\bm a_*}(\bm a_n)p_{\bm r^\text{o}_\text{BH}}(\bm r^\text{o}_n) \mathrm d^3\bm a_n \mathrm d^3 \bm r^\text{o}_n$.

Therefore, \cref{eq:conditional_expectation_value} can be thought of as originating from an integration over a $6+3+1$ dimensional phase space $\int \mathrm d^3 \bm r^\text{o}_n \mathrm d^3 \bm v_n^\text{o} \mathrm d^3\bm a_n \mathrm dM_n$. This is reminiscent of a kinetic theory description of a gas of classical point particles, additionally characterized by two continuously distributed ``internal'' physical degrees of freedom, the mass $M_n$ and a (classical) continuous spin vector $\bm a_n$. This perspective conjures up an image of a classical ``BH gas'' situated at the Galactic Center and radiating away scalar emissions to spatial infinity. In particular, it is clear that such an integration can be performed not only in the observer's frame, but any other frame of convenience, such as the Galactic frame (\cref{fig:galactic_frame}). As shown below, exploiting frame independence allows for a more straightforward computation of some of the integrals.
\begin{figure}[h!]
    \centering
    \includegraphics[width=1\linewidth]{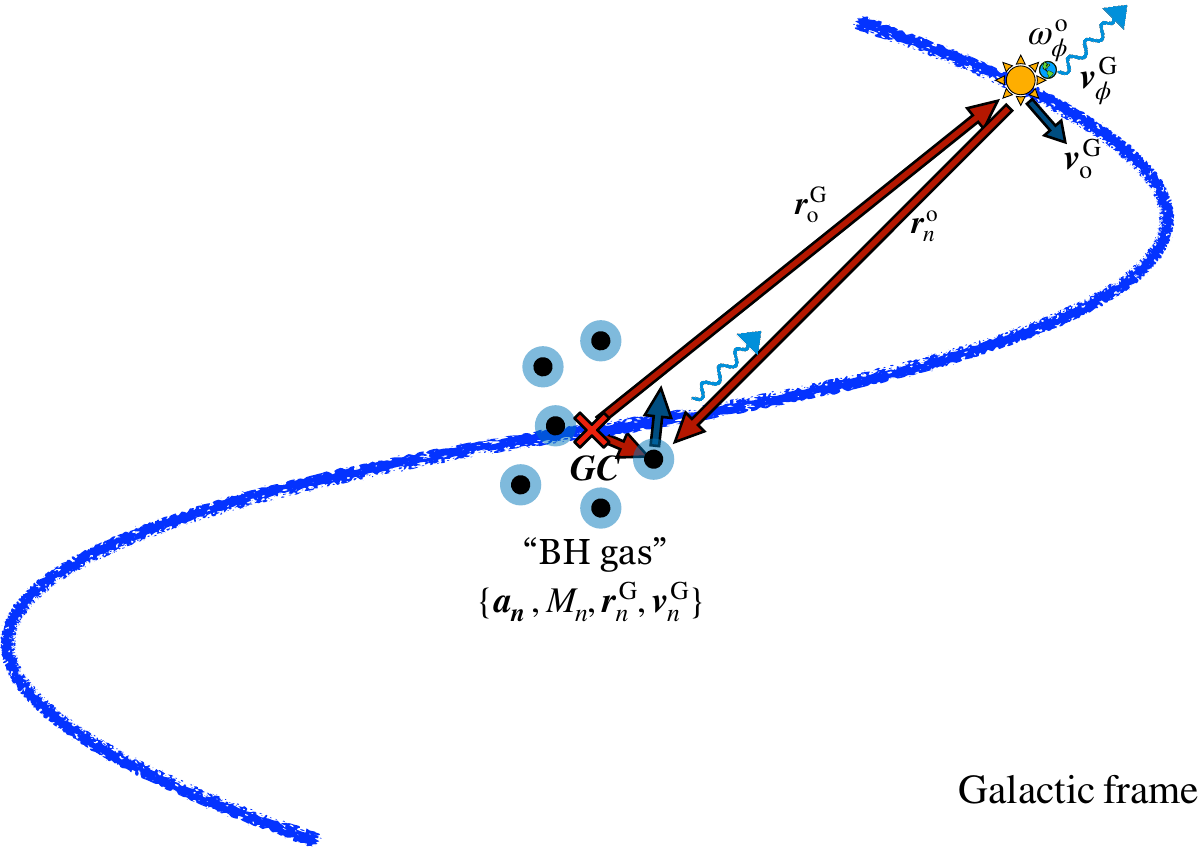}
    \caption{In the Galactic frame, with one of its arms represented as a blue line, we compute the sum total scalar emission of the set of $\sim 10^8$ BHs near the Galactic Center. The power spectral density, \cref{eq:power_spectral_density}, is obtained from the incoherent sum of coherent emitters. It can be recast in the form of a kinetic theory calculation of the radiative power emanating from a ``gas'' of BHs at the Galactic Center, with mass and (continuous, classical) spin internal degrees of freedoms. Because both the observer and the BHs are bound to the Galaxy, their velocity is small relative to that of the emitted scalars of suitably large mass $\mu$. Position vectors are in red, velocity vectors are in blue-green.}
    \label{fig:galactic_frame}
\end{figure}

Evaluating \cref{eq:conditional_expectation_value} requires specifying the distribution of BH and observer parameters. In particular, it readily simplifies if the BH spins are uniformly distributed in the Galactic frame
. In this case, the \emph{observer} is equally likely to be in any solid angle around the BH, defined relative to the BH spin axis: $p_{\bm a_\star}(\bm a_n) = p_{a_\star}(a_n)(4\pi)^{-1} \implies p_{\hat r^\text{BH}_\text{o}}(\hat r_\text{o}^n) = (4\pi)^{-1}$. 
The integral over the angular factor is then done by using the fact that we have already normalized it (\cref{sec:sirens_signal}): $\int|\mathcal A|^2\mathrm d^2\Omega_\text{o}^n = 1$, and
\begin{align}
\begin{split}
{}&E_{\omega_j}\left(s_n^2\right) = (4\pi)^{-1}\int  \mathrm dM_n  \mathrm da_n \mathrm d r_\text{o}^n \mathrm d^3 \bm v^n_\text{o}\delta\left(\tilde\omega^\text{o}_\phi -\tilde\omega_j\right)\,A^2 S^2\\{}&\times  p_M(M_n)p_{a_*}(a_n)p_{r_\text{o}^\text{BH}}(r_\text{o}^n)p_{\bm v_\text{o}^\text{BH}}(\bm v^n_\text{o}) \qquad \text{(isotropic spins)}.
\end{split}
\end{align}

Further, both the position of the observer and the spatial distribution of BHs (e.g.,\,Eq.\,\eqref{eq:spatial_distribution}) are most easily described in the Galactic frame. The remaining spatial integral over $A^2\propto (\bm r_\text{o}^n\cdot \bm r_\text{o}^n)^{-1}$ is therefore,
\begin{align}
\begin{split}
\label{eq:spatial integral}
{}&\frac{1}{R^2} \equiv \int \frac{p_{r_\text{o}^\text{BH}}(r_\text{o}^n) \mathrm d r_\text{o}^n }{\bm r_\text{o}^n\cdot \bm r_\text{o}^n}= \int \frac{p_{\bm r_\text{BH}^\text{G}}( \bm r_n^\text{G})\mathrm d^3 \bm r_n^\text{G}}{|\bm r_n^\text{G}-\bm r_\text{o}^\text{G}|^2}\\
\approx{}& \int  \frac{p_{r_\text{BH}^\text{G}}(r_n^\text{G})\mathrm dr^\text{G}_n
}{\left||r_\text{o}^\text{G}|^2 - |r_n^\text{G}|^2\right|}\\
\approx{}& \frac{1}{|r_\text{o}^\text{G}|^2}\left(1+\frac{2\int |r_n^\text{G}|^2 p_{r_\text{BH}^\text{G}}(r_n^\text{G})\mathrm dr_n^\text{G}}{|r_\text{o}^\text{G}|^2}+\dots\right),
\end{split}
\end{align}
where, in the second line, the observer is assumed to be within the thin Galactic disk. The integral $1/R^2 \approx |r_\text{o}^\text{G}|^{-2}$ is dominated by the distance $r_\text{o}^\text{G} \approx 8\,\kpc$ between the Earth and the Galactic Center, which is much greater than the thickness of the bulge, $\mathcal O(0.2)\,\kpc$ \cite{wegg2013mapping}, justifying the thin-disk approximation. 
The last line further exhibits that the signal has little dependence on the precise radial distribution of BHs within the Galactic disk as long as it is predominantly towards the Galactic Center. Note that, in principle, the integral diverges if a BH is very close to the observer, $1/|\bm r_n^\text{G}-\bm r_\text{o}^\text{G}|\rightarrow \infty$. For that reason, we have included a near-BH cutoff distance. When numerically calculating the spatial component, we assume that there is no BH closer than Gaia BH1, the closest detected BH to date; 0.478\,kpc away \cite{El-Badry:2022zih}. Varying the cut-off to one thousandth of this distance only changes the numerical integral by $\mathcal O(1\%)$. The full calculation gives an effective distance of $R_\text{eff}=7.3$\,kpc.

All in all, under these simplifying assumptions, \cref{eq:conditional_expectation_value} obtains the form
\begin{align}
\begin{split}
{}&E_{\omega_j}(s_n^2) \approx 2.1^2\times \frac{Gf^2}{4\pi R^2}\int  \alpha_n^4  GM^2_n S^2(\alpha_n,a_n)\\\times{}& p_M(M_n)p_{a_*}(a_n)p_{\bm v^\text{o}_\text{BH}}(\bm v^\text{o}_n)\mathrm dM_n \mathrm da_n \mathrm d^3 \bm v^\text{o}_n \delta\left(\tilde \omega_\phi^\text{o}-\tilde \omega_j\right).
\label{eq:5D integral}
\end{split}
\end{align}
Note that if we were to relax the assumption of isotropic orientation of BH spins relative to the Galactic frame, the spectral properties of the signal would not change, nor the scaling of the intensity with $f^2$, $M_n^2$, or $\alpha_n$; the prefactor $1/4\pi R^2$ and the directionality of the signal (\cref{sec:gradient series}), however, would be modified.

\subsubsection{Population variables, bandwidth, and Doppler broadening from source-observer motion}
\label{sec:population variables}

The isotropy in the BH spin distribution and the relative thinness of the Galactic plane have allowed us to simplify what is, in principle, a 10-dimensional integration to a 5-dimensional one, as described by \cref{eq:5D integral}.
The remaining integration $\int[\dots]\,p_{\bm v^\text{o}_\text{BH}}(\bm v^\text{o}_n)\mathrm d^3 \bm v^\text{o}_n$ accounts for the spread in BH, or equivalently, observer velocities. This leads to Doppler broadening and an overall Doppler shift of the spectrum via the velocity dependence of the observed frequency: $\delta(\omega^\text{o}_\phi-\omega_j)=\delta(\omega^\text{o}_\phi(\bm v^n_\text{o})-\omega_j)$, through \cref{eq:Doppler_formula}. For concreteness, one might consider a Maxwell-Boltzmann velocity distribution for the BH gas globally at rest in the Galactic frame. Then,
\begin{equation}
p_{\bm v^\text{o}_\text{BH}}(\bm v^\text{o}_n) = \frac{1}{\pi^{3/2}\sigma_s^3}\exp\left[-\frac{(\bm v_n^\text{o}-\bm v_\text{o}^\text{G})^2}{\sigma_s^2}\right],
\end{equation}
where $\sigma_s^2$ is the velocity dispersion of the BHs in the Galactic frame.

However, as we have already argued, the velocities of both Galactic BHs and Galactic observers are, by definition, below the Galactic escape velocity scale, while the stochastic signal computed in this section is only valid for scalars for which \cref{eq:escape_velocity} holds. That is, relative to the above, the parametric separation $\{\sigma_s,v_\text{o}^\text{G}\}\ll v_\text{esc} \ll v^n_\phi$ holds. The effects of Doppler broadening from the relative motion of the BHs and Earth in the Galactic frame are therefore subdominant. 

This conclusion is further affirmed, \emph{a posteriori}, once the signal of an ensemble of BHs nearly at rest is conveniently parametrized in terms of \emph{population variables}. For BHs at rest relative to the observer, the velocity distribution is simply $p_{\bm v^\text{o}_\text{BH}}(\bm v^\text{o}_n)\rightarrow \delta^3(\bm v^\text{o}_n)$,  such that $\delta(v_\phi^\text{o}-v_j) \rightarrow \delta(v_\phi^\text{BH}-v_j)$. The integration $\int[\dots] \mathrm dM_n$ then readily yields

\begin{align}
\begin{split}
E_{v_j}(s_n^2) \approx{}&158.8\times\frac{f^2}{4\pi}\frac{1}{R^2\mu^2}\alpha_j^4 M_jv_j p_M(M_j)\mathcal S^2(\alpha_j)\,,
\end{split}
\end{align}
where 
\begin{equation}
\label{eq:spin factor}
\mathcal S^2(\alpha_j) \equiv \int  S^2(\alpha_j,a_n) p_{a_*}(a_n)\ \mathrm da_n\,,
\end{equation}
and the \emph{bin mass} $M_j$ is defined through
\begin{equation}
v_j = \frac{GM_j\mu}{6},
\end{equation}
and $\alpha_j = GM_j\mu$.
In other words, in the limit of vanishingly small BH velocities, there is a one-to-one correspondence between the observed velocity (observed frequency) and the mass of the BH siren emitter. The intrinsic spread of the power spectral density, \cref{eq:power_spectral_density}, is therefore tied to the spread of the mass distribution of Galactic BHs.

\begin{figure}

    \includegraphics[width=
\linewidth]{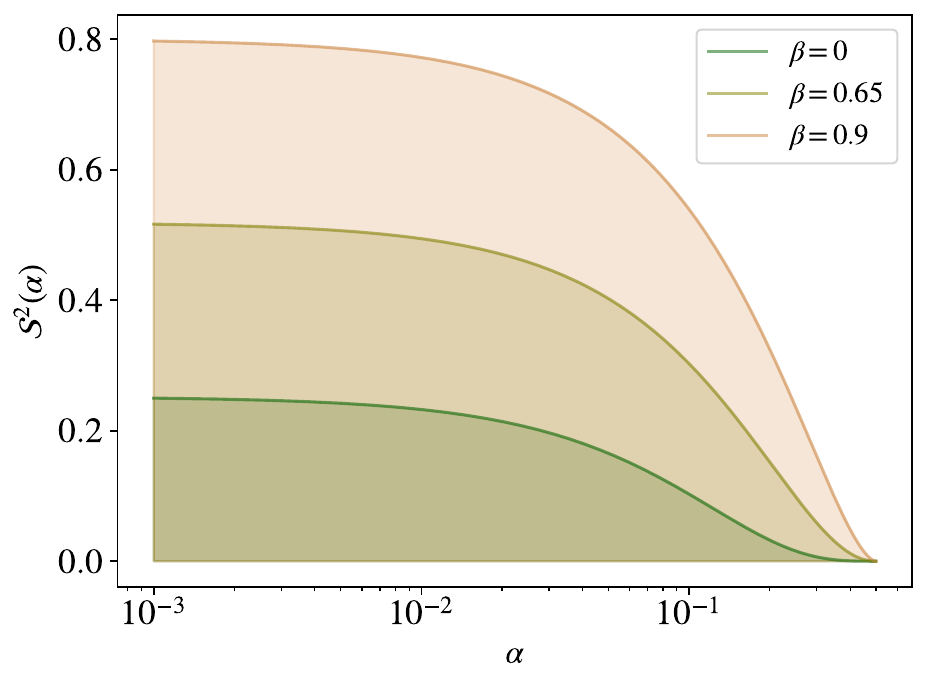}
    \caption{The effective ``spin'' factor $\mathcal{S}^2(\alpha)$ (Eq.\,\ref{eq:spin factor}) of a BH siren ensemble assuming a power-law distribution of spins (Eq.\,\ref{eq:spindistro}). Astronomical observations \cite{Draghis:2023vzj} favor a ``pitch'' parameter $\beta\approx0.65$. $\mathcal S^2 (\alpha)$ is nearly independent of $\alpha$ at small $\alpha$ with $\beta$ determining the value there. However, as $\alpha$ increases, the SR condition (Eq.\,\ref{eq:SR_condition_2}) becomes stricter; the average radiation is reduced as the subpopulation of BHs with low $a_\star$ are no longer sirens.
    }
    \label{fig:S(omega)vsalpha}
\end{figure}

For fixed $\mu$, the numbers $\alpha, v^\text{BH}$ and $\omega^\text{BH}$ characterize the emission of a single BH siren. It is desirable to define corresponding population-level numbers for the characterization of \emph{populations} of BHs. By dimensional analysis, the BH mass distribution $p_M(M_n) \sim M_s^{-1}\hat p_M(M_n/M_s)$ can be written as a unitless distribution $\hat p_M$ that must be characterized by \emph{some} mass scale $M_s$; this is true of \cref{eq:mass_distribution}, but also more generally. We can use the value $M_s$ to define a \emph{scale} $\alpha_s$, \emph{scale velocity} $v_s$ and \emph{scale frequency} $\tilde \omega_s$ that characterize the stellar-mass BH population through
\begin{equation}
v_s = \frac{\alpha_s}{6} = \frac{GM_s\mu}{6},
\end{equation} 
and
\begin{equation}
\tilde \omega_s = \frac{1}{2}\mu v_s^2.
\end{equation} 
Quantitatively, 
\begin{equation}
v_s  \approx (5\times 10^3\,\,\km/\s) \left(\frac{\alpha_s}{0.1}\right),
\end{equation}
while
\begin{equation}
\tilde\omega_s \approx (1.4\times 10^{-2}\text{ Hz})\times \left(\frac{\alpha_s}{0.1}\right)^2 \left(\frac{\mu}{100\text{ Hz}}\right).
\end{equation}
Then, we can re-express
\begin{align}
\begin{split}
\label{eq:scale power density}
{}&E_{v_j}(s_n^2)\\ {}&\approx158.8\times\frac{f^2}{4\pi}\frac{G^2M_s^2}
{R^2}\alpha_s^2v_s\left(\frac{v_j}{v_s}\right)^6\hat p_M\left(\frac{v_j}{v_s}\right)\mathcal S^2(\alpha_j)\,.
\end{split}
\end{align}
The spin factor $\mathcal S^2(\alpha_j)$ predominantly accounts for the fact that a decreasing fraction of BHs -- those with spins larger than $a_\text{crit}(\alpha_j)$ (Eq.\,\ref{eq:a crit}) --
can achieve the SR condition for larger $\alpha_j \propto v_j$. Conversely, as long as $\alpha_j \ll 0.5$, a BH can be a siren regardless of spin, and $\mathcal S^2(\alpha_j)$ is nearly constant. This is illustrated in \cref{fig:S(omega)vsalpha}.

The
\emph{shape} of the spectrum at small $\alpha_j$ is therefore predominantly set by the velocity scale $v_s$ (equivalently the frequency scale $\tilde \omega_s$) and the shape of BH mass distribution $\hat p_M$. Assuming the exponential form, \cref{eq:mass_distribution}, for the mass distribution, a reasonable estimate for the FWHM linewidth $\Gamma_v$ as a function of $\{\mu,M_s\}$ is therefore
\begin{equation}
\label{eq:velocity_linewidth}
\Gamma_v\,(\mu,M_s) \approx
v_s\times \min\left[5.8,\,\frac{1}{2\alpha_s}\right] \,, 
\end{equation}
where $\alpha_s \leq 0.5.$

Mathematically, the effect of the BH velocity distribution in \cref{eq:conditional_expectation_value} is to obfuscate the definite relationship that exists, for BHs at rest, between the bin velocity and the corresponding bin mass through the delta function $\delta(\tilde \omega_\phi^\text{o}(\bm v_n^\text{o})-\tilde\omega_j)$. Thereby, it smears the signal beyond the inherent linewidth, \cref{eq:velocity_linewidth}. Thus, \emph{a posteriori}, we find Doppler broadening (and the Doppler shift due to the motion of the Earth relative to the Galactic frame) can indeed be ignored as long as the velocity spread of the BH distribution and the Earth's velocity in the Galactic frame are less than the inherent breadth of the signal, \cref{eq:velocity_linewidth}. In light of the prefactor in \cref{eq:velocity_linewidth}, this yields the (slightly relaxed) criterion $\alpha_s \gtrsim 2\times 10^{-3}$. Moreover, for $M_s \sim 10\,M_\odot$, the requirement that SR growth of the cloud be fast relative to Galactic timescales $\mathcal O(1-10\,\Gyr)$ gives $\alpha_n \gtrsim 10^{-2}$ or $v^n_\phi \gtrsim 500\,\km/\s$, and is therefore in fact slightly more constraining (\cref{eq:SR timescale}).

\begin{figure}[htb]
     \centering
    
         \includegraphics[width=.5\textwidth]{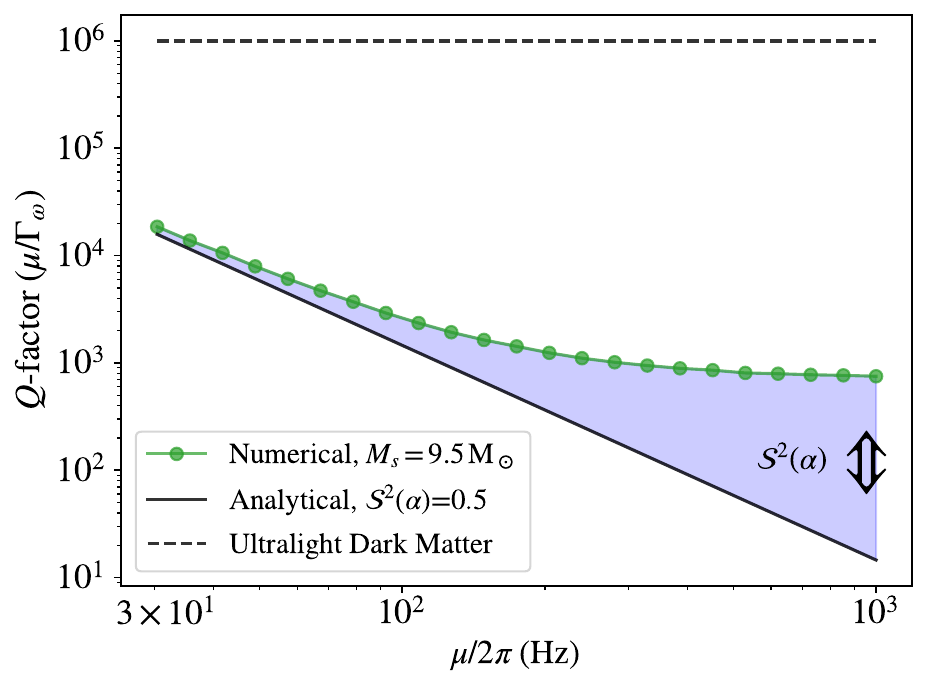}
        
        \caption{$Q$-factor $\mu/\Gamma_\omega$ of the ensemble scalar emission of stellar-mass BH scalar sirens in the Milky Way, shown here for fixed $M_s =9.5M_\odot$. The linewidth $\Gamma_\omega$ scales as the scale frequency $\tilde\omega_s(\mu)$ of the population. At low $\alpha_s(\mu)$ (low $\mu$), the linewidth is well approximated by $58\,\tilde \omega_s$ (Eq.\,\ref{eq:frequency linewidht estimate}). Towards higher $\alpha_s(\mu)$, only the ensemble becomes further restricted by the fraction of BHs with spins large enough to undergo SR. This is quantified through the effect of the spin factor $\mathcal S^2(\alpha_j)$, \cref{eq:spin factor}, which is considered in the numerical calculation of  $\mathcal S^2(\alpha_j)$.
        }
    \label{fig:freqdeltafreq}
 \end{figure}

The frequency-space power spectral density can be obtained using \cref{eq:velocity to frequency}:
\begin{align}
\begin{split}
\label{eq:scale power spectral density frequency}
{}&E_{\omega_j}(s_n^2)\\ {}&\approx\frac{158.8}{\mu}\times\frac{f^2}{4\pi}\frac{G^2M_s^2}{R^2}\alpha_s^2\left(\frac{\tilde \omega_j}{\tilde \omega_s}\right)^{5/2}\hat p_M\left(\sqrt{\frac{\tilde \omega_j}{\tilde \omega_s}}\right)\mathcal S^2(\alpha_j).
\end{split}
\end{align}
In that case, the FWHM linewidth as a function of $\{\mu,M_s\}$ is (\cref{fig:freqdeltafreq})
\begin{equation}
\label{eq:frequency linewidht estimate}
\Gamma_\omega\,(\mu,M_s) \approx \tilde\omega_s\min\left[58,\,\frac{1}{4\alpha_s^2}\right].
\end{equation}
Rewriting the linewidth as the $Q$-factor ($\mu/\Gamma_\omega$) allows for a more straightforward comparison to ALP DM. ALP DM must be virialized in the galaxy, and it is expected to follow a Maxwell-Boltzmann distribution with characteristic speed $v_\mathrm{vir}\approx 300\,\mathrm{km/s}\approx 10^{-3}$, which, in turn, gives $Q_\mathrm{DM}\approx 1/(v_\mathrm{vir})^2\approx10^6$, independently of $\mu$. This is in contrast to $Q_\mathrm{BH}$, the Q-factor of the BH siren ensemble, which scales with $\mu$ as $Q\propto \mu/\tilde\omega_s \approx 1/\mu^2$, until the spin factor $\mathcal S^2$ becomes significant and suppresses the signal at higher frequency, or equivalently, higher BH mass and $Q_\mathrm{BH}$. Figure \ref{fig:freqdeltafreq} shows the $Q$-factor of the BH scalar ensemble as a function of $\mu$, together with the  $Q$-factor of ALP DM.

Over the whole explored mass range, the BH scalar siren signal is significantly broader in frequency space than DM. This affects the signal-to-noise ratio (SNR) for an experiment with a fixed measurement time $T$. In particular, the coherence time $\tau$ of a signal is $\propto 1/\Gamma$. Below the coherence time, the SNR scales coherently as $\sqrt{T}$, while above the scaling is to the fourth-root. This means that, for $T=n\tau$, with $n$ an integer, the scaling is $\propto \tau^{1/2}n^{1/4}\,$\cite{budker_proposal_2014,Gavilan-Martin:2024nlo}. Comparing the BH-scalar-siren ensemble SNR scaling to that of ALP DM yields
\begin{equation}
\label{eq:linewidthSNRscaling}
    \frac{\mathrm{SNR}_\mathrm{BH}}{\mathrm{SNR}_\mathrm{DM}} \propto \frac{\tau_\mathrm{BH}^{1/2} n_\mathrm{BH}^{1/4}}{\tau_\mathrm{DM}^{1/2} n_\mathrm{DM}^{1/4}}=\left( \frac{\tau_\mathrm{BH}}{\tau_\mathrm{DM}}\right)^{1/4}=\left( \frac{\Gamma_\mathrm{DM}}{\Gamma_\mathrm{BH}}\right)^{1/4}.
\end{equation}

For frequency-tuned resonant searches, the natural frequency step is set by the sensitive bandwidth. This bandwith is the larger of the detector bandwidth and the signal linewidth (see, for example, Ref.\,\cite{zhang2024frequency}). 
Because the BH-scalar-siren ensemble signal is intrinsically broader than the expected signal from scalar DM (Fig.\,\ref{fig:freqdeltafreq}), one may take correspondingly larger frequency steps without risking that the scan ``hops over'' the line and misses the signal. 
However, a broader linewidth also implies a shorter coherence time ($\tau_c \sim 1/\Gamma_\omega \sim Q/\omega$) and therefore reduced coherent averaging within each step, as \cref{eq:linewidthSNRscaling} shows. 
For example, in frequency-scanning analyses of spin-based NMR haloscopes (such as the Cosmic Spin Precession Experiment (CASPEr) \cite{budker_proposal_2014,aybas_casper-e_2021} as considered in Ref.\,\cite{zhang2024frequency}), the two effects largely compensate: while fewer steps are required, the sensitivity per step improves only slowly once the integration time exceeds the signal coherence time. 
For a broadband experiment, there is no trade-off, since there is no frequency-scanning. In this case, lower $Q$ strictly translates to worse SNR.

\subsubsection{Gradient series}
\label{sec:gradient series}
We have so far focused on the stochastic time series produced by the sum of scalar field amplitudes from Galactic BHs. Another observable of experimental interest is the projection of spatial gradients of the field onto a certain direction in the lab. Let $\bm J$ be a (classical) vector specifying this direction. Reproducing the steps analogous to those above, but using \cref{eq:gradient} instead of \cref{eq:observed_field} yields
\begin{subequations}
\label{eq:gradient PSD}
\begin{align}
\begin{split}
\left\langle \left(\bm J \cdot \bm \nabla \Phi_\text{rad,o}(t)\right)^2\right\rangle = \sum_j \Delta \omega_j \frac{N_\text{BH}E_{\omega_j}\left((\bm J\cdot \mu \bm v^\text{o}_\phi)^2s_n^2\right)}{2}\,,
\end{split}
\end{align}
for the stochastic series
\begin{align}
\begin{split}
\bm J \cdot \bm \nabla \Phi_\text{rad,o}(t)
\equiv \sum^{N_\text{BH}}_{n\,\in\,\{\text{galactic BHs}\}} \bm \nabla_\text{obs} \phi_{\text{o},n}.
\end{split}
\end{align}
\end{subequations}
A key difference relative to the previous case is the additional presence of the observed velocity of the scalar $\bm v^\text{o}_\phi$ in the integrand \emph{outside} the spectral delta function $\delta\left(v^\text{o}_\phi-v_j\right)$. Indeed, the coupling between the gradient of a scalar field and an experimental apparatus is sometimes called the ``wind'' coupling due to this \emph{directional} nature \cite{stadnik2014axion,graham2018spin}.

In the case of scalar DM, the ``DM wind'' is generated by the relative motion between the Earth and the local DM density; the DM wind ``blows'' primarily in the direction of the flow of DM relative to the Earth. In fact, the DM wind signal is non-zero even when there is no bulk motion of the Earth relative to the DM frame because of the directional spread of the DM particles' velocities \cite{lisanti_stochastic_2021, Gavilan-Martin:2024nlo}.

In the case which interests us, namely that of BH sirens towards the Galactic Center, the scalar wind primarily blows from the Galactic Center, with a small dispersion around that direction coming from the spatial spread of the BHs (in addition to the velocity spread of the BHs, the effects of which are, as above, sub-dominant for large enough $\alpha_s$).

\begin{figure}
    \centering
    \includegraphics[width=1\linewidth]{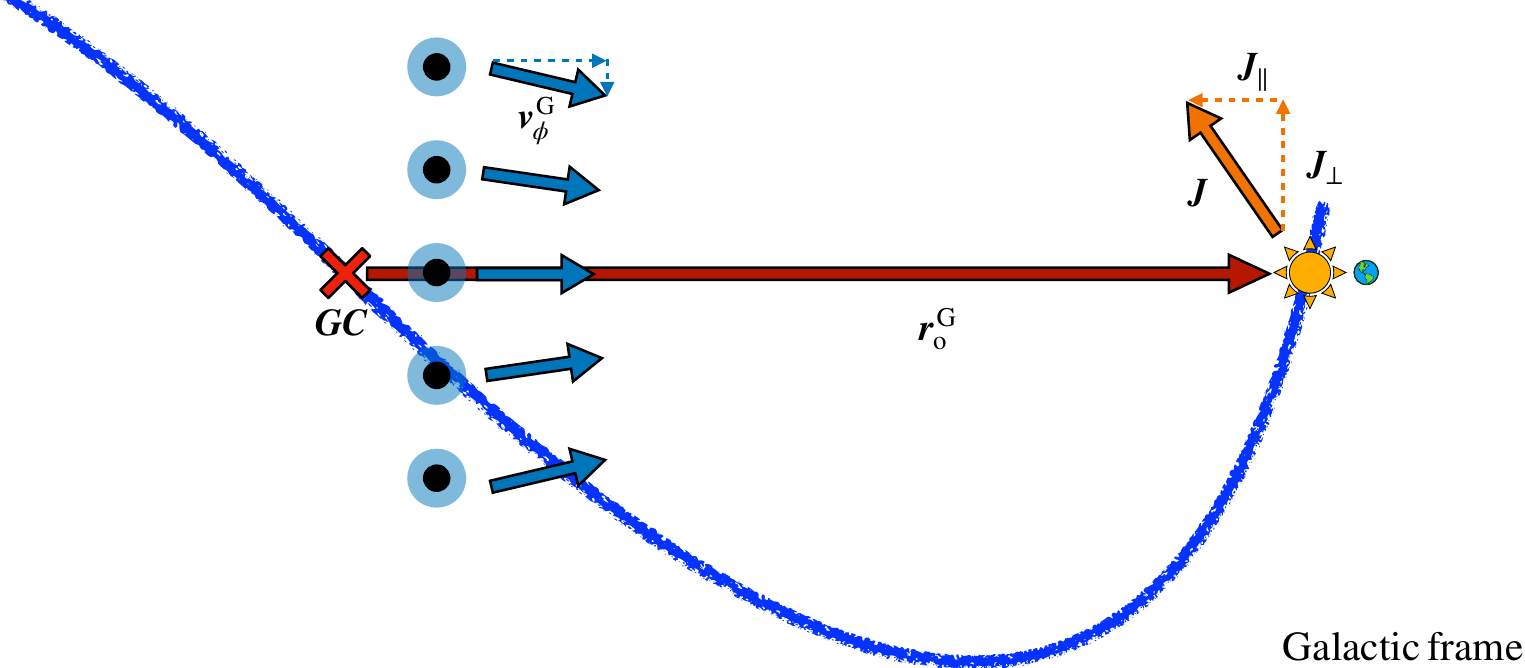}
    \caption{Schematic depiction for the calculation of the wind signal. The vector $\bm J$ (orange) represents the directional axis of an experimental apparatus on Earth, such as set of polarized nuclear or atomic spins. $\bm J_\parallel$ is the component of $\bm J$ parallel to the line of sight to the Galactic Center, while $\bm J_\perp$ is the component perpendicular to that line of sight. The finite angular spread  of the spatial distribution of BHs  relative to the line of sight  causes a spread in the direction of the constituents of the ``scalar wind''. The angular spread is $\sim r_s/r_\text{o}^\text{G}$ for $\bm J_\perp$ lying in the Galactic plane and suppressed by the relative thickness of the Galactic disk for $\bm J_\perp$ perpendicular to the Galactic plane. As a result, there is a non-zero signal for every direction of the detector vector $\bm J$. The sum of projections leads to \cref{eq:R_para,eq:R_perp}.}
    \label{fig:BH gradient dispersion}
\end{figure}

It is therefore useful to define the projections $\bm J_\parallel = (\hat {\bm r}_\text{o}^\text{G}\cdot \bm J) \,\hat {\bm r}_\text{o}^\text{G}$, and $\bm J_\perp = \bm J-\bm J_\parallel$ of the vector $\bm J$ in the directions parallel, and plane perpendicular to $\bm r_\text{o}^\text{G}$ (\cref{fig:BH gradient dispersion}). For BHs distributed symmetrically around the Galactic plane, then
\begin{align}
\begin{split}
{}&\left\langle \left(\bm J \cdot \bm \nabla \Phi_\text{rad,o}(t)\right)^2\right\rangle \\={}& \left\langle \left(\bm J_\parallel \cdot \bm \nabla \Phi_\text{rad,o}(t)\right)^2\right\rangle + \left\langle \left(\bm J_\perp \cdot \bm \nabla \Phi_\text{rad,o}(t)\right)^2\right\rangle \,.
\end{split}
\end{align}
For $v_s$ larger than the dispersion in BH velocities,
\begin{subequations}
\begin{align}
\begin{split}
E_{\omega_j}\left(\left(\bm J_\parallel \cdot \mu \bm v_\phi^\text{o}\right)^2s_n^2\right) \approx \left(\mu v_j\right)^2 J_\parallel^2 \frac{R^2 E_{\omega_j}(s_n^2)}{R_\parallel^2}\,,
\end{split}
\end{align}
where
\begin{align}
\begin{split}
\label{eq:R_para}
\frac{1}{R_\parallel^2}\equiv{}& \int \frac{p_{\bm r_n^\text{G}}(\bm r_n^\text{G})\mathrm d\bm r_n^\text{G}}{|\bm r_n^\text{G}-\bm r_\text{o}^\text{G}|^2}\left(\frac{\bm r_\text{o}^\text{G}-\bm r_n^\text{G}}{|\bm r_n^\text{G}-\bm r_\text{o}^\text{G}|}\cdot \hat {\bm r}_\text{o}^\text{G}\right)^2\\
\approx{}& \int\frac{2 |r_{\text{o}}^\text{G}|^6 - r_n^2 |r_{\text{o}}^\text{G}|^4 +4 r_n^4 |r_{\text{o}}^\text{G}|^2  -r_n^6 }{2 |r_{\text{o}}^\text{G}|^2 ( |r_{\text{o}}^\text{G}|^2-r_n^2)^3} p_{r_\text{BH}^\text{G}}(r_n)\mathrm dr_n\\
\sim{}& \frac{1}{|r_\text{o}^\text{G}|^2}\,,
\end{split}
\end{align}
\end{subequations}
where the second line uses the approximation of a thin Galactic disk and the last line emphasizes the parametrics. Similarly, 
\begin{subequations}
\begin{align}
\begin{split}
E_{\omega_j}\left(\left(\bm J_\perp^2 \cdot \mu \bm v_\phi^\text{o}\right)^2s_n^2\right) \approx\left(\mu v_j\right)^2J_\perp \frac{R^2E_{\omega_j}(s_n^2)}{R_\perp^2}\,,
\end{split}
\end{align}
where
\begin{align}
\begin{split}
\label{eq:R_perp}
\frac{1}{R_\perp^2} \equiv{}& \int \frac{p_{\bm r_n^\text{G}}(\bm r_n^\text{G})\mathrm d\bm r_n^\text{G}}{|\bm r_n^\text{G}-\bm r_\text{o}^\text{G}|^2}\left(\frac{\bm r_\text{o}^\text{G}-\bm r_n^\text{G}}{|\bm r_n^\text{G}-\bm r_\text{o}^\text{G}|}\cdot \hat {\bm J}_\perp\right)^2\\
\approx{}& \int \frac{r_n^2}{2|r_\text{o}^\text{G}|^4\left(1+ \frac{r_n^2}{|r_\text{o}^\text{G}|^2}  \right)}p_{r_\text{BH}^\text{G}}(r_n)\mathrm dr_n\\
\sim{}& \frac{1}{|r_\text{o}^\text{{G}}|^2} \left(\frac{r_s}{r_\text{o}^\text{{G}}}\right)^2\,,
\end{split}
\end{align}
\end{subequations}
where in the second line $\bm J_\perp$ is assumed to be in the Galactic plane. Thus, even if $\bm J$ points precisely perpendicular to the line of sight to the Galactic Center (i.e.\,$J^2_\parallel = 0$), the signal remains nonzero but is rather suppressed by the finite width of the Galactic disk: $\left(r_s/r_\text{o}^\text{G}\right)^2 \approx 0.14$ for $\bm J_\perp$ lying within the Galactic plane. This is analogous to the signal from the DM wind, where the spread in the direction of incidence of individual particles makes the signal non-zero even in the absence of a bulk motion of the Earth in the Galactic frame.

Because the amplitude and gradient signals are related by two powers of the velocity, the approximate FWHM linewidths of the gradient signal are 
\begin{subequations}
    \begin{equation}
\Gamma_v\,(\mu,M_s) \approx
v_s\times \min\left[6.7,\,\frac{1}{2\alpha_s}\right]\,,
\end{equation}
\begin{equation}
\Gamma_\omega\,(\mu,M_s) \approx
\tilde \omega_s\times \min\left[93.5,\,\frac{1}{4\alpha_s^2}\right]\,.
\end{equation}
\end{subequations}

\section{Observing stellar-mass MW BH scalar sirens}
\label{sec:observational}

\subsection{Density distribution of scalars in the Galaxy}
The power spectral density (Eq.\,\ref{eq:power_spectral_density}) characterizes the energy density of scalars at a location through the relation 
\begin{equation}
\label{eq:energy density}
\langle\rho_\text{rad,o}(t)\rangle = \mu^2 \langle \Phi_\text{rad,o}^2(t)\rangle,
\end{equation}
valid for non-relativistic scalars. Scalars from Milky Way BH sirens would thus contribute a non-primordial component to the local dark matter density at the Sun–Earth position. We can estimate its size by integrating \cref{eq:scale power density} with the exponential estimate (Eq.\,\ref{eq:mass_distribution}) for the BH mass distribution and ignoring the dependence of the spin factor $\mathcal S^2(\alpha_j)$ on $\alpha_j$ (which is valid for suitably small $\alpha_j$):
\begin{equation}
\label{eq: rho}
\langle\rho_\text{rad,o}(t)\rangle \approx (3.2\times 10^1)\times N_\text{BH}\,\alpha_s^6\frac{f^2}{R^2}.
\end{equation}

\begin{figure}[htb]
    \centering
    
         \includegraphics[width=.5\textwidth]{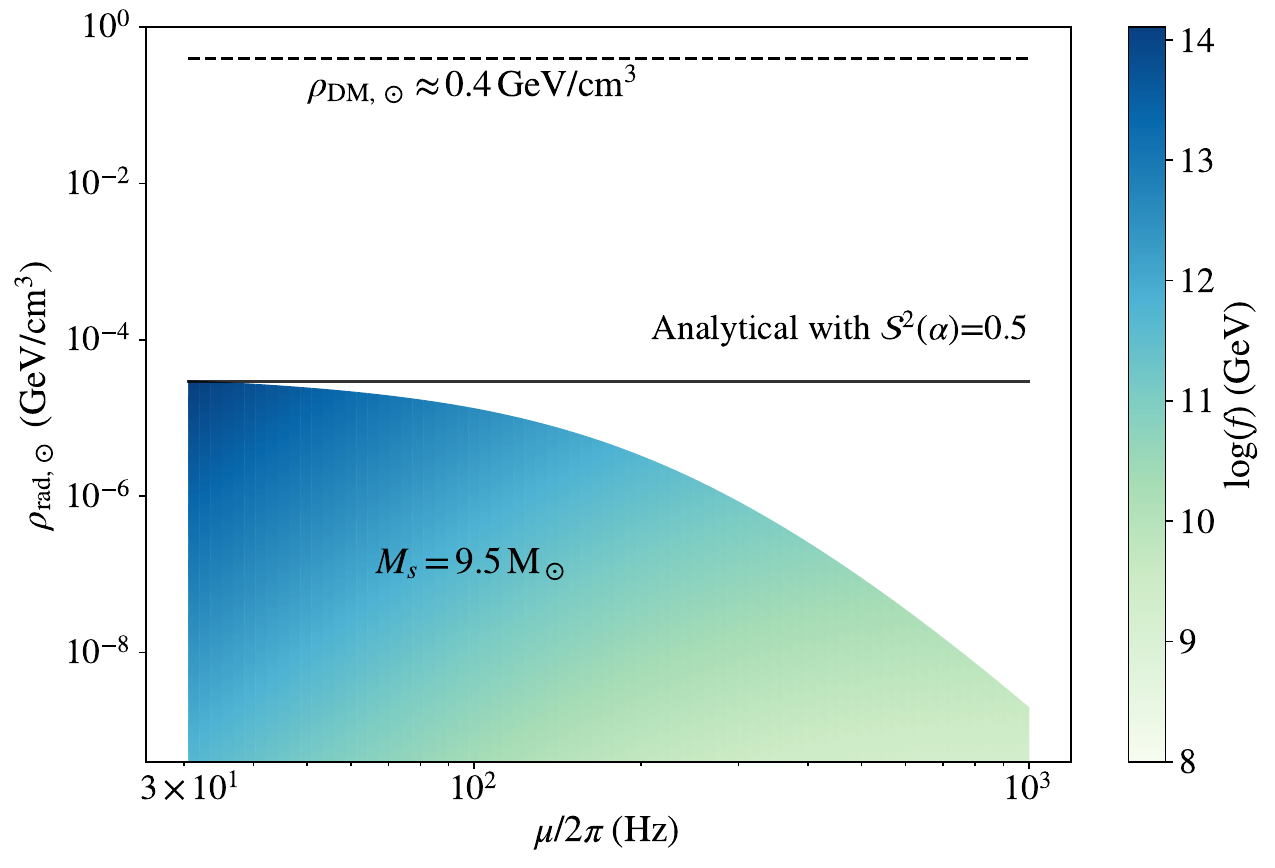}
        
         \caption{Energy density $\langle \rho_{\odot}(t)\rangle$ of scalars emitted by the MW population of stellar-mass BH scalar sirens ($M_s = 9.5 M_\odot$) evaluated at the Sun-Earth location $|\bm r_\odot^\text{G}| \sim 8 \,\kpc$ and different values of $f$ (heat map). For each scalar mass $\mu$, the energy density is maximized for the largest value of $f$ for which sirens have effectively infinite lifetime, \cref{eq:threshold f}. Towards small values of $\alpha_s$, almost all BHs have enough spin to trigger the SR instability and the energy density is well approximated by \cref{eq:approximate energy density}.}
    \label{fig:rhorad}
 \end{figure}

The size of $f$ is limited by \cref{eq:threshold f}, as we require the extraction of angular momentum to be slow enough that every MW BH is still emitting until today. Therefore, an upper bound on the abundance is
\begin{align}
\begin{split}
\label{eq:approximate energy density}
\langle\rho_\text{rad,o}(t)\rangle \lesssim{}&10^{-5}\,\GeV/\cm^3\\\times{}&\left(\frac{N_\text{BH}}{10^8}\right)\left(\frac{8\,\kpc}{R}\right)^2\left(\frac{M_s}{10\,M_\odot}\right)\left(\frac{10 \,\Gyr}{T}\right)\,.
\end{split}
\end{align}
The energy density at the Sun-Earth position is displayed on \cref{fig:rhorad}, as a function of $\mu$ and for $M_s = 9.5 M_\odot.$ As for the linewidth (\cref{fig:freqdeltafreq}), the energy density towards larger $\mu$ (large $\alpha_s$) is further suppressed by the spin factor $\mathcal S^2$.

Equation \eqref{eq:approximate energy density} not only captures the local abundance of scalars at the Sun-Earth location for the purpose of direct detection, but also the spatial profile of the energy density of the emitted scalars throughout the Galaxy. Per \cref{eq:spatial integral}, spatial loci of equal density are approximately spherical for $|r_\text{o}^\text{G}| \gg r_s$ and become more oblate closer to the Galactic Center. Within the Galactic plane, parametrically,
\begin{align} 
\begin{split}
-\frac{1}{2}\frac{\mathrm d \ln \langle \rho_\text{rad,o}(t)\rangle}{\mathrm d\ln \left|r_\text{o}^\text{G}\right|} \sim \begin{cases}
1, & \text{for}\,\,\frac{\left|r_\text{o}^\text{G}\right|}{r_s}\gg 1,\\
\left(\left|r_\text{o}^\text{G}\right|/r_s\right)^2, &\text{for}\,\,\frac{\left|r_\text{o}^\text{G}\right|}{r_s}\ll 1,\\
\end{cases}
\end{split}
\end{align}
flattening towards the center.

Therefore, the flux of scalars emitted from BHs in the Galaxy constitutes a faint, cored companion structure to the baryonic Galactic disk, somewhat reminiscent of the ``dark disks'' known to form in models of DM with dissipative channels \cite{Fan:2013yva,Fan:2013tia,McCullough:2013jma,Randall:2014kta,Roy:2023zar} and excluded in the MW at the $\sim \GeV/\cm^3$ level \cite{Schutz:2017tfp}. The oblate features of the density profile of emitted scalars are not tied, however, to any dissipation in the dark sector, but rather to their originating from BH remnants of stellar matter in the baryonic disk.

\subsection{Standard Model observables}
The flux of scalars from Milky Way BH sirens is of interest to existing and future haloscope experiments sensitive to a local abundance of (non-relativistic) scalars in the light or ultralight mass ranges ($<10\,\mathrm{eV}$) \cite{JacksonKimball:2017elr,Garcon:2019inh,Kimball:2023vxk,Garcon:2017ixh,Jiang:2021dby,Walter:2025ldb,Gavilan-Martin:2024nlo,GNOME:2023rpz,Safronova:2017xyt,Tretiak:2022ndx,Bloch:2019lcy,Bloch:2023wfz,Sikivie:1983ip,ADMX:2019uok,ADMX:2018gho,Sushkov:2023fjw,DMRadio:2022pkf,Caldwell:2016dcw, Millar:2016cjp,Baryakhtar:2018doz,lee_laboratory_2023,wei_dark_2023,Arvanitaki:2024taq}. While such experiments are often designed to look for the scalar as the primordial dark matter, they remain, in principle, capable of searching for a local abundance of scalars with non-primordial origins, such as a solar basin \cite{VanTilburg:2020jvl,Giovanetti:2024rxi,Budker:2023sex}, or the scalars produced by the Galactic BH scalar sirens considered in this work. 

All such experiments rely on there being a coupling between the scalar and one or more (effective) SM operators. For definiteness, we consider here only \emph{linear} interactions:
\begin{equation}
\label{eq:SM operators}
\mathcal L \supset g_{\phi\mathcal O} \phi\, \mathcal O_\mathrm{SM}\,,
\end{equation}
where $\mathcal{O}$ is a generic SM operator and $g_{\phi\mathcal O}$ is the coupling strength. Interactions involving higher-powers of $\phi$ could also probe the scalar siren background.

From an effective field theory (EFT) standpoint, the dimensionful coupling ought to further be expressed relative to the same decay constant $f$ characterizing the strength of self-interactions:
\begin{equation}
g_{\phi\mathcal O} = \frac{\mathcal C_{\phi\mathcal O}}{f}\,,
\end{equation}
for some dimensionless Wilson coefficient $\mathcal C_{\phi \mathcal O}$. For the purpose of tabletop experiments probing low energies \cite{Graham:2013gfa,JacksonKimball:2017elr,Safronova:2017xyt}, 
\begin{align}
\begin{split}
\mathcal O_\text{SM}\in{}& \Big\{F^2, m_\Psi \bar \Psi \Psi, F_{\mu\nu}\tilde F^{\mu\nu},-\frac{i}{2m_\Psi}\bar \Psi\sigma_{\mu\nu}\gamma^5\Psi F^{\mu\nu},\\ {}&\partial_\mu \left(\bar\Psi \gamma^\mu \gamma^5 \Psi\right)\Big\}\,,
\end{split}
\end{align}
where $F^{\mu\nu}$ is the electromagnetic field strength tensor, $\tilde F^{\mu\nu}$ its (Hodge) dual, $\Psi$ is the fermion field of a nucleon, nuclei, atom or electron with particle mass $m_\Psi$, and $\gamma$ and $\sigma$ are the standard Dirac matrices. 

 The first two couplings in \cref{eq:SM operators}, to $F^2$ and $\bar \Psi\Psi$, correspond to $\phi$ modulating the local value of the fundamental electric charge or the mass of $\Psi$, respectively. The  \emph{linear} form of this interaction mediates long-range, spin-independent ``fifth'' forces which are stringently constrained to have sub-gravitational strength. The open fifth-force parameter space forces $|\mathcal C_{\phi\mathcal O}| \lesssim 10^{-4}(f/\mpl)$. Considering that the siren regime forces $f < f_\text{siren}\ll \mpl$, BH sirens are only phenomenologically relevant to models where the Wilson coefficients of those operators are exponentially small. In contrast, linear couplings to the latter three operators in \cref{eq:SM operators}, mediate spin-dependent forces, which are much less constrained \cite{cong2025spin}.

For all but the last operator in \cref{eq:SM operators}, the experimental signal strength associated with incident scalar radiation external to the experiment scales $\propto g_{\phi\mathcal O}^2\left\langle \Phi_\text{rad,o}(t)^2\right\rangle$ and is therefore also characterized by the power spectral density, \cref{eq:power_spectral_density}. Meanwhile the last SM operator can be integrated by parts in the Lagrangian, leading to the so-called pseudo-vector, derivative ``wind'' coupling: $\phi \,\partial_\mu \bar\Psi \gamma^\mu \gamma^5 \Psi \rightarrow - (\partial_\mu \phi) \bar \Psi \gamma^\mu\gamma^5 \Psi$. The signal from a wind of non-relativistic scalars then scales as $g_{\phi\mathcal O}^2\left\langle\left(\bm J \cdot \bm \nabla \Phi_\text{rad,o}(t)\right)^2\right\rangle$, where $\bm J = \langle \bm\sigma\rangle$ is the ensemble averaged spin polarization of the experimental sample of species $\Psi$ \cite{JacksonKimball:2017elr}. The relevant power spectral density for such experiments is therefore \cref{eq:gradient PSD}.

For both derivative and non-derivative interactions, the relevant measure of scalar abundance from a BH siren population scales $\propto f^2$, but only enters the strength of experimental signals when multiplied by $g_{\phi\mathcal O}^2$. Therefore, experimental sensitivities to the signal from BH scalar sirens have no net dependence on $f$ when written in terms of the Wilson coefficient $\mathcal C_{\phi\mathcal O}$. This is because, for fixed $\mathcal C_{\phi\mathcal O}$, smaller $f$ result in smaller clouds and smaller radiated power, but also larger expected couplings to the experiments. Therefore, it is useful to think of the \emph{laboratory signal per unit $\mathcal C^2_{\phi\mathcal O}$.}

Integrating our estimates for \cref{eq:power_spectral_density} and \cref{eq:gradient PSD} yields
\begin{subequations}
\label{eq:theta and wind}
    \begin{align}
    \begin{split}
    \label{eq:theta}
        \langle \theta_\text{rad,o}^2(t)\rangle  \equiv f^{-2}{}&\langle \Phi_\text{rad,o}^2(t)\rangle\approx(3.3\times 10^{-14})^2\\\times{}&\left(\frac{M_s}{10\,M_\odot}\right)^2 \frac{N_\text{BH}}{10^8}\left(\frac{\alpha_s}{0.1}\right)^4\left(\frac{8\,\kpc}{R}\right)^2,
    \end{split}
    \end{align}
    and
    \begin{align}
    \begin{split}
    \label{eq:wind}
        f^{-2}\langle\bm{\hat J_\parallel} \cdot \bm \nabla \Phi_\text{rad,o}^2(t)\rangle \approx{}&(5.6\times 10^{-27}\,\eV)^2\\\times{}&
\,\frac{N_\text{BH}}{10^8}\left(\frac{\alpha_s}{0.1}\right)^8\left(\frac{8\,\kpc}{R}\right)^2.
    \end{split}
    \end{align}
\end{subequations}
Towards larger $\alpha$, the spin factor $\mathcal S^2$ suppresses the signal relative to this estimate. Results obtained numerically are displayed on \cref{fig:sirensvsrelic}. 

Because the production of the scalars at the BH does not require a coupling to the SM particles, the experimental scenario in which BH siren emission is detected with haloscopes directly probes the dimensionless Wilson coefficients $\mathcal C_{\phi\mathcal O}^2 \sim g_{\phi\mathcal O}^2f^2$. Siren production shares this property with cosmological production mechanisms like pre-inflationary cosmic misalignment, and the decay of cosmic string networks \cite{marsh_axion_2016,Dror:2021nyr}. 

In contrast, most astrophysical bounds (e.g.\,\cite{Brockway:1996yr,Grifols:1996id,Payez:2014xsa, DeRocco:2022jyq, Hoof:2022xbe,Dolan:2022kul,Hardy:2016kme, Bottaro:2023gep,OHare:2020wah,Fiorillo:2025zzx,Fiorillo:2025gnd,Candon:2025sdm}, as well as \cite{ParticleDataGroup:2024cfk} and references therein), laboratory fifth-force type experiments (e.g.\,\cite{Schlamminger:2007ht,Adelberger:2003zx,Mostepanenko:2020lqe,Lee:2020zjt,Kapner:2006si,Yang:2012zzb}), and light-shining-through-walls (e.g.\,\cite{CAST:2024eil,Ehret:2010mh}) experiments probe the dimensionful coupling $g_{\phi\mathcal O}^2$ directly, while the DM/haloscope experiments probe the combination $g^2_{\phi\mathcal O}(\Omega_{\phi,\odot}/\Omega_{\text{DM},\odot})^2$. This fact can make it difficult to visualize the reach of BH-siren/haloscope scenarios alongside other probes of the coupling $g_{\phi\mathcal O}$, as $f$ must then be specified independently. 

\begin{figure}[htb]
     \centering
    
        \includegraphics[width=.5\textwidth]{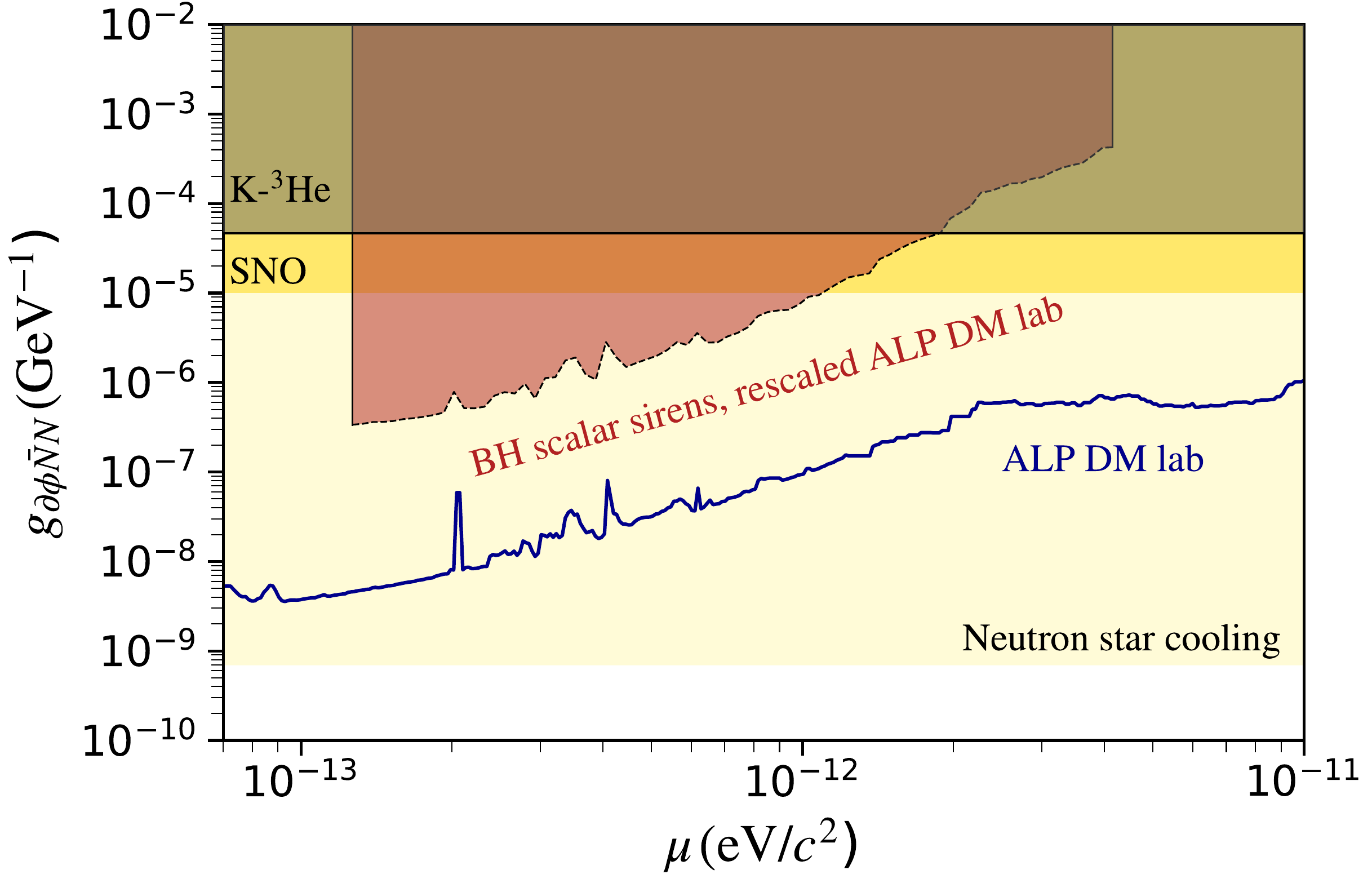}
        
         \caption{
         Estimated sensitivity of existing ALP DM searches to BH scalar sirens in the Milky Way for derivative coupling to neutron spins based on rescaling of ALP DM laboratory limits. The rescale factor is given by \cref{eq:rescale nuclear coupling}. Even though the energy density of the emission is orders of magnitude lower than the estimated energy density of halo DM, \cref{eq:energy density}, the larger velocity of BH scalar sirens emission compensates partially for this. In the red shaded region, the bosonic radiation is much faster than the virial velocity and it escapes the galaxy. Note that the projections do not require any assumptions about the DM nature. The assumptions in this case are related to the properties of the BH population, and the self-interaction scale of the scalar, determined by $f$, which is chosen as the maximum value that fulfills \cref{eq:SR timescale}, making BH sirens effectively infinitely long-lived. As a reference, other limits are shown. Due to the nucleon spin content of the samples used in ALP DM laboratory searches, the limits correspond to couplings to neutron spins; the limits on proton spins are weaker. It is a combination of the limits of NASDUCK \cite{bloch_new_2022} and ChangE \cite{wei_dark_2023}; astronomical constraints from solar bosons (SNO) \cite{bhusal_searching_2021} and neutron star cooling \cite{buschmann_upper_2022}; fifth force search with K-$^3$He comagnetometer \cite{lee_improved_2018}.
         Limits are extracted from \cite{AxionLimits}.
         Note that the sensitivity line for stellar-mass BH sirens should not be understood as an exclusion, as proper reanalysis of existing data would be needed.
         }
     \label{fig:projections}
 \end{figure}

Figure \ref{fig:projections} situates the reach of the BH scalar siren radiation compared to axion DM limits via the nuclear spin coupling. In particular, we display the projected sensitivity for neutron spins. To be able to compare BH siren radiation to other searches, a particular value of $f$ must be chosen. In analogy with DM haloscope searches, which usually set $\Omega_{\phi,\odot}=\Omega_{\text{DM},\odot}$, its maximum allowed value, we set $f$ to the maximum that allows for virtually infinitively-lived BH sirens. 
The scale $f$ is maximized for each individual mass according to \cref{eq:SR timescale}, and its mass scaling is reflected in this figure.
This includes an additional scaling with mass that is not present for fixed $f$, where the maximum reach of the signal tends to peak at higher masses, right before the spin factor $\mathcal S^2(\alpha)$ starts significantly suppressing the signal of slower rotating BHs. The projections assume a BH population with a typical spread of distance from the GC of $R_s=3\,\mathrm{kpc}$, a mass scale of $M_s=9.5\,\mathrm{M}_\odot$, and $N_\mathrm{BH}\approx 10^8$.

We restrict projections to the mass range $\mu\approx[30,1000]\,\Hz\approx[10^{-13}, 2\times10^{-12}]\,\eV$.
In the $\alpha_s>0.15$ (\cref{sec:larger alpha ensemble}) regime multiple levels of the BH siren might be radiating scalars with different velocities and the ensemble of BH sirens contains polychromatic sources of scalars. For $\alpha_s<10^{-2}$, $\tau_\mathrm{SR}$, \cref{eq:SR timescale}, is comparable or larger than the age of the Milky Way.

The projected sensitivity to BH scalar siren radiation, assuming current experimental sensitivity, is shown in red. The estimation is obtained by rescaling the ALP DM limits (in blue) to account for the difference in the signal characteristics. In particular, these are three: the energy density near Earth $\rho_{\text{rad},\odot}$, the linewidth $\Gamma_\omega$ and the typical velocity of the scalars $v_s$. This gives a rescaling factor $\xi_N$ for nuclear spin coupling
\begin{align}
\begin{split}
\label{eq:rescale nuclear coupling}
    \xi_N&=\left(\frac{\rho_\mathrm{rad,\odot}}{\rho_\mathrm{DM}}\right)^{1/2}\left(\frac{Q_\mathrm{BH}}{Q_\mathrm{DM}}\right)^{1/4} \left(\frac{v_s}{v_\mathrm{vir}}\right)\\
    &\approx 10^{-1} \left(\frac{\rho_\mathrm{rad,\odot}}{10^{-5}\mathrm{GeV/cm^3}}\right)^{1/2}\left(\frac{Q_\mathrm{BH}}{10^{3}}\right)^{1/4} \left(\frac{v_s}{0.1}\right)\,,
    \end{split}
\end{align}
which depends on $\mu$. The rescaled projection line is obtained by multiplying $\xi_N$ by ALP DM lab exclusions, as displayed in \cref{fig:projections}. The scalar sirens sensitivity line then corresponds to values of the dimensionless Wilson coefficient $\mathcal C_{\partial \phi \bar NN}\approx [10^9-10^6]$, much larger than the most naive expectation from effective field theory, $\mathcal C_{\phi\mathcal O}=\mathcal O(1)$. The ALP DM line also corresponds to large values of $\mathcal C_{\partial \phi \bar NN}$ if the scalar is taken to be a pre-inflationary cosmically misaligned scalar in this mass range \cite{marsh_axion_2016,Arvanitaki:2019rax} (see also \cref{sec:cosmic misalignment}).

For a coupling to photons, the rescaling is independent of the velocity, and it is
\begin{align}
\begin{split}
    \xi_\phi&=\left(\frac{\rho_{\mathrm{rad},\odot}}{\rho_\mathrm{DM}}\right)^{1/2}\left(\frac{Q_\mathrm{BH}}{Q_\mathrm{DM}}\right)^{1/4}\\
    &\approx 10^{-3} \left(\frac{\rho_{\mathrm{rad},\odot}}{10^{-5}\mathrm{GeV/cm^3}}\right)^{1/2}\left(\frac{Q_\mathrm{BH}}{10^{3}}\right)^{1/4} \,.
    \end{split}
\end{align}
Note that this rescaling exercise is only intended to show the potential reach of BH scalar sirens in probing the scalar-SM interactions. They are not intended to be exclusions of such radiation. In order to properly exclude BH scalar sirens, a specific reanalysis of existing data would be necessary, since it is likely that a hypothetical signal might have been assumed as noise due to differences in the quality factor $Q$ between ALP DM and BH scalar siren radiation.

Spin couplings are among the best portals to search for stellar-mass BH sirens because of their derivative nature and, thus, direct dependence on the velocity of the scalar. This compensates, although not completely, the comparatively low energy density (\cref{fig:rhorad}) with respect to DM. 

In the presence of a coupling to photons, one might ask whether the (coherent) Galactic magnetic field $|\vec B_\text{G}| \sim 10^{-6}\,\G$ can lead to the conversion of a scalar to a photon when traveling the $\sim 8\,\kpc$ separating the Galactic Center from the Sun. For propagation in a background, ``small-scale'' random magnetic field \cite{brown2010magnetic,ohno1993} whose coherence length $\mathcal O(10\,\pc)$ is much longer than the DeBroglie wavelength but much smaller than the propagation length, the conversion probability
per unit length is\footnote{The Galactic magnetic field has a coherence length much greater than $\mu^{-1}$, but much less than the total length of propagation $\sim 8\kpc$ of the scalars from the BH to the Sun. The probability of conversion per scalar over a singular volume of coherent, non-oscillatory magnetic field is $g_{\phi\gamma\gamma}B^2\mu^{-2}$ \cite{Sikivie:1983ip,Kimball:2023vxk}, valid when this is $\ll 1$. For the Galactic magnetic field, this is added incoherently over the coherence patches of the field along $L$, the number of which is bounded above by $L\mu$. Thus, the total probability of conversion is at most $g_{\phi\gamma\gamma}^2B^2\mu^{-1}L$.}
\begin{align}
\begin{split}
{}&\frac{\mathrm d \ln \langle \phi^2\rangle}{\mathrm dL} \lesssim 10^{-18}\,\kpc^{-1}\\{}& \left(\frac{g_{\phi\gamma\gamma}}{10^{-11}\,\GeV^{-1}}\right)^2 \left(\frac{|\vec B|}{10^{-6}\,\G}\right)^2\frac{10^{-11}\,\eV}{\mu}\,,
\label{eq:axion-conversion-to-photon}
\end{split}
\end{align}
where $g_{\phi\gamma\gamma} \in \{g_{\phi F^2}, g_{\phi F\tilde F}\}$. Requiring this to be small enough that the propagation of the scalars is unaffected places a limit on $g_{\phi\gamma\gamma}$.

Tree-level spontaneous decay to two photons in the equilibrium cloud proceeds at a very slow rate relative to the SR timescale (which sets the rate of equilibrium processes):
\begin{align}
\begin{split}
\Gamma_{\phi\mathcal \gamma\gamma} \approx {}&8\times 10^{-46}\,\Gyr^{-1}\\{}& \left(\frac{g_{\phi\mathcal \gamma\gamma}}{10^{-11}\,\GeV^{-1}}\right)^2\left(\frac{\mu}{10^{-11}\,\eV}\right)^3,
\end{split}
\end{align}
with additional suppression factors $\sim\alpha_\text{EM}(\mu/m_\Psi)^2$ if this decay needs to go through a loop of heavy electromagnetic charged fermions $\Psi$ \cite{Resnick:1973vg,ParticleDataGroup:2024cfk}. Because the scalars in the cloud have high phase-space occupation numbers however, the decay rate of singular particles can be enhanced via a parametric resonance effect (a manifestation of Bose-enhanced, stimulated decay) \cite{Tkachev:1986tr,Hertzberg:2018zte,Rosa:2017ury,PhysRevLett.58.171,PhysRevD.52.3226}, thus producing BHs that ``shine'' low energy electromagnetic waves, rather than non-relativistic scalars. This occurs when a second decay is statistically likely to take place within the cloud before the photons from a first, seed decay have left it. Quantitatively, given the number of particles in the cloud at quasi-equilibrium, this happens for $\mathcal C_{\phi\gamma\gamma} \gtrsim 1$ \cite{Baryakhtar:2020gao}. Note however that, for interactions with photons, EFT expectations set $
\mathcal C_{\phi\gamma\gamma} \sim \alpha_\text{EM}/2\pi \approx 10^{-3}$.

\begin{figure*}[htb]
     \centering
    
        \includegraphics[width=\textwidth]{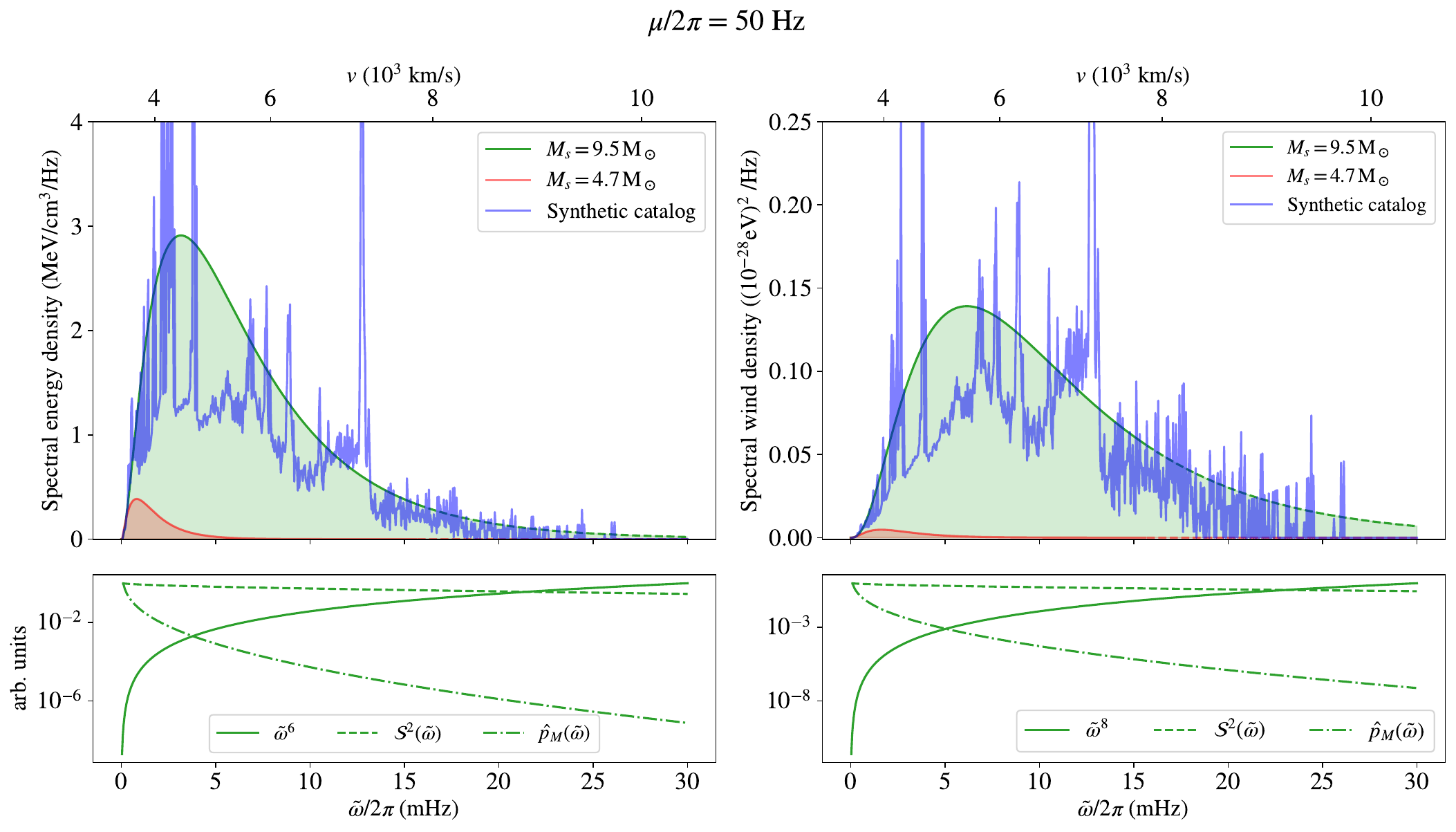}
        
         \caption{
         Spectral energy density $\frac{\mathrm d}{\mathrm d\tilde \omega}\langle\mu^2 \Phi_{\text{rad,}\odot}^2(t)\rangle$ (upper left) for $f_\text{siren}=7\times 10^{13}$\,GeV, and spectral wind density $f^{-2}\frac{\mathrm d}{\mathrm d\tilde \omega} \langle\left(\bm J_\parallel \cdot \bm\nabla\Phi_{\text{rad,}\odot}(t)\right)^2\rangle$ for unit $\bm J_\parallel$ (upper right) at the Sun-Earth location. Both lower panels represent the frequency dependence of the functional components that, together, form the non-trivial lineshape. The signal scales with frequency, but is suppressed by the ``spin'' factor $\mathcal S^2(\tilde\omega)$ and the number of BH sirens $\hat p_M(\omega)$ emitting at higher frequencies.
        The synthetic BH catalog \cite{Olejak:2019pln} mass distribution presents spikes, which are translated to the lineshape since the frequency of the emission depends on the BH siren mass (\cref{eq:velocity}). This is intended to represent a more realistic scenario where the BH mass distribution is not a perfect exponential distribution.
        The spectral density for $\bm J_\perp$ can be obtained by rescaling by $\left(r_s/r_\text{o}^\text{G}\right)^2 \approx 0.14$. The lineshapes are dashed in the regime $\alpha>0.15$, where polychromatic emission might be present, see \cref{sec:larger alpha ensemble}. 
         }
     \label{fig:powerspectra}
 \end{figure*}

On \cref{fig:powerspectra}, we show the spectral densities of $\mu^2\langle\Phi_\text{rad,o}(t)\rangle$ and $f^{-2}\langle\bm J \cdot \bm\nabla\Phi_\text{rad,o}(t)\rangle$ for unit $\bm J$, evaluated at the Sun-Earth location, \cref{eq:power_spectral_density,eq:gradient PSD}, under the assumptions discussed in \cref{sec:MW sirens} and for representative values of $\mu$ and $M_s$. Both spectra have similar shapes. At the lower frequencies/velocities, they rise polynomially from the power law prefactor, whose power depends on how many factors of the velocity enter the amplitude. The spectrum then decays towards larger frequencies/velocities because of the suppression in the number of BHs which are both heavy enough to be associated with larger orbital velocity scales (and therefore larger velocities of expelled scalars), \emph{and} have fast enough spins that the SR condition, \cref{eq:SR_condition_2}, remains satisfied at those larger velocities.

An interesting feature of the ``wind'' signal is the dependence of the strength of the signal on the direction of $\bm J$. For an experiment undergoing periodic rotation of its sensitive axis, the amplitude of the signal would be modulated in time, going down up to $\mathcal O(90\%)$ from its maximum size, depending on the relative angle of the modulation plane respect to the Galactic disk. This is particularly relevant to experiments for which $\bm J$ is fixed relative to the rotating Earth. 
This daily modulation can be exploited to both increase the detectability of the signal and to distinguish it from a putative scalar component of a different origin, such as a primordial dark matter component. A daily modulation of the DM signal from Earth's rotation is also expected, but while the ``DM wind'' blows in the direction opposite the motion of the Earth relative to the Galactic frame, the wind of Galactic BH sirens primarily blows from the direction of the Galactic Center.

\begin{figure}[htb!]
    \centering
    \includegraphics[width=\linewidth]{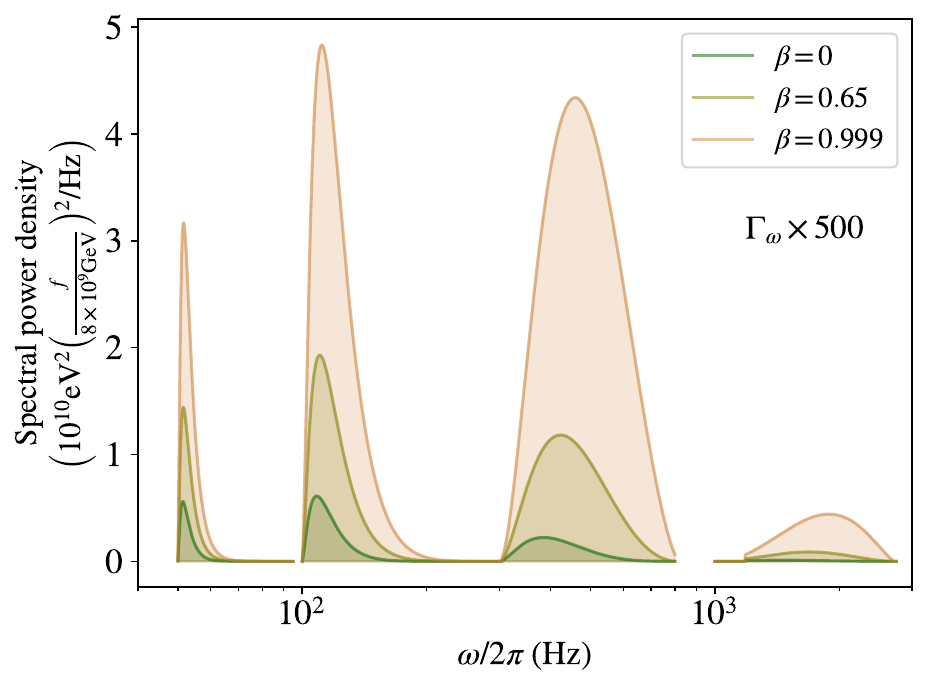}
    
    \caption{Spectral power density $\frac{\mathrm d}{\mathrm d \omega }\langle \Phi_{\text{rad, }\odot}^2(t)\rangle$ of an ensemble of MW stellar-mass BH sirens with $M_s = 9.5 M_\odot$ (Eq.\,\ref{eq:mass_distribution}) and different values of $\beta$ (\cref{eq:spindistro}) for scalar masses $\mu/2\pi = \{50,100,300, 1000\}\,\Hz$. The linewidth $\Gamma_\omega$ is exaggerated 500 times here for clarity. The value $f \approx 8\times 10^9\,\GeV$ is chosen to be the largest possible such that \cref{eq:SR timescale} is obtained for the largest scalar mass plotted here. This shows how the signal of a single scalar changes as its mass $\mu$ varies, while the BH parameters are kept fixed.  Alternatively, this illustrates how the MW ensemble of stellar-mass BHs could look like in the presence of an ``axiverse'' of scalars close in mass and characterized by a (narrow range of) parametrically small $f$. For stellar-mass sirens, the low end of the ``visibility'' range of scalar masses is set by the requirement that the SR cloud grows fast relative to Galactic timescales; the high end is determined by the interplay between the SR condition and the spin distribution at the high end. The effects of varying the pitch factor $\beta$ of the spin distribution should be compared to \cref{fig:S(omega)vsalpha}.
    }
    \label{fig:Powerspin}
\end{figure}

Because the shapes of the spectral densities are fixed in units of the population scale velocity scale $v_s$ (see \cref{eq:scale power density}), the dependence of the lineshape on $\mu$ is easiest to understand in terms of the population scale variables $\alpha_s$ and $v_s$. As $\mu$ is taken higher (lower) for fixed BH population mass $M_s$, both $\alpha_s$ and $v_s$ become correspondingly higher (lower), the lineshape moves to higher (lower) values, and broadens (narrows) in relative velocity or frequency units (\cref{fig:freqdeltafreq}). This is further  illustrated on \cref{fig:Powerspin}. Starting towards small $\alpha_s$, the peak amplitude of the signal first scales with some positive power of $\alpha_s$.
As $\alpha_s$ gets larger, however, even the lightest BHs in the distribution correspond to large orbital velocities at the cloud and the overall signal becomes suppressed by the spin factor $\mathcal S^2.$ Together, these features reflect at the population level the fact that the signal of a single siren peaks around $\alpha \simeq 0.5$ and extremal spin $a_\star = 1$.
Conversely, when $\alpha_s$ is too small, one of two things happens. Either the SR timescales \cref{eq:SR timescale} becomes too long, or the slow emitted scalars fall below the Galactic escape velocity and the calculation does not hold. 

\subsubsection{Experimental outlook}
\label{sec:experiments}
A useful benchmark for spin-based searches for signals from BH scalar sirens is the effective energy scale associated with the scalar-wind observable. For our fiducial Milky-Way stellar-mass population, the wind signal at the Sun-Earth location reaches $f^{-1}\sqrt{\langle(\bm{\hat J_\parallel} \cdot \bm \nabla \Phi_\text{rad,o}(t))^2\rangle} \approx  10^{-27}$\,eV, see \cref{fig:sirensvsrelic}. 
Interpreted as a target for comagnetometer/spin-precession experiments, this sits below current demonstrated energy-resolution benchmarks \cite{zhang2025helium,zhang2023search,su2025limits,Padniuk2024Mar,aybas2021quantum}: for example, recent Ramsey clock-comparison comagnetometry reports sub-nHz frequency sensitivity \cite{zhang2025helium} and emphasizes that nuclear-spin precession techniques can reach absolute energy resolutions $\lesssim 10^{-25}$\,eV. 
This corresponds to a remaining gap of order $10^2-10^3$ in amplitude, depending on the integration and analysis strategy.
Recasting existing constraints to estimate sensitivity to stellar-mass BH sirens signals, see \cref{fig:projections}, the expected sensitivity reaches only values of the Wilson coefficients $\mathcal C_{\phi \mathcal O} \gg 1$, substantially larger than the minimal expectations from effective field theories. 
Current laboratory sensitivity is typically probing parameter space that is already constrained by astrophysical considerations (e.g., stellar-cooling bounds on the underlying dimensionful couplings), as seen in Fig.\,\ref{fig:projections}. 
At the same time, there is intense ongoing experimental work to improve the sensitivity of spin-based sensors \cite{jackson2023probing}.
For example, levitated ferromagnets \cite{jackson2016precessing,vinante2021surpassing,fadeev2021ferromagnetic,kalia2024ultralight} provide a complementary route to axion wind searches with high mechanical $Q$ and realistic projections for improved dissipation, temperature, and readout coupling. 
These developments motivate treating BH scalar sirens as a well-defined long-term target for the next generation of broadband and resonant spin-based sensors.

Complementary experimental avenues in this same ultralight mass band include axion-to-photon searches using the terrestrial magnetic field and earth-ionosphere system as a transducer \cite{arza2022earth,sulai2023hunt,friel2024search,taruya2025hunting}. 
These efforts exploit the fact that an axion field coupled to photons in the presence of a static magnetic field sources an oscillating magnetic field at the Compton frequency, enabling a direct search for narrow spectral lines in magnetic field data collected in quiet environments. 
Existing searches reach up to axion masses of $\mu \sim 10^{-13}$\,eV, and there is continuing work to extend the range up to $\mu \sim 10^{-11}$\,eV \cite{bloch2024curl}.
In the context of BH scalar sirens, these results benchmark the achievable sensitivity to any coherent magnetic signature sourced via $g_{\phi\gamma\gamma}$ in the same frequency range, even when the underlying scalar field is not a substantial dark matter component, motivating a dedicated reanalysis of existing and future data taking into account the siren signal morphology and amplitude.

\subsubsection{Comparison to pre-inflationary cosmic misalignment}
\label{sec:cosmic misalignment}
A well motivated target for experiments seeking to detect a local abundance of scalars is the conservative prediction of a pre-inflationary cosmically misaligned scalar \cite{ParticleDataGroup:2024cfk,preskill_cosmology_1983,abbott_cosmological_1983,dine_not-so-harmless_1983,Turner:1985si,Hui:2016ltb,Hu:2000ke,Linde:1987bx,Cyncynates:2024bxw,Cyncynates:2024ufu}. In such scenarios, physics of the early-universe (e.g.\,a phase transition, inflation, coupling to the SM bath,\,etc.) effectively initialize the long-wavelength modes of the scalar to some value $\theta_i$. At fixed $\theta_i$, the energy-density of the scalar relic at late times scales $\propto f^2$, just like that of scalars produced by BH sirens. For a cosmically misaligned scalar, the local, present-day value of scalar observables today can be estimated from the cosmic, present-day abundance:
\begin{subequations}
\label{eq:misalignment}
\begin{align}
\begin{split}
\left\langle\theta^2_{\odot, \text{mis}}\right\rangle  ={}& \frac{\langle\rho^\text{cosmic}_{\text{mis},0}\rangle}{\mu^2f^2} \mathcal B \\\approx{}&\theta_i^2\,\frac{(9\,\Omega_r)^{3/4}}{2}\left(\frac{H_0}{\mu}\right)^{3/2}\mathcal B\\
\approx{}& (6.5\times 10^{-15})^2 \times \theta_i^2\left(\frac{10^{-12}\,\eV}{\mu}\right)^{3/2},
\end{split}
\end{align}
and
\begin{align}
\begin{split}
{}&f^{-2}\langle \bm {\hat J}\cdot \bm\nabla \Phi_{\text{mis},\odot}^2\rangle  \\
\approx{}& \mu^2 \langle v^2\rangle \left\langle\theta^2_{\odot, \text{mis}}\right\rangle\\
\approx{}&\left(\frac{6.5\times 10^{-30}\,\eV}{\sqrt 2}\right)^2\times \left(\frac{v_\text{vir}}{10^{-3}}\right)^2\theta_i^2\left(\frac{\mu}{10^{-12}\,\eV}\right)^{1/2},
\end{split}
\end{align}
\end{subequations}
where the present-day cosmic abundance resulting from an initial angle $\theta_i$ is $\langle\rho^\text{cosmic}_{\text{mis},0}\rangle$  \cite{marsh_axion_2016}, $H_0 \approx (4448\,\Mpc)^{-1}$ \cite{Planck:2018vyg} is the present-day Hubble constant, $\Omega_r\approx 9.2 \times 10^{-5}$ is the radiation density parameter of the Universe, $\mathcal B\sim 10^{5.5}$ is the approximate ``boost'' ratio relating the local and cosmic DM densities, and the velocity distribution of DM is assumed to be Maxwell-Boltzmann $p(v)\propto e^{-v^2/v_\text{vir}^2}$ with a velocity dispersion $\langle v^2\rangle = v^2_\text{vir}/2$.

As displayed on \cref{fig:sirensvsrelic}, at equal values of scalar model building parameters $\mu$, $f<f_\text{siren}$, and $\{\mathcal C_{\phi\mathcal{O}}\}$, the scalar background expected by BH scalar sirens in the MW may produce \emph{larger} local scalar observables than those expected for a cosmically misaligned scalar with generic initial misalignment angle, within the mass range corresponding to stellar-mass BH sirens.

In both cases, an initial value of $\theta$ (locally at the cloud, or uniformly in the early-Universe) is discounted by a measure corresponding to a distance in spacetime. For scalar sirens, this is the explicit spatial flux dilution factor $1/R^2$. For a cosmic scalar, this is the cosmological dilution due to the expansion of space, which affects the scalar from the time when $H \simeq \mu$ to today. Because the spacetime distance separating us from scalar siren emissions is smaller than that separating us from the cosmic sphere corresponding to $H\simeq \mu$, these ``geometric'' effects favor sirens. It is, however, entirely fortuitous that our Galaxy is expected to contain enough BHs that their finite number of localized emissions competes with the diffuse, cosmic initial conditions.

Thus, not only do BH scalar sirens provide a target for experiments that is independent from the primordial abundance of the scalar, but this target can be more attainable than the expected cosmic relic for scalars in the mass range $10^{-13}$--$10^{-11}\,\eV$ and $f < f_\text{siren}$, under conservative assumptions.

\subsubsection{Larger $\alpha$ sirens in an ensemble}
\label{sec:larger alpha ensemble}

It is straightforward at this stage to assess how the polychromatic spectra of individual sirens at $0.15 \lesssim \alpha < 0.5$ discussed in \cref{sec:larger alpha sirens} would affect the spectral densities observed. For a polychromatic object, a unique observer frequency $\tilde\omega_\phi^\text{o}(M_n,\mu,\bm v^n_\text{o})$ is no longer singularly specified by the BH parameters at fixed $\mu$. In appendix \cref{app:polychromatic}, we argue that in such a case the polychromatic or finite-width spectrum of the siren in the BH's rest frame effectively enters as a convolution kernel for the expression of a monochromatic signal. Thus, analogously to the BH velocity distribution, the rest frame spectrum of the source smears the ensemble signal over the frequency range corresponding to its width. 

Even in the worst possible scenario, the spectrum corresponding to a series of emission processes $n_1\ell_1m_1\times n_2\ell_2m_2\rightarrow n_3\ell_3m_3\times \infty$ may extend from $\tilde\omega_\phi^\text{BH} \rightarrow 0$ up to at most $\tilde\omega_\phi^\text{BH} = \mu\alpha^2/2$, the ground state energy of the BH scalar siren system. This is on the order of, but slightly smaller, than the ensemble width $\Gamma_\omega$ described by \cref{eq:frequency linewidht estimate}. 

This smearing only occurs for sirens with $\alpha_n \gtrsim 0.15$, so comparing $\Gamma_\omega$ to $\mu\alpha^2/2$ slightly 
overstates its impact on the spectrum. For values of $\alpha_s \lesssim 0.15$, the higher frequencies of the observed spectrum will be affected. We display the affected region with dashed lines on \cref{fig:powerspectra} for $\mu/2\pi = 50\,\Hz$. For $\alpha_s\gtrsim 0.15$, polychromaticity may more significantly obfuscate the relationship between the observed lineshape and the BH mass distribution $\hat p_M$. In this part of parameter space however, the effect of the spin factor $\mathcal S^2(\alpha_j)$ already suppresses the signal significantly, see \cref{fig:Powerspin}.

Additional line features in the resting-frame spectrum are unlikely to modify the frequency-integrated signal—the figure of merit in \cref{fig:sirensvsrelic}. Nevertheless, strong theoretical predictions for the polychromatic emission spectrum (amplitude and shape) of high-$\alpha$ sirens remain desirable: a well-characterized lineshape would extend experimental reach by enabling Wiener-type optimal filters to boost search SNR. In practice, however, the BH mass distribution is unknown beyond educated guesses, leaving searches vulnerable to \emph{overfitting}. The broadband signal shape shown in \cref{fig:powerspectra} may therefore represent the most conservative search strategy, even if that template degrades towards higher frequencies and overall for populations with higher $\alpha_s$.

Conversely, should the scalar signal be detected, accurate modeling of the polychromatic rest-frame spectrum would enable deconvolution methods to resolve degeneracies between the intrinsic siren spectrum and $p_M$ -- particularly if the former indeed consists of sharp peaks.

\section{Non-stellar-mass BH sirens in the MW and nearby}

Aside from primordial dark matter, the ensemble of stellar-mass BH scalar sirens in the Milky Way is the most compelling target for the detection of non-relativistic scalar bosons in the mass range $\sim 10^{-13} - 10^{-11}\,\eV$, accounting both for theoretical priors and experimental limitations. 
However, a compelling feature of BHSR as a probe of BSM physics is its broad generality: it requires only the gravitational interaction between a new, sufficiently long-lived bosonic field and a rapidly rotating BH. As discussed above, the efficiency of BHSR is governed primarily by the dimensionless parameter $\alpha$, which can be interpreted as the ratio of the boson’s Compton wavelength to the black hole’s gravitational radius. Consequently, although we have so far observed only stellar-mass BHs of astrophysical origin, the BHSR mechanism could in principle operate for heavier (lighter) BHs and  lighter (heavier) bosons, as dictated by \cref{eq:alpha}, with the relation saturating, in principle, at $\mpl$.

The prospect of a phenomenological alignment between, on the one hand, the mass spectra of BSM bosons and, on the other hand, that of non-astrophysical BHs becomes all the more interesting in light of theoretical expectations for a richly populated spectrum of bosons, such as the axion-like fields anticipated in the “string axiverse” \cite{Arvanitaki:2009fg}. Indeed, in the seminal work reviving the recent interest of BHSR as a probe for BSM, the authors of \cite{Arvanitaki:2010sy} are already driven by the idea of probing the axiverse with (astrophysical) BHs through the BHSR process.

This raises the intriguing (if speculative) possibility that, in theories predicting a densely populated scalar spectrum, one might conversely use BHSR-induced signals to probe or constrain non-astrophysical or otherwise inaccessible BH populations. In fact, because the expected MW population of $\sim 10^8$ isolated stellar-mass BHs is extremely challenging to detect directly, one might actually see it part of this paradigm. Consider, for example, that the detection of the signal we describe would not only reveal a new light bosonic degree of freedom, but could also constitute the most direct observational confirmation of the otherwise elusive population of isolated BHs in the MW. Of course, the same could be said of the detection of the ensemble signal of GWs for scalars without significant self-interactions \cite{Zhu:2020tht}. While there are stringent constraints on the abundance of non stellar-mass BHs in the MW, stellar mass BHs themselves are expected to constitute no more than $10^{-3}$ of the total MW mass.

Using the MW stellar-mass BH population as a benchmark, we therefore summarily assess here prospective signals from other populations of BH scalar sirens, of both astrophysical and non-astrophysical origins. First, we generalize our perspective on previous results, allowing $M_s$ to float.

\subsection{Scalar luminosity}
\label{sec:scalar luminosity}
Before making more particular considerations, two general points are worth stressing. Firstly, for a local scalar DM component, one typically quantifies the abundance of the scalar in terms of the local density of DM, the putative value of which is fixed to be $\rho_{\text{DM},\odot}\approx 0.4\,\GeV/\cm^3$ \cite{ParticleDataGroup:2024cfk}. On the other hand, because scalars from BH scalar sirens are free streaming from a collection of compact sources, the local abundance of scalars from a putative population of BH scalar sirens is easier to quantify in terms of the total mass of that population $M_s^\text{total}\sim N_\text{BH}M_s$. (In much the same way, the visible luminosity of astrophysical objects (galaxies or clusters) is quantified relative to the total mass of the baryons which produce that radiation.) This mass can in turn be given as a fraction of the total mass of the host galaxy. For example, stellar-mass BH sirens in the MW obtain $M_s^\text{total} \approx 10^{-3}M_\text{MW}$, in terms of the approximate MW mass $M_\text{MW}\approx 10^{12}\,M_\odot$. For populations inside the MW, a strict upper bound on scalar ``luminosity'' can be estimated by setting $M_s^\text{total} \approx M_\text{MW}$. Stronger constraints exist at specific specific mass scales (e.g.\,\cite{Mroz:2024mse}).

Secondly, for fixed couplings $g_{\phi\mathcal O}$, non-derivative interactions such as the interactions with the photon are sensitive to the local \emph{field amplitude squared}\,$ \langle \Phi_\text{rad,o}^2(t)\rangle$ (corresponding to the \emph{number} of scalars \emph{per unit $\mu$}). In contrast, derivative interactions like the scalar wind coupling are sensitive to the \emph{momentum} flux of scalars, itself proportional to the \emph{energy} density: $\sim \mu^2v_s^2 \langle \Phi_\text{rad,o}^2(t) \rangle \sim v_s^2\langle \rho_\text{rad,o}(t)\rangle$. This distinction has implications for the parametric dependence of the signal on the mass scale $M_s$. 

At fixed $f$ and $\alpha_s$, as is apparent from \cref{eq:theta and wind}, the amplitude squared of the radiation field scales as $N_\text{BH}M_s^2 \sim M_s^\text{total}M_s$.  On the other hand, the \emph{number} of scalars produced at fixed $f$ scales as $\mu\langle \Phi_\text{rad,o}^2(t)\rangle \sim N_\text{BH}M_s \sim M_s^\text{total}$. Finally, the ``wind'' $f^{-2}\langle\rho_\text{rad,o}(t)\rangle$, \cref{eq: rho}, scales as $N_\text{BH} \sim M_s^\text{total}/M_s$. We summarize these scalings in \cref{tab:observable scalings}.

As a result, the total number density of free streaming scalars produced by a population of sirens is set by the total mass of the population relative to, say, the total mass of the MW. Per \cref{eq:alpha}, bigger BHs produce lighter scalars. The resulting energy density produced is therefore lower, but yields comparatively larger field excitations, giving a relative advantage to detection channels involving non-derivative couplings, such as that to the photon. Conversely, smaller BHs produce heavier scalars carrying comparatively more momentum (since the velocity $v_s$ is fixed at fixed $\alpha_s$), which advantages detection through derivative couplings.

Accounting for the fact that $f<f_\text{siren}$ must be obtained, the maximum energy density $\langle\rho_\text{rad,o}(t)\rangle$ scales as $\sim N_\text{BH}M_s \sim M_s^\text{total}$, \cref{eq:energy density}.

Of course, scaling of the signal with distance must be also additionally be accounted for. Finally, we limit ourselves to brief considerations about the astrophysical \emph{input} signals; a comprehensive evaluation of \emph{observability} would necessitate explicit consideration of the experimental apparatus, associated technologies, and measurement strategies, which are frequently frequency-specific.

\begin{table*}
    \centering
    \renewcommand{\arraystretch}{1.25}
    \setlength{\tabcolsep}{10pt}
    \begin{tabular}{c|c|c|c}
        operator  & observable & probe & scaling ($\alpha_s$ fixed)\\
        \hline
        \hline
        $\mu \,\langle \Phi_\text{rad,o}^2(t)\rangle$  
            & number density (fixed $f$) 
            & --
            & $M_s^\text{total}$, \cref{eq: rho}  \\
        $f^{-2}\langle \Phi_\text{rad,o}^2(t)\rangle$  
            & $\theta^2$/number per unit $\mu^{-1} f^{-2}$ (\cref{eq:theta definition})
            & ND couplings
            & $M_s^\text{total} M_s$, \cref{eq:theta}\\
        $f^{-2}\langle \left(\hat{\bm J}\cdot \bm \nabla \Phi_\text{rad,o}\right)^2(t)\rangle$ 
            & ``wind'' / force per unit $f^{-2}$ area $\perp$ to $\hat{\bm J}$ 
            & D couplings to spins 
            & $M_s^\text{total}/M_s$, \cref{eq:wind} \\
        \hline
        $\mu^2\langle \Phi_\text{rad,o}^2(t)\rangle$  
            & max.\,\,energy density (largest possible $f$) 
            & gravity 
            & $M_s^\text{total}$, \cref{eq:approximate energy density} \\
    \end{tabular}
    \caption{
    Scalings of scalar-field observables produced by a BH scalar siren population as the scale mass $M_s$ and total mass $M_s^\text{total} \sim N_\text{BH}M_s$ are changed, all at fixed $\alpha_s$, as discussed in \cref{sec:scalar luminosity}.
    Operators proportional to $\langle \Phi_\mathrm{o}^2(t) \rangle$ are listed together with their physical interpretation, the experimental or observational probes to which they couple, and their parametric scaling with $M_s$ and $M_s^\mathrm{total}$. ND and D stand for ``non-derivative'' and ``derivative'', respectively.
    }
    \label{tab:observable scalings}
\end{table*}

\

\subsection{Individual supermassive BHs}
\label{sec:supermassive BH}
Scalars at the lower end of the ultralight mass range (ultralow frequency), around $10^{-19}-10^{-16}\,\eV$, can trigger the SR instability for the MW's singular ($N_\text{BH} = 1$) supermassive BH, Sagittarius A$^{*}$ ($M_\text{Sgr} \approx 10^6 M_\odot \approx 10^{-6}\,M_\text{MW}$). Compared to the ensemble stellar-mass MW BH sirens, Sgr A$^*$ yields a maximum energy density $10^{-3}$ times smaller, and a wind $10^{-8}$ times smaller. On the other hand, the amplitude squared of the field produced by Sgr is $10^2$ times larger. At such low frequencies, methods involving atomic \cite{BACON:2020ubh,filzinger2023improved} or nuclear clocks \cite{fuchs2025searching,delaunay2025probing}, long-baseline optical interferometers \cite{arvanitaki2018search}, atom interferometers \cite{geraci2016sensitivity,du2022atom}, and space-based interferometry \cite{Antypas2022Mar,huang2025hunting,yao2025prospects} constitute promising detection avenues. The signal from such a singular, nearby supermassive BHs is also much \emph{narrower}, its power spectral density being, formally, $\delta(v_j- G\mu M_\text{Sgr}/6)$.

The next closest supermassive BHs is M31$^*$, at the center of the Andromeda galaxy. Even though Andromeda is at $744\,\kpc$ \cite{Vilardell_2010} from Earth, the mass of M31${^*}$ is 100 times that of Sgr A$^{*}$  \cite{Melchior_2024}. Since field-amplitude-squared produced by a single siren scales as $\propto \left(\frac{M_\text{BH}}{R}\right)^2$, the increase in mass compensates for the 100-fold increase in distance. The mass range of the boson for superradiance would be shifted to even lower masses,  $10^{-21}-10^{-18}\,\eV$. Note that Andromeda is close enough that the lookback time of the signal remains negligible relative to the requirement that the scalars be produced above the galactic escape velocity.

\subsection{Extragalactic ensemble of stellar-mass BHs}
For the purpose of setting an upper bound, we may assume that every galaxy in the Universe hosts a population of stellar-mass BHs comparable to that of the MW, up to a few orders of magnitude. While the signal from individual distant galaxies ($\gtrsim 10\,\Mpc)$ suffers a strong $1/r^2$
suppression, one might wonder whether this is compensated by their large number. Taking $N_\text{BH}$ as a representative number of stellar-mass BHs in a galaxy, the cumulative signal from the $N_\text{gal}$ galaxies contained in the visible universe scales as
\begin{align}
\begin{split}
\sim N_\text{gal}\left\langle \frac{N_\text{BH}}{r^2_\text{gal}}\right\rangle \sim{}& N_\text{BH}\int \frac{\mathrm dn_\text{gal}}{r^2_\text{gal}}\\
\sim{}& N_\text{BH}\int \frac{\rho_\text{gal} \,r_\text{gal}^2\mathrm dr_\text{gal}}{r_\text{gal}^2}\\
\lesssim{}& N_\text{BH}\,\rho_\text{gal}\,H_0^{-1}\\
\sim{}& 10^{-4}\times \frac{\rho_\text{gal}}{10^{-2}\,\Mpc^{-3}}  \frac{N_\text{BH}}{(8\,\kpc)^2},
\end{split}
\end{align}
where $\rho_\text{gal}\approx 10^{-2}\,\Mpc^{-3}$ is the approximate cosmological number density of galaxies (Schechter function) \cite{Longair:2008gba}, and $H_0^{-1}$ is used as the characteristic size of the visible Universe. Because scalars from sirens are non-relativistic, this estimate is a strict upper bound; in actuality, the visible volume is limited to distances such that the lookback time, \cref{eq:lookback time}, compares to $\mathcal O(10\,\Gyr)$. Additionally, Hubble friction (cosmological redshift) acts to slow down non-relativistic particles as they travel over cosmological time scales. Thus, the extragalactic signal suffers \emph{at least} a $10^{-4}$ suppression relative to the Galactic signal. 

Estimating the signal from the more nearby Local Group of galaxies does not yield a significant enhancement (most galaxies in the Local Group are in fact smaller than the MW).

\subsection{Exotic black hole populations in the MW}
\label{sec:exotic}

Stellar-mass and supermassive BHs have been observed to exist, and as such they should be the primary targets for BH scalar siren phenomenology. Some evidence, however, both theoretical and observational, motivates the existence of ``exotic'' BHs outside these ranges. If supermassive BHs form via ``seeds,'' it is plausible that some BHs exist whose evolution has been delayed, and that persist today in the intermediate range $10^2$--$10^5\,M_\odot$. The existence of at least one such IMBH has recently been inferred from a LIGO event of extragalactic origin \cite{LIGOScientific:2020iuh}.
On the other hand, light BHs are a longstanding DM candidate \cite{Green:2020jor} and are generically theorized to have early-universe (i.e.\ primordial) origins. While asteroid-mass BHs ($\sim 10^{-18}$--$10^{-16}\,M_\odot$) remain viable candidates for the totality of DM, other BH masses are constrained to contribute no more than $\sim\, 10^{-3}$ of the DM density across much of parameter space \cite{Carr:2026hot}, which is comparable to the expected mass fraction of stellar-mass BHs in the MW. Both IMBHs and light PBHs could, if they exist, give rise to BH scalar siren populations.

A number of studies \cite{Bernal:2022oha,Dent:2024yje,Jia:2025vqn,Manno:2025dhw,Rosa:2017ury,Calza:2023rjt} have considered the possibility of joint phenomenology between light primordial BHs (PBHs) and one or many new scalars. Many \cite{Bernal:2022oha,Dent:2024yje,Jia:2025vqn} expound a version of the idea that a cosmic population of PBHs, as either a primary or subdominant component of primordial DM, would, via superradiance, produce a partner, minimal abundance of the heavy scalar. While this joint phenomenology is interesting, it is plagued by degeneracy with any initial abundance of the scalar, through misalignment, inflationary production, direct reheating, or some other myriad mechanism of cosmic production. Unless the PBHs themselves can be independently observed, this degeneracy obfuscate any mutual information one might leverage between the scalar and PBH abundances. Concretely, in such scenarios, the experimental detection today of a scalar component of (virialized) DM, and a characterization of its fundamental mass $\mu$, still hardly singles out a PBH mass scale because the (mutual information) parameter $\alpha$ cannot be determined independently. Similarly, even if the scalar abundance were to be precisely determined, it would not single out a PBH abundance. While the possibility of joint constraints between $\mu$, $f_\text{BH}$ and (likely continuous distribution of) $M_\text{BH}$ remains, a characterization of the BH mass scale and spectrum independent of total abundance would be desirable.

In contrast, the spectrum of scalars from BH sirens preserves some information about the progenitor BH population: from \cref{eq:scale power density,eq:scale power spectral density frequency} (as well as \cref{app:polychromatic} for the case of polychromatic emissions in the rest-frame), we see that the observed spectrum encodes the mass distribution of the BH sirens population, up to a simple monomial that depends on the portal to the SM in the detection. While the lower end of the frequency spectrum singles out $\sim \mu$, the width and shape are primarily determined by the BH mass distribution. Note that the primary condition underlying this correspondence between the spectral properties of observed scalars and the BH mass spectrum is that the BH distance and mass distributions be independent, such that the factor $1/R^2$ can be extracted as an overall amplitude prefactor.

It has also been suggested that the decay of individual scalars around BHs to two energetic photons (with energies $\sim\mu/2$) could produce an observable photon flux and thereby provide an alternative to detecting the ultrafaint Hawking radiation \cite{Dent:2024yje,Ferraz:2020zgi}.  Unlike the direct detection of a virialized scalar DM population, the decay of a BH's SR cloud to photons can be directional, and may carry location information about the BH population. 

The signal from photon emission in an individual BH siren is equally affected by the overall reduction in the number of scalars in the equilibrium cloud: the detectability of the scalar and photon emissions from a BH in the scalar siren regime scale $\propto g_{a\gamma\gamma}f^2$. Moreover, unlike the coherent scalar emission, the decay of individual scalars in the cloud to photons is summed incoherently for each individual sirens, leading to slower scaling with the number of particles in the cloud $N_\text{cloud}^\text{eq}\gg 1$, and therefore with the BH mass scale $M_s$. Similarly, the lack of coherence between individual scalar decays also implies that the photon flux from each siren has an inherent spread $\Delta E_\gamma/E_\gamma = \mathcal O(\alpha)$, since the spectrum of each decaying scalar is Doppler shifted by its orbital velocity in the cloud. This ultimately reduces the sensitivity of the observed flux on the BH mass spectrum of the siren population (since there is no longer a one-to-one correspondence between the mass of a BH at rest and the observed frequency of the radiation). For this reason, for BHs in the scalar siren regime, the conversion of scalars to photons at the cloud's location is both generally harder to detect and less informative than the conversion of the emitted scalars in an Earth-based experiment.

\subsection{Low $\alpha$ scalars bound to the Galaxy}\label{sec:low_alpha_bound}
Before concluding this section, we comment on the regime in which scalars emitted by black hole scalar sirens fail to exceed the Galactic escape velocity and therefore do not stream freely out of the Galactic potential. In this case, as already noted, the waveform in \cref{eq:radiation} is substantially modified: the Galactic potential effectively acts as a reflecting boundary at the classical turnaround radius. The emitted scalars are thus gravitationally bound, remain trapped in the Galaxy, and accumulate over time. This “Galactic basin’’ scenario is analogous to the more familiar solar (or, more generally, stellar) basins of axions \cite{VanTilburg:2020jvl,DeRocco:2022jyq,Giovanetti:2024rxi,Budker:2023sex}. 
For the MW stellar-mass black hole scalar sirens primarily considered in this work, this is about as constraining as the requirement that the SR timescale, \cref{eq:SR timescale}, be parametrically much shorter than the age of the population of stellar-mass BHs in the MW ($\mathcal O(1-10)\,\Gyr$). Thus, stellar-mass BHs in the siren regime necessarily produce scalars that escape the Galaxy, except for maybe a very narrow range of $\alpha$ which is hard to estimate given uncertainties on e.g.\,the escape velocity. However, smaller values of $\alpha$ can be considered for lighter BHs of non-stellar origin (see \cref{sec:exotic}).

We can estimate the maximum total amount of energy that can be trapped in the Galaxy in the form of scalar particles produced by a BH siren population. As explained in \cref{sec:BH-superradiance}, the maximum amount of energy that can be extracted from an individual BH by the SR process in the absence of self-interactions is $\sim \alpha  M_\text{BH}$, for small $\alpha \lesssim 10^{-1}$. This remains true in the BH scalar siren regime. The total \emph{number} of particles that can be produced is set by the initial angular momentum budget of the BH, $J_\text{BH} = GM_\text{BH}^2$. Since each particle carries energy $\mu$ and each particle carries $\mathcal O(\text{few}\times\hbar)$ of angular momentum, lighter particle exhaust the angular momentum  of the BH at a lower cost to its energy. Thus, at the ensemble level, at most $\mathcal O(\alpha)$ of the total mass $M_\text{total} \sim N_\text{BH}M_s$ of the Galactic population of BH sirens considered can end up in the form of scalar radiation trapped in the Galaxy. Considering that we are now interested in scalars produced below the Galactic escape velocity, and that $v\approx \alpha$, only at most a small amount $\mathcal O(v_\text{esc}) \approx \mathcal O(0.1\%)$ of the mass budget allotted to a BH population in the MW may be in the form of trapped scalar radiation. 

Values of $f$ exist that are parametrically larger than \cref{eq:threshold f}, but still small enough that the BHs angular momentum is primarily depleted through scalar radiation \cite{Baryakhtar:2020gao}. While this work is primarily concerned with the study of effectively infinitely long-lived sirens, the limit estimated above would actually be saturated for such intermediate $f$ and sirens whose lifetime are short enough to have effectively been ``fully-depleted'' within the $T_\text{MW}\sim 10\,\Gyr$ elapsed since the formation of the MW.

The de Broglie wavelength of the emitted scalars is set by the size of the black hole emitter and is therefore short, even for small values of $\alpha$, compared to the length scales over which the Galactic potential varies. This separation of scales fails only when $\alpha$ becomes so small that the de Broglie wavelength approaches Galactic dimensions; however, the strong suppression of the SR rates at such $\alpha$ makes this regime altogether irrelevant. Consequently, the propagation of emitted scalars through the Galactic potential at sub-escape velocities should be well described by an eikonal/WKB treatment. In the eikonal \cite{MCP} regime, the waveform is obtained from the classical trajectories of classical point particles in the Galactic gravitational potential. For bound emissions, these generically correspond to eccentric, non-closed orbits in the Galaxy. For an individual BH scalar siren emitting continuously until today over many orbital times of the scalar, reflecting boundary conditions yield the associated wave to build up in the MW. Because a siren moves significantly during the course of a scalar orbital time, this buildup is incoherent and applies to the density (rather than to the wave amplitude). This buildup of the field density might enhance the detection possibilities compared to if the same radiation were to escapes the Galactic potential. A detailed analysis is beyond the scope of this work.

\section{Discussion and conclusion}
\label{sec:conclusion}
The primary goal of this study is to compute and characterize the collective emissions of BSM scalar particles by the expected population of old, isolated stellar-mass BHs in the Milky Way. To do so, we defined in \cref{sec:sirens} a new class of long-lived astrophysical scalar emitters which we called BH scalar sirens. Scalar sirens are obtained when a BSM scalar exists whose Compton wavelength is parametrically comparable to the size of the BHs in a population, and the self-interactions of the scalar, quantified by a dimensionful decay constant $f$, are large enough that the otherwise rapid extraction of angular momentum by the black hole superradiance (BHSR) process is insignificant over the course of Galactic history. For small $f$ ($\lesssim 10^{14}$ -- $10^{9}\,\GeV$)  and $\alpha \sim 0.1$, a newly born, spinning, stellar-mass BH will quickly (relative to the Galactic timescale) become enveloped in a cloud of BSM scalars via the BHSR process. Today, the energy and angular momentum of the BH are still being channeled, via this cloud, to spatial infinity in the form of radiated scalars, a process akin to continuous autoionization in which the BH gradually sheds its spin through scalar emission (\cref{fig:self ionization}). 

In \cref{sec:MW sirens} we summed the emissions of the ensemble of stellar-mass BH scalar sirens in the MW disk (\cref{fig:galactic_frame}). Our analysis supports the existence of a background of non-relativistic scalars with astrophysical, rather than early-universe, origin. We characterize the size of the local amplitude and the spatial derivative of this background. While we further argue that such a background could arise for a broad range of putative BH masses, stellar-mass MW sirens in the disk privilege scalars in the mass range $10^{-13}\,$---$\,10^{-11}\,\eV$ ($10^{1}\,$---$\,10^{3}\,\Hz$).
Interestingly, this mass range is the same that
would be \emph{disfavored} by observations of high-spin BHs \emph{in the absence of strong self-interactions} \cite{Arvanitaki:2010sy,Arvanitaki:2014wva, Baryakhtar:2020gao,Stott:2018opm,Fernandez:2019qbj,Ng:2019jsx,Mehta:2021pwf,Hoof:2024quk, Witte:2024drg, Caputo:2025oap}. This highlights the key complementarity between the regimes of small and large particle self-interactions. 

Scalars emitted from BH scalar sirens have velocities $\sim \alpha$, comparable to the orbital velocity of the scalars in the SR cloud itself. Further, they are only free to reach us unimpeded if traveling at velocities greater than the Galactic escape velocity $\sim 600\,\km/\s \sim 10^{-3}$. This is in exact complementarity to DM particles virialized in the Galactic halo, which, by definition, must be gravitationally bound to the Galaxy.

The astrophysical scalar background from stellar-mass MW BH scalar sirens would make up a small ($\lesssim 10^{-5}\,\GeV/\cm^3$ at the Sun-Earth location; \cref{fig:rhorad}), but distinct contribution to the invisible matter budget of the Galaxy: the profile of outgoing scalars would be unvirialized and inherit a slight oblateness from the Galactic disk from which it originates. 

Prospects for detectability of the scalar background at Earth-based experiments in the presence of non-gravitational couplings to SM states are assessed in \cref{sec:observational}. For fixed scalar mass $\mu$, the \emph{shape} of the frequency spectra / the distributions of  kinetic energies of the observed scalars is primarily informed by the shape of the underlying BH mass distribution (\cref{fig:powerspectra}), while the \emph{amplitude} of the signal spectrum is sensitive to the pitch of the spin distribution (\cref{fig:Powerspin}). Interestingly, the increased velocity of scalars from BH scalar sirens, relative to DM, compensates for their reduced energy density at Earth, when considering the 
scalar-wind observable (\cref{fig:projections}). Moreover, the wind from BH scalar sirens points primarily along the line of sight to the Galactic Center, with a calculable directional spread (\cref{fig:BH gradient dispersion}), and therefore undergoes daily modulation with the Earth's rotation due to the change in orientation of the detector. This can be leveraged for detection. 

We stress that the detection of the \emph{unvirialized} scalars produced by BH scalar sirens provides a sharply quantifiable correspondence between BH and scalar phenomenology. The spectral features of the signal directly encode the BH population properties. In particular, every observable is independent of the relic abundance of the scalar. The observed spectrum directly reflects the BH population \emph{today} and can be clearly distinguished from a scalar component of primordial DM. The spectrum amplitude, shape, frequency spread, and directional dependence strongly constrain degeneracies between the ensemble BH abundance, mass, position, and spin distribution, while the siren regime itself infers a lower bound on the size $\lambda$ of particle self-interactions (equivalently, an upper bound on the microphysical scale $f$).

The sharp characterization of ensembles of isolated BHs possible via observation of scalar emissions is only rivaled by that achievable through gravitational waves from SR clouds, with the crucial distinction that GWs are known to exist. These, however, represent complementary but mutually exclusive regimes of BHSR phenomenology: large self-interactions trade GW observables for scalar emissions. The key difference between GW and scalar emissions in the context of BHSR is the independent parameter $f$. This parameter controls the lifetime of the source at fixed $\mu$ and allows us to define effectively infinitely-long-lived scalar sirens, greatly simplifying the analysis relative to finitely-long-lived GW emitters.

The scalar background expected from stellar-mass MW sirens constitutes a target for experiments independent of the cosmological abundance of the scalar. At equal values of scalar model-building parameters, BH sirens can produce amplitude fluctuations of local scalar observables that are up to two orders-of magnitude larger than the expected relic abundance from a cosmically misaligned scalar (\cref{sec:cosmic misalignment} and \cref{fig:sirensvsrelic}). 
This target might come within reach of near future experiments. 
Current avenues of detection employing nuclear spin samples are expected to reach energy resolutions of $<10^{-25}\,\mathrm{eV}$, two orders of magnitude away from the expected wind signal, assuming Wilson coefficients of $\mathcal C_{\phi \mathcal O}=\mathcal{O}(1)$.
Particularly in the frequency range where stellar-mass BH would emit radiation,
the experimental effort includes not only spin-precession-type experiments, but also levitated ferromagnets and direct magnetic measurement that would use the terrestrial magnetic field and earth-ionosphere system as a transducer for scalars (see \cref{sec:experiments}).

If one assumes, agnostically of a cosmological production mechanism, that the totality of local DM is scalars, the signal from BH scalar sirens remains only one-to-two-orders of magnitude smaller than of scalar DM when considering the wind signal specifically, because the larger velocities compensate for the reduced local energy density. 
Scalar observables from sirens are independent of $f$ (as long as $f < f_\text{siren}$). Assuming strong priors on $\mathcal C_{\phi\mathcal O}$, confident failure to observe the stellar-mass MW BH siren target would exclude \emph{all} (minimal) parameter space with $f < f_\text{siren}$ within this mass range, up to uncertainties in the expected population of MW stellar-mass BHs. This is complementary to how confident observations of singular fast spinning BHs rule out all values of $f$ with negligible self-interactions, for which the gravitational BHSR process proceeds.

Successfully observing the scalar background would greatly inform our knowledge of isolated stellar-mass BHs in the MW, and could be considered the first direct observation of this expected, but largely invisible population. In this spirit, a secondary objective of this study has therefore been to extend our analysis to other BH populations and articulate the prospect of discovering BH populations with scalars, see \cref{sec:exotic}.

\begin{figure}[h]
     \centering
    
        \includegraphics[width=.5\textwidth]{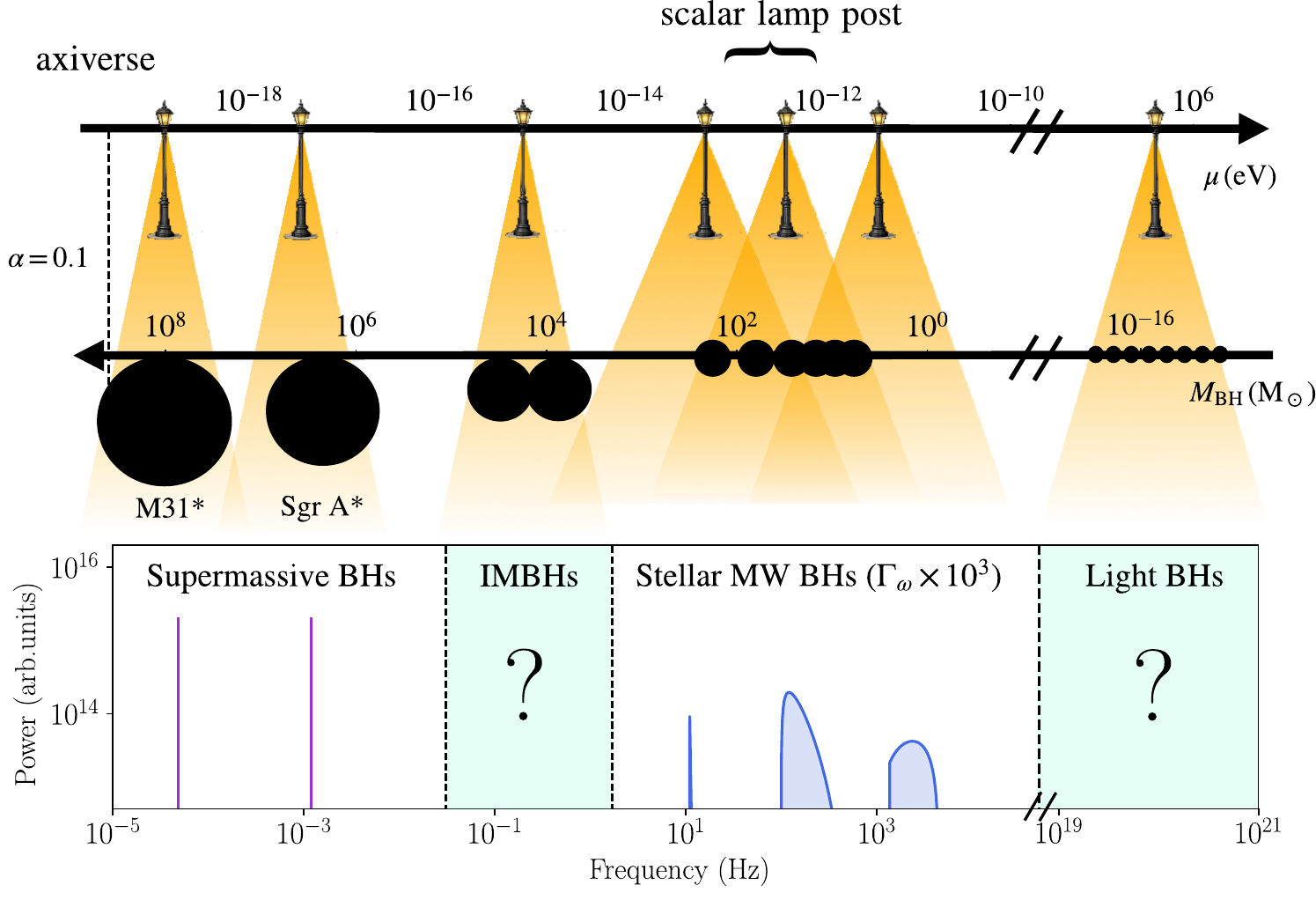}
        
         \caption{
         BH astronomy with the axiverse (sketch). BH sirens offer a unique opportunity not only for the detectability of a new bosonic scalar field, but for the astronomical possibilities after its discovery, including, for example, characterizing the mass and spatial distributions of the BH population. In this illustration, we show how multiple new scalars (\textit{scalar lamp posts}) could shed light into BH populations, going from supermassive BH, which would be monochromatic in frequency due to the limited amount of sources near Earth, to expected stellar-mass BH population which has not been directly observed and studied yet, except for a relatively small number of binary systems \cite{Reynolds:2020jwt}. The latter population would have a frequency distribution of their scalar radiation that can be obtained based on their spin, mass, and spatial distribution (Eq.\,\ref{eq:conditional_expectation_value}), and, if detected, allows to reconstruct some of its properties. Finally, hypothetical populations of intermediate-mass or light BHs, including primordial BHs, could also be studied and detected with a \textit{lamp post} of the right mass. 
         The idea of multiple \textit{scalar lamp posts} existing simultaneously becomes even more appealing in the context of the axiverse \cite{Arvanitaki:2009fg,Demirtas:2021gsq,Gendler:2023kjt}.
         The \textit{scalar lamp post} mass and the BH mass axes are adjusted to correspond to a value of $\alpha=0.1$.
         }
     \label{fig:axiverse}
 \end{figure}
 
Our analysis extends not only to supermassive BHs (\cref{sec:supermassive BH}), but also to putative MW populations of non-stellar-mass BHs, including IMBHs, and light BHs, whether primordial or not. The prospect of using scalars to learn about BHs is enticing precisely because it remains rather agnostic about the origin of a BH population. Lighter scalar backgrounds produced by heavier BHs are comparatively most visible through non-derivative interactions, such as conversion into photons. Backgrounds of heavier scalars produced by lighter BHs carry greater momentum density and are therefore most easily seen by experiments leveraging derivative, wind-type couplings (\cref{tab:observable scalings}). This BH-mass-dependent detection strategy becomes especially powerful when multiple scalars exist across a range of masses.
 
Such a scalar-BH correspondence becomes particularly powerful in the context of theories predicting a densely populated mass spectrum of scalars, such as string axiverse constructions \cite{Arvanitaki:2009fg,Demirtas:2021gsq,Gendler:2023kjt,Baryakhtar:2026}. If such a spectrum exists with sufficiently large self-interactions -- either through a universal parametrically small $f$, or through values of $f$ that scale inversely with scalar mass to maintain $f<f_\text{siren}$ -- each scalar may act as a \emph{lamp post} potentially revealing the corresponding BH population in the MW (\cref{fig:axiverse}). Thus, while it has been aptly suggested that we use BHs to learn about scalars \cite{Arvanitaki:2010sy}, this study motivates the complementary proposition: given a dense spectrum of scalars, we may use Earth-based scalar detection experiments to learn about BHs.
 
We take care not to overstate this proposition. There could be an axiverse with no BHs of the corresponding mass (in which case the lamp posts of \cref{fig:axiverse} have nothing to illuminate), or there could be exotic BH populations with no scalars at the corresponding mass (in which case, segments of the lower axis of \cref{fig:axiverse} remain unilluminated).
Our study establishes, however, that a universe with \emph{both} BHs and scalars holds meaningfully enhanced discovery potential over one where either exists independently, thereby exhibiting genuine joint phenomenology between BHs and scalars.

Future developments in scalar detection technology, the modeling of individual BH scalar sirens, and BH population studies will be crucial for realizing this joint discovery potential.

\section*{Acknowledgments}
The authors would like to thank Masha Baryakhtar, Nataniel L. Figueroa, Marios Galanis, Mariangela Lisanti, Grzegorz Łukasiewicz, Ken Van Tilburg, Zilu (Tony) Zhou for useful discussions. This work has been supported by the Cluster of Excellence ``Precision Physics, Fundamental Interactions, and Structure of Matter'' (PRISMA++ EXC 2118/2) funded by the
German Research Foundation (DFG) within the German Excellence Strategy (Project
ID 390831469), and by the COST Action within the project COSMIC WISPers
(Grant No. CA21106). OS is supported by a Fellowship from the Princeton Center for Theoretical Science. DFJK was supported by the U.S. National Science Foundation under grant PHYS-2510625. LLM artificial intelligence agents were used to assist the editing of this manuscript, but all scientific results, concepts and claims represent original work by the human authors listed.

\bibliography{Bibliography}

\clearpage

\appendix
\section{Radiated power, waveform and amplitude}
\label{app:power waveform}
In this appendix, we review the rate of energy emission in a BH scalar siren from the analysis of \cite{Baryakhtar:2020gao} and extract the amplitude of the waveform. 
Formally, the rate of the process $322\times 322\rightarrow 211 \times \infty$ is computed by looking at the flux of energy of the outgoing wave in the far away radiation zone. This in turn involves computing the overlap integral between the hydrogenic states waveforms $|211\rangle$ and $|322\rangle$ onto the outgoing ``Coulomb'' wavefunction \cite{Landau:1991wop}. In \cite{Baryakhtar:2020gao}, the (time-averaged) total outgoing power is found to be
\begin{equation}
P_{\infty} \approx GM_\text{BH}^2\mu^2 \kappa^\infty\alpha^8\left(\mpl/f\right)^4\varepsilon_{322}^2\varepsilon_{211},
\end{equation}
where 
$\kappa^\infty \approx 1.1\times 10^{-8}$ and $\varepsilon_{n\ell m}$ quantify the occupation of the states $|211\rangle$ and $|322\rangle$. In the BH siren regime, the occupations obtain their quasi-equilibrium values:
\begin{subequations}
\begin{equation}
\varepsilon_{211}^{\mathrm{eq}}
\approx
\frac{2}{\sqrt{3}}\,
\frac{\sqrt{\kappa^\infty \kappa^{\mathrm{SR}} \!\left(a_* - 2\alpha \tilde r_+\right)}}{\alpha^3 \kappa^{\mathrm{BH}} \tilde r_+}
\left(\frac{f}{\mpl}\right)^2\,,
\end{equation}
\begin{equation}
\varepsilon_{322}^{\mathrm{eq}}
\approx
\sqrt{
\frac{1}{3}\,
\frac{\kappa^{\mathrm{SR}}\!\left(a_* - 2\alpha \tilde r_+\right)}{\kappa^\infty}
}\,
\left(\frac{f}{m_{\mathrm{pl}}}\right)^2.
\end{equation}
\end{subequations}
where $\kappa^\text{SR} \approx 4\times 10^{-2}$, $\kappa^\text{BH} \approx 4.3\times 10^{-7}$ and $\tilde r_+ = 1+\sqrt{1-a_\star}$. Accordingly, the power emitted at equilibrium is
\begin{equation}
\label{eq:equilibrium power}
P^\text{eq}_\infty\approx\frac{2}{3\sqrt{3}}\,
\alpha^7\,\sqrt{\kappa^\infty}\,
\frac{\big[\kappa^{\rm SR}(a_* - 2\alpha\tilde r_+)\big]^{3/2}}
{\kappa^{\rm BH}\,\tilde r_+}\,
f^2.
\end{equation}
The total power is related to the time averaged of the squared field amplitude by $\langle \phi^2\rangle$ is related by the (non-relativistic) Poynting-flux relation 
\begin{equation}
\label{eq:Poynting}
P_\infty =\int\mu^2 \langle\phi^2\rangle v^\text{BH}_\phi r^2\mathrm d\Omega\,.
\end{equation}
By conservation of energy and momentum, in the radiation zone $\bm r \rightarrow \infty$,
\begin{equation}
\label{eq:ansatz}
\phi_\text{rad}(\bm r,t) = \frac{\mathcal N}{2}\frac{\mathcal  A(\hat r)}{r}e^{-i(\omega^\text{BH}_\phi t-\mu \bm v^\text{BH}_\phi \cdot \bm r - 3\varphi +\delta'_\text{BH})}+\text{c.c.}\,,
\end{equation}
where the normalized angular ``wavefunction" $\mathcal A(\hat r)$ defined in \cref{eq:angular_dependence} is fixed by Clebsch-Gordan decomposition of products of the spherical harmonics of the source and $\mathcal N$ is a number-valued normalization constant.
Putting the ansatz \cref{eq:ansatz} in \cref{eq:Poynting}, and comparing with \cref{eq:equilibrium power}, one recovers
\begin{equation}
\mathcal N = \alpha^2 GM_\text{BH}f\left[\frac{4}{\sqrt 3}\frac{\sqrt{\kappa^\infty}\left[\kappa^\text{SR}\right]^{3/2}}{\kappa^\text{BH}}\right]^{1/2} S(\alpha,a_\star)\,,
\end{equation}
\begin{equation}
\left[\frac{4}{\sqrt 3}\frac{\sqrt{\kappa^\infty}\left[\kappa^\text{SR}\right]^{3/2}}{\kappa^\text{BH}}\right]^{1/2} \approx 2.1.
\end{equation}
\,
\section{Frequency drifts}
\label{sec:drift}
As the schematic depiction of \cref{fig:self ionization} suggests, the state of the BH, characterized by $\{M_\text{BH}, a_\star\}$, after each ionization event remains very nearly unchanged while at quasi-equilibrium. The conservation laws demand, however, that the energy and angular momentum carried away by the scalar emission is lost by the BH. The BH numbers are therefore not truly static, but rather obtain slow time-dependence: $\{M_\text{BH}, a_\star\} \rightarrow \{M_\text{BH}(t), a_\star(t)\}$. Because the frequency (one-particle energy) of the emitted scalars is tied to the BH parameters, this in turn means that the frequency of emissions is itself slowly changing with time. Here, we show that, in the BH scalar siren regime, the frequencies of detection scalar drift negligibly over any experimentally relevant period of observation.

We go over the analysis in App.\,H of \cite{Baryakhtar:2020gao}, with a more particular focus on the phenomenological siren regime, defined as
\begin{equation}
\frac{\mathrm d \ln J_\text{BH}}{\mathrm d t} =  \frac{1}{a_\star} \frac{\dot J}{GM_\text{BH}^2} \lesssim (10\,\Gyr)^{-1}\,,
\end{equation}
where $J_\text{BH} = a_\star GM_\text{BH}^2$ is the angular momentum of the BH. This rate of angular momentum extraction acts as a ``master rate'' in the siren regime; we exhibit how all rates for drifts are even further suppressed relative to that rate.

In the siren regime, we further have
\begin{equation}
\frac{\mathrm d M_\text{BH}}{\mathrm d J_\text{BH}} = \frac{\mu}{3}\,,
\end{equation}
as follows from the mass-to-angular momentum ratio of the outgoing radiation. From this, it follows that
\begin{equation}
\frac{\mathrm d \ln \alpha}{\mathrm dt} = \frac{\dot M_\text{BH}}{M_\text{BH}} =  \frac{\alpha\,a_\star}{3}\frac{\mathrm d \ln J_\text{BH}}{\mathrm d t} \ll (10\,\Gyr)^{-1},
\end{equation}
and
\begin{equation}
\frac{\mathrm d \ln a_\star}{\mathrm d t} = \left(1-2\frac{\alpha\,a_\star}{3}\right)\frac{\mathrm d \ln J_\text{BH}}{\mathrm dt}  \ll (10\,\Gyr)^{-1}.
\end{equation}
To leading order, the energy of each bound state in the cloud is further corrected by the self-binding energy of the cloud, both from the self-interaction $\lambda$ and from self-gravity:
\begin{equation}
E_{n\ell m} = \mu + E_{n\ell m}^\alpha + E_{n\ell m}^\lambda + E^G_{n\ell m} + \dots
\end{equation}
where
\begin{equation}
E_{n\ell m}^\alpha  \propto -\frac{\alpha^2\mu}{2n}\,,
\end{equation}
\begin{equation}
E_{n\ell m}^\lambda  \propto -\mu \alpha^5 \left(\frac{\mpl}{f}\right)^2\varepsilon_{n'\ell' m'}^\text{eq}\,,
\end{equation}
and
\begin{equation}
E_{n\ell m}^G  \simeq -\mu\alpha^3 \varepsilon_{n'\ell' m'}^\text{eq}\,.
\end{equation}
From the estimates in \cite{Baryakhtar:2020gao},
\begin{subequations}
\label{eq: fractional energies}
\begin{equation}
\frac{E_{n\ell m}^\lambda}{E_{n\ell m}^\alpha}   \simeq10^{-2}\times  \begin{cases}
1 , &n\ell m = 211,\\
\alpha^3,& n \ell m = 322,
\end{cases}
\end{equation}
and
\begin{align}
\begin{split}
\frac{E_{n\ell m}^G}{E_{n\ell m}^\alpha} \simeq{}& 10^{-1}\,\alpha\,  \varepsilon_{n\ell m}^\text{eq}\,.
\end{split}
\end{align}
\end{subequations}
In the siren quasi-equilibrium regime, $\varepsilon_{n\ell m}^\text{eq} \ll 1$, by definition. Thus, both corrections to the energy are fractionally small relative to the principal kinetic energy contribution.

The frequency of radiation emitted through $322\times 322 \rightarrow 211 \times \infty$ is $\omega_\phi^\text{BH} = 2E_{322}-E_{211}$, and so has a similar decomposition
\begin{equation}
\omega_\phi^\text{BH} = \mu + E_\text{rad}^\alpha + E^\lambda_\text{rad} + E^G_\text{rad} + \dots\,,
\end{equation}
with
\begin{equation}
E^i_\text{rad} = 2E^i_{322}-E^i_{211}\,,
\end{equation}
where $i=\{\alpha, \lambda, G\}$. We are interested in the quantity
\begin{align}
\begin{split}
{}&\frac{\omega_\phi^\text{BH}}{\tilde\omega_\phi^\text{BH}}\frac{\mathrm d \ln \omega_\phi^\text{BH}}{\mathrm d t} 
\approx{} \frac{2\dot E_{322}}{E_\text{rad}^\alpha} - \frac{\dot E_{211}}{E^\alpha_\text{rad}}\\
={}& \sum_{i} \left(\frac{2E_{322}^i}{E_\text{rad}^\alpha} \frac{\partial \ln E^i_{322}}{\partial t}-\frac{E_{211}^i}{E_\text{rad}^\alpha} \frac{\partial \ln E^i_{211}}{\partial t}\right)\,.
\end{split}
\end{align}

The rate of frequency drift of the radiation is therefore set by the rate of the drift of the internal energies of the cloud, multiplied by the fractional contribution of that energy to the total kinetic energy of the radiation. That fractional contribution is generally suppressed, \cref{eq: fractional energies}, except for $i=\alpha$, which yields $E^\alpha_{n\ell m}/E_\text{rad}^\alpha = \mathcal O (1)$. In this case,
\begin{align}
\begin{split}
\frac{\mathrm d \ln E^\alpha_{n\ell m}}{\mathrm d t}  ={}& 2\frac{\mathrm d \ln \alpha}{\mathrm dt}\,.
\end{split}
\end{align}
Moreover,
\begin{align}
\begin{split}
\frac{\mathrm d \ln E^\lambda_{n\ell m}}{\mathrm d t}  ={}& 5\frac{\mathrm d\ln \alpha}{\mathrm dt} + \sum_{s\,=\,\alpha,a_\star}\frac{\mathrm d \ln \varepsilon_{n\ell m}^\text{eq}}{\mathrm d \ln s}\frac{\mathrm d\ln s}{\mathrm d t},
\end{split}
\end{align}
and similarly
\begin{align}
\begin{split}
\frac{\mathrm d \ln E^G_{n\ell m}}{\mathrm d t}  ={}& 3\frac{\mathrm d\ln \alpha}{\mathrm dt} + \sum_{s\,=\,\alpha,a_\star}\frac{\mathrm d \ln \varepsilon_{n\ell m}^\text{eq}}{\mathrm d \ln s}\frac{\mathrm d\ln s}{\mathrm d t}.
\end{split}
\end{align}
Except for BHs that whose spin is very close to the critical value, \cref{eq:a crit}, we have that
\begin{equation}
\frac{\mathrm d\ln \varepsilon_{n\ell m}^\text{eq}}{\mathrm d\ln a_\star} = \frac{1}{2}\,,
\end{equation}
while
\begin{equation}
\frac{\mathrm d\ln \varepsilon_{n\ell m}^\text{eq}}{\mathrm d \ln \alpha} = \begin{cases}
-3,{}&n\ell m = 211,\\
0,{}&\, n \ell m = 322.
\end{cases}
\end{equation}
Thus, we can conclude that the rates of the frequency drifts are at most comparable to the rate of angular momentum extraction, which is itself much longer than observational lifetimes in the siren regime.\\

\section{Ensembles with polychromatic sirens}
\label{app:polychromatic}
Consider the field of a \emph{polychromatic} siren evaluated at an observer location:
\begin{align}
\begin{split}
\phi(\bm r_\text{o}(t),t) = \sum_k s^ke^{-i\left(\mu +\tilde\omega_k^\text{BH}\right)t + i\bm k \cdot \bm r^\text{BH}_\text{o}} + \text{c.c.},
\end{split}
\end{align}
where $\{\mu+\tilde\omega_k^\text{BH}\}$ are the set of emission frequencies in the BH rest frame, and $s^k = s^k(\mu,M_\text{BH},f,a_\star;\bm r_\text{o}^\text{BH})$ is the amplitude of the different emission frequencies as a function of the siren parameters. Define the \emph{spectrum} of the $n^\text{th}$ siren in its rest frame as
\begin{equation}
    t_n(\tilde \omega) = \sum_k \left|s_n^k\right|^2 \delta(\tilde\omega-\tilde\omega_k^\text{BH}).
\end{equation}
Then, the conditional expectation value \cref{eq:conditional_expectation_value} undergoes the change
\begin{align}
\label{eq:polychromatic formula}
s_n^2\delta(\tilde\omega_\phi^\text{o}-\tilde \omega_j)\rightarrow \int \mathrm d\tilde\omega \,t_n(\tilde\omega)\delta\left(\tilde \omega^\text{o}_\phi(\tilde \omega)-\omega_j\right),
\end{align}
where $\tilde \omega^\text{o}_\phi(\tilde \omega) = \tilde \omega^\text{o}_\phi(\tilde \omega,\bm v_\text{o}^n,\bm r_\text{o}^n)$ is the Doppler relation, \cref{eq:Doppler_formula}, mapping a rest frame frequency to an observed frequency.

For the purpose of indexing, let $ \tilde\omega_0^n  \equiv \mu\alpha_n/6$ and shift the dummy variable $\tilde \omega \rightarrow \tilde\omega + \tilde \omega_0^n$. Then, \cref{eq:polychromatic formula} becomes
\begin{equation}
\int \mathrm d\tilde\omega \,\tilde t_n(\tilde \omega) \delta\left(\tilde \omega^\text{o}_\phi(\tilde \omega+\tilde \omega_0^n)-\omega_j\right),
\end{equation}
where $\tilde t(\tilde \omega) = t_n(\tilde\omega+\tilde\omega_0^n)$ is the spectrum centered on $\tilde \omega_0^n$. For small observer velocities $\bm v_\text{o}^n$, then $\tilde \omega^\text{o}_\phi(\tilde \omega) \approx \tilde \omega$, so that the expression becomes
\begin{equation}
\int \mathrm d\tilde\omega \,\tilde t_n(\tilde \omega) \delta\left(\tilde \omega + \tilde \omega_0^n-\omega_j\right).
\end{equation}
This form makes it clear then that the finite-width spectrum of the siren enters as a \emph{convolution kernel} $\tilde t_n(\tilde\omega)$ in frequency space for the equivalent, monochromatic signal normalized at the frequency $\tilde\omega_0^n$.
\end{document}